\documentclass[a4paper,11pt]{article}
\pdfoutput=1
\usepackage{jheppub}
\usepackage[T1]{fontenc}
\usepackage{xcolor}
\usepackage{booktabs}
\usepackage{multirow}
\usepackage{makecell}
\usepackage{placeins}

\newcommand{\del}{{\mathrm d}}
\newcommand{\qb}{\bar{q}}
\newcommand{\qp}{q'}
\newcommand{\qbp}{\bar{q}'}
\newcommand\id[1]{#1^{\rm id.}}

\newcommand{\XTO}[2]{\mathcal{#1}_3^{1,{\rm id.}#2}}
\newcommand{\XFZ}[2]{\mathcal{#1}_4^{0,{\rm id.}#2}}
\newcommand{\XTZ}[2]{\mathcal{#1}_3^{0,{\rm id.}#2}}
\newcommand{\XTZd}[3]{\mathcal{#1}_{3,#3}^{0,{\rm id.}#2}}
\newcommand{\XTZe}[3]{\mathcal{#1}_{4,#3}^{0,{\rm id.}#2}}
\newcommand{\XTZf}[3]{\mathcal{#1}_{3,#3}^{1,{\rm id.}#2}}

\newcommand{\e}{\epsilon}

\newcommand{\der}{\mathrm{d}}
\def\hsig{\hat{\sigma}}

\def\S{{\rm S}}

\def\T{{\rm T}}
\def\U{{\rm U}}

\newcommand{\wt}[1]{\widetilde{#1}}
\newcommand{\wh}[1]{\widehat{#1}}
\newcommand{\XIFint}[4]{\mathcal{#1}_{#2,#4}^{#3}}
\newcommand{\deltaone}{\delta_1}

\newcommand{\Sgtoq}{S_{g\to q}}
\newcommand{\Sqtog}{S_{q\to g}}

\newcommand{\boelite}{\frac{b_0}{\epsilon}}

\newcommand{\bFoelite}{\frac{b_{0,F}}{\epsilon}}

\newcommand{\QQslite}[1]{\left(\frac{|s_{#1}|}{\mu_r^2}\right)^{-\epsilon}}

\newcommand{\J}[1]{J_{2}^{(#1)}}
\newcommand{\Jh}[1]{\hat{J_{2}}^{(#1)}}
\newcommand{\Jt}[1]{\tilde{J_{2}}^{(#1)}}
\newcommand{\Jht}[1]{\hat{\tilde{J_{2}}}^{(#1)}}
\newcommand{\Jhh}[1]{\hat{\hat{J_{2}}}^{(#1)}}
\newcommand{\Jb}[1]{\bar{J_{2}}^{(#1)}}
\newcommand{\Jtb}[1]{\tilde{\bar{J_{2}}}^{(#1)}}

\newcommand{\Jic}[2]{J_{2,#2}^{(#1)}}
\newcommand{\Jhic}[2]{\hat{J}_{2,#2}^{(#1)}}
\newcommand{\Jtic}[2]{\tilde{J}_{2,#2}^{(#1)}}
\newcommand{\Jhtic}[2]{\hat{\tilde{J}}_{2,#2}^{(#1)}}

\newcommand{\Gammaoneconv}[2]{\Gamma^{(1)}_{#1}(#2)}
\newcommand{\Gammaone}[2]{\Gamma^{(1)}_{#1}\left(#2\right)}

\newcommand{\Gammatwo}[2]{\overline{\Gamma}^{(2)\text{id.}}_{#1}\left(#2\right)}

\newcommand{\GammaoneFconv}[2]{\wh{\Gamma}^{(1)}_{#1}(#2)}
\newcommand{\GammaoneF}[2]{\wh{\Gamma}^{(1)}_{#1}\left({#2}\right)}
\newcommand{\GammatwoF}[2]{\wh{\overline{\Gamma}}^{(2)\text{id.}}_{#1}\left(#2\right)}
\newcommand{\GammatwoFt}[2]{\wh{\wt{\overline{\Gamma}}}^{(2)\text{id.}}_{#1}\left(#2\right)}
\newcommand{\GammatwoFF}[2]{\wh{\wh{\overline{\Gamma}}}^{(2)\text{id.}}_{#1}\left(#2\right)}
\newcommand{\Gammatwot}[2]{\wt{\overline{\Gamma}}^{(2)\text{id.}}_{#1}\left(#2\right)}
\newcommand{\Gammatwott}[2]{\wt{\wt{\overline{\Gamma}}}^{(2)\text{id.}}_{#1}\left(#2\right)}

\newcommand{\iGammatwo}[2]{\overline{\Gamma}^{(2)}_{#1}\left(#2\right)}

\newcommand{\iGammatwoF}[2]{\wh{\overline{\Gamma}}^{(2)}_{#1}\left(#2\right)}
\newcommand{\iGammatwoFt}[2]{\wh{\wt{\overline{\Gamma}}}^{(2)}_{#1}\left(#2\right)}
\newcommand{\iGammatwoFF}[2]{\wh{\wh{\overline{\Gamma}}}^{(2)}_{#1}\left(#2\right)}
\newcommand{\iGammatwot}[2]{\wt{\overline{\Gamma}}^{(2)}_{#1}\left(#2\right)}
\newcommand{\iGammatwott}[2]{\wt{\wt{\overline{\Gamma}}}^{(2)}_{#1}\left(#2\right)}

\newcommand{\ourIop}[3]{\mathcal{I}^{(#1)}_{#2}\left(#3\right)}
\newcommand{\poles}{\mathcal{P}oles}

\newcommand{\Jfull}[1]{\mathcal{J}_{2}^{(#1)}}
\newcommand{\Jfullb}[1]{\overline{\mathcal{J}}_{2}^{(#1)}}
\newcommand{\Jfullic}[2]{\mathcal{J}_{2,#2}^{(#1)}}

\def\ba{\begin{eqnarray}}
\def\ea{\end{eqnarray}}

\def\eps{\epsilon}
\def\nn{\nonumber}
\def\b#1{\bar{#1}}

\allowdisplaybreaks

\title{Antenna subtraction for processes with identified particles at hadron colliders}
\author[a]{Leonardo Bonino,}
\author[a]{Thomas Gehrmann,}
\author[b]{Matteo Marcoli,}
\author[a]{Robin Sch\"urmann,}
\author[c]{Giovanni Stagnitto}
\affiliation[a]{Physik-Institut, Universit\"at Z\"urich,
  Winterthurerstrasse 190, CH-8057 Z\"urich, Switzerland}
\affiliation[b]{Institute for Particle Physics Phenomenology, Physics Department, Durham University, Durham, DH1 3LE, UK}
\affiliation[c]{Universit\`{a} degli Studi di Milano-Bicocca \& INFN, Piazza della Scienza 3, I-20126 Milano, Italy}
\emailAdd{leonardo.bonino@physik.uzh.ch}
\emailAdd{thomas.gehrmann@uzh.ch}
\emailAdd{matteo.marcoli@durham.ac.uk}
\emailAdd{robins@physik.uzh.ch}
\emailAdd{giovanni.stagnitto@unimib.it}


\preprint{{\raggedleft%
		IPPP/24/32\\
		ZU-TH 29/24\\
}}

\abstract{Collider processes with identified hadrons in the final state are widely studied 
 in view of determining details of the proton structure and of understanding hadronization. 
 Their theory description 
 requires the introduction of fragmentation functions, which parametrise the transition 
 of a produced parton into the identified hadron. 
To compute higher-order perturbative corrections to these processes requires a
subtraction method for 
infrared singular configurations. We extend the antenna
  subtraction method to hadron fragmentation processes in hadronic collisions
  up to next-to-next-to-leading order (NNLO) in QCD by computing the 
  required fragmentation antenna functions in initial-final kinematics. 
  The integrated antenna functions retain their dependence 
  on the momentum fractions of the incoming and fragmenting partons.  }

\begin{document} 
\maketitle
\flushbottom

\section{Introduction}
\label{sec:intro}

The transition from a parton into a hadron (fragmentation) is a non-perturbative process, whose probability is 
described by a fragmentation function (FF,~\cite{Field:1976ve,Field:1977fa}). Like parton 
distributions functions (PDFs), these FFs fulfil 
perturbative evolution equations~\cite{Altarelli:1977zs} in their resolution scales, whose 
non-perturbative initial conditions can not be determined from 
first principles. Instead, FFs are determined from global fits~\cite{Albino:2008gy,Metz:2016swz,Albino:2008fy,deFlorian:2014xna,Sato:2016wqj,Anderle:2015lqa,Bertone:2017tyb,Borsa:2022vvp,AbdulKhalek:2022laj} to data on the production of a selected hadron species 
in various collider processes. The FF evolution kernels are known~\cite{Furmanski:1980cm,Almasy:2011eq} 
to next-to-next-to-leading order (NNLO) in QCD.

In terms of hadron production observables, one distinguishes semi-inclusive observables, which are differential in 
the hadron momentum but fully inclusive in all other particles in the event, and exclusive observables, where the hadron is 
identified in final states that have been selected based on specific properties, such as the identification of a jet or a gauge boson.
 Hadron production in $e^+e^-$ annihilation or deep-inelastic scattering has typically been studied 
in terms of semi-inclusive observables. In contrast, hadron production cross sections at hadron colliders are often 
of exclusive type, such as gauge-boson-plus hadron production or identified hadron spectra inside jets. 

The coefficient functions for semi-inclusive hadron production are known to NNLO accuracy 
in compact analytical form for $e^+e^-$ 
annihilation~\cite{Altarelli:1979kv,Rijken:1996ns,Mitov:2006ic} and for deeply-inelastic lepton-proton 
scattering~\cite{Altarelli:1979kv,deFlorian:1997zj,Goyal:2023xfi,Bonino:2024qbh,Bonino:2024wgg,Goyal:2024tmo}, 
thereby allowing for the consistent inclusion of the respective data sets into global FF fits at NNLO. 

Semi-inclusive hadron production at hadron colliders~\cite{Aversa:1988vb} as well as any type of exclusive 
identified hadron production cross section can not be computed to higher orders using the methods that yielded 
the respective semi-inclusive coefficient functions in $e^+e^-$ annihilation or lepton-proton scattering. Instead, these 
processes require a numerical implementation of all parton-level subprocess contributions to a given perturbative order,
using an appropriate method to identify and recombine infrared singular contributions among different subprocesses.

 Generic subtraction methods for
NLO~\cite{Catani:1996vz,Frixione:1995ms} and
NNLO~\cite{Gehrmann-DeRidder:2005btv,Currie:2013vh,DelDuca:2016ily,Catani:2007vq,Czakon:2010td,Czakon:2014oma,Gaunt:2015pea,Cacciari:2015jma,Caola:2017dug,Magnea:2018hab,Bertolotti:2022aih,Devoto:2023rpv}
calculations are available and have been used widely for jet cross sections.
For processes involving hadron fragmentation, any subtraction method requires an
extension in order to keep track of parton momentum fractions in unresolved
emissions, which are usually integrated over. Such an extension is available at
NLO for dipole subtraction~\cite{Catani:1996vz}. At NNLO, recent work towards
fragmentation processes yielded results for heavy hadron production in top quark
decays~\cite{Czakon:2021ohs} in the residue subtraction
method~\cite{Czakon:2010td} and photon
fragmentation~\cite{Gehrmann:2022cih,Chen:2022gpk} in the antenna subtraction
method~\cite{Gehrmann-DeRidder:2005btv,Daleo:2006xa,Currie:2013vh}. 

Subsequently, 
the antenna subtraction method was extended~\cite{Gehrmann:2022pzd} 
to accommodate generic 
hadron fragmentation processes by deriving the required phase space 
factorizations and mappings and by devising the construction of the antenna subtraction terms. 
The newly introduced fragmentation antenna functions contain one parton
with identified momentum fraction in the final state and another radiator parton in the final state 
(final-final fragmentation antenna) 
or in the initial state (initial-final fragmentation antenna). The unintegrated forms of these antenna functions are 
identical to their inclusive counterparts~\cite{Gehrmann-DeRidder:2005btv}, while their integration must keep 
proper track of the identified parton's momentum fraction. 

In~\cite{Gehrmann:2022pzd}, the full set of integrated final-final fragmentation
antenna functions was derived. A subset of integrated initial-final
fragmentation antenna functions was computed in~\cite{Gehrmann:2022cih},
focusing on identified photons in the final state.
In this paper, we complete the construction of the NNLO 
antenna subtraction method for processes with identified hadrons.
We derive the full set of integrated initial-final fragmentation antenna
functions and we describe how the integrated antenna functions can be
systematically combined with mass factorization counter-terms for PDFs and FFs
to yield so-called dipole operators~\cite{Currie:2013vh,Gehrmann:2023dxm}.
Such dipole operators conveniently combine with purely virtual corrections,
thereby ensuring the infrared finiteness of the numerical implementation of all
parton-level contributions to a given observable.

In Section~\ref{sec:ant}, we briefly recapitulate the formulation of the antenna subtraction method 
for fragmentation processes~\cite{Gehrmann:2022cih,Gehrmann:2022pzd}.  The integration of the 
fragmentation antenna functions in initial-final kinematics is described in detail in Section~\ref{sec:intant},
with the relevant master integrals tabulated  in Appendix~\ref{app:MI}.  The resulting integrated antenna functions are combined with mass factorization counter-terms, described
in Appendix~\ref{app:MFkern},
to yield integrated 
fragmentation dipoles in Section~\ref{sec:dipoles}. We conclude with an outlook in Section~\ref{sec:conc}.

\section{Antenna subtraction for fragmentation processes}
\label{sec:ant}

The generic form of a cross section with an identified final-state hadron is given as convolution over parton-level 
cross sections with PDFs and FFs: 
\begin{equation}
\del \sigma^H = \sum_{i,j,p} \del \hat{\sigma}_{ij}^p(\xi_1,\xi_2,\eta,\mu_r,\mu_f,\mu_a) \otimes f_i(\xi_1,\mu_f) \otimes f_j(\xi_2,\mu_f) \otimes D^H_p(\eta,\mu_a),
\end{equation}
where $f_{i,j}$ denote the parton distributions in the two incoming hadrons, defined at 
the factorization scale $\mu_f$ and $D^H_p$ is the fragmentation function 
describing the transition of parton $p$ into hadron $H$, defined at factorization scale $\mu_a$. The parton-level 
cross section $\del \hat{\sigma}$ for the  process $i+j \to p+X$ 
contains the full definition of the 
selected final state, including specific cuts on $X$, which can select other particles (such as gauge bosons) 
accompanying the identified hadron or involve a jet reconstruction. 
It can be computed in perturbative QCD as an expansion in the renormalized coupling
constant $\alpha_s$, defined at the renormalization scale $\mu_r$:
 \begin{equation}
 \del \hat{\sigma} =  \del \hat{\sigma}_{{\rm LO}} + 
\left(\frac{\alpha_s}{2\pi}\right)  \del \hat{\sigma}_{{\rm NLO}} +  
 \left(\frac{\alpha_s}{2\pi}\right)^{2}  \del \hat{\sigma}_{{\rm NNLO}} + \ldots \;,
\end{equation}
with the leading order (Born-level) cross section
\begin{eqnarray}
\del \hat\sigma_{{\rm LO}} &= &{\cal N}_B \int \del \Phi_n(k_1,\ldots,k_p,\ldots,k_{n};p_i,p_j)  \nonumber \\ &&
\times \frac{1}{S_n}M^0_n( k_1,\ldots,k_p,\ldots,k_{n};p_i,p_j) 
J(k_1,\ldots,k_p,\ldots,k_{n};p_i,p_j;\xi_1,\xi_2,\eta)\;,
\end{eqnarray}
 where $ {\cal N}_B$ is the Born-level normalization factor (potentially containing further powers of $\alpha_s$), 
$\Phi_n$ is the Born-level phase space,
$S_n$ is the symmetry factor appropriate for the final state, $M^0$ represents the Born-level squared matrix element and 
$J$ the final state definition, based on the Born-level kinematics of 
the parton-level process $\{k_1,\ldots,k_p,\ldots,k_{n};p_i,p_j\}$ and the 
incoming ($p_i/\xi_1$, $p_j/\xi_2$) and outgoing ($\eta k_p$) hadron momenta. 

At higher orders in perturbation theory, the parton-level cross sections receive contributions from 
real and virtual corrections (or combinations thereof), which 
contain infrared singularities and 
are defined on phase-spaces of varying multiplicity ($n+i$ for $i$ real emissions relative to the Born-level process). 
These singularities are made explicit by working in dimensional regularization in $d=4-2\epsilon$ dimensions. 
The PDFs and FFs are redefined by mass factorization (absorption of initial- and final-state real radiation singularities)
 at each order. Only the 
sum of all contributions at a given order is infrared-finite and well-defined. In order to extract the infrared 
singularities from each subprocess, one typically applies a subtraction scheme which introduces
subtraction terms that enable to define finite remainders at the level of each phase space multiplicity. 

The parton-level cross section at NLO reads as follows:
\begin{eqnarray}
\del \hat{\sigma}_{{\rm NLO}} & = & \phantom{+}
\int_{n+1} \left( \del \hat{\sigma}_{{\rm NLO}}^R - \del \hat{\sigma}_{{\rm NLO}}^S 
\right) 
+ \int_{n} \left( \del \hat{\sigma}_{{\rm NLO}}^V - \del \hat{\sigma}_{{\rm NLO}}^T
\right) \,,
\label{eq:nlomaster}
\end{eqnarray}
where $\del \hat{\sigma}_{{\rm NLO}}^{R,V}$ denote the real and virtual NLO corrections to the Born-level 
process and $\del \hat{\sigma}_{{\rm NLO}}^{S,T}$ are the respective subtraction terms. These fulfil
\begin{equation}
\del \hat{\sigma}_{{\rm NLO}}^{T} =- \del \hat{\sigma}_{{\rm NLO}}^{MF}  -\int_1 \del \hat{\sigma}_{{\rm NLO}}^{S} 
\end{equation}
with the NLO mass factorization counter-term contribution $\del \hat{\sigma}_{{\rm NLO}}^{MF}$ and $\int_1$ 
denoting the analytical integration over the phase space relevant to the emission of one extra parton. 
Likewise, at NNLO we have: 
 \begin{eqnarray}
\del \hat{\sigma}_{{\rm NNLO}} & = & \phantom{+}
\int_{n+2} \left( \del \hat{\sigma}_{{\rm NNLO}}^{RR} - \del \hat{\sigma}_{{\rm NNLO}}^S 
\right)  \nonumber \\
&& 
+ \int_{n+1} \left( \del \hat{\sigma}_{{\rm NNLO}}^{RV} - \del \hat{\sigma}_{{\rm NNLO}}^T
\right) \nonumber \\
&&
+ \int_{n} \left( \del \hat{\sigma}_{{\rm NNLO}}^{VV} - \del \hat{\sigma}_{{\rm NNLO}}^U
\right)
 \,,
\label{eq:nnlomaster}
\end{eqnarray}
 with $\del \hat{\sigma}_{{\rm NNLO}}^{RR,RV,VV}$ being the double-real, real-virtual and double-virtual 
 corrections to the Born-level process and $\del \hat{\sigma}_{{\rm NNLO}}^{S,T,U}$ the respective subtraction terms.
 These are related as follows:
 \begin{eqnarray}
  \del \hat{\sigma}_{{\rm NNLO}}^S  &=&  \del \hat{\sigma}_{{\rm NNLO}}^{S,1} +  \del \hat{\sigma}_{{\rm NNLO}}^{S,2}
  \nonumber \\
     \del \hat{\sigma}_{{\rm NNLO}}^T  &=& \del \hat{\sigma}_{{\rm NNLO}}^{VS,1} 
     -\del \hat{\sigma}_{{\rm NNLO}}^{MF,1} - \int_1   \del \hat{\sigma}_{{\rm NNLO}}^{S,1}  \nonumber \\
    \del \hat{\sigma}_{{\rm NNLO}}^U  &=&   -\del \hat{\sigma}_{{\rm NNLO}}^{MF,2} - \int_1
     \del \hat{\sigma}_{{\rm NNLO}}^{VS,1} - \int_2  \del \hat{\sigma}_{{\rm NNLO}}^{S,2}\;.
 \end{eqnarray}
In here, $\del \hat{\sigma}_{{\rm NNLO}}^{MF,1}$ and $\del \hat{\sigma}_{{\rm NNLO}}^{MF,2}$
are the mass factorization terms for one and two unresolved emissions and  $\int_{1,2}$ are analytical 
integrations over the single- and double-real emission components of the phase space. 
 Each integration in (\ref{eq:nlomaster}) and (\ref{eq:nnlomaster})  
 is numerically well-defined and finite. It can be implemented in a 
 numerical parton-level event generator code that analyses each parton-level momentum configuration according 
 to the observable definition $J$. 
 
 The antenna subtraction
method~\cite{Gehrmann-DeRidder:2005btv,Daleo:2006xa,Currie:2013vh} describes a systematic procedure 
to construct the subtraction terms up to NNLO in QCD. The method is based on exact factorizations of the 
$(n+1)$ and $(n+2)$ particle phase spaces into a reduced $n$-particle phase space and an antenna phase space, which 
allows to build real radiation subtraction terms from 
 products of antenna 
functions with lower-multiplicity matrix elements, evaluated over the reduced phase space. Owing to the phase space 
factorization, each antenna function can be integrated analytically over its corresponding antenna phase space, 
thus entering the construction of the virtual subtraction terms. 
An antenna function 
encapsulates all unresolved radiation between two colour-ordered hard radiator partons. For subtraction at 
 NLO, only three-parton 
(two hard radiators, one unresolved parton) tree level 
antenna functions $X_3^0$ are required. At NNLO, these are supplemented by four-parton tree-level antenna functions 
$X_4^0$ and three-parton one-loop antenna functions $X_3^1$ as well as by large-angle single-soft antennae $S$~\cite{Currie:2013vh}.
The antenna functions depend on the parton species of the two hard radiators 
(quark-antiquark~\cite{Gehrmann-DeRidder:2004ttg}, quark-gluon~\cite{Gehrmann-DeRidder:2005svg} or 
gluon-gluon~\cite{Gehrmann-DeRidder:2005alt}) and are derived from matrix elements for the decay of a 
colour-neutral current to a three- or four-parton system. Alternatively, they can also be constructed in an 
iterative manner from the unresolved behaviour in all limits~\cite{Braun-White:2023sgd,Braun-White:2023zwd,Fox:2023bma}.
An extension of antenna subtraction to N3LO, involving up to five-parton antenna functions, has been put forward most 
recently~\cite{Jakubcik:2022zdi,Chen:2023fba,Chen:2023egx}. 
Each of the two radiators can be in the initial or in the final state, yielding different integrated antenna functions 
for final-final~\cite{Gehrmann-DeRidder:2005btv}, initial-final~\cite{Daleo:2009yj}, and 
initial-initial~\cite{Gehrmann:2011wi,Gehrmann-DeRidder:2012too} cases. 
These integrated antenna functions are fully inclusive in the final-state radiation, but differential in the
momentum fractions of the initial-state radiation. They can be combined with the mass factorization terms of the PDFs 
to yield integrated antenna dipoles~\cite{Currie:2013vh,Gehrmann:2023dxm} $J_2^1$ and $J_2^2$ that make the infrared pole structure 
of the integrated subtraction terms explicit and allow for an analytical cancellation of $\epsilon$-poles between 
subtraction terms and virtual corrections to the matrix elements. 

The extension of the antenna subtraction method to incorporate final-state hadron fragmentation requires 
to keep track of the momentum fraction of the fragmenting final-state
parton in the phase space factorization and mappings, as 
well as in the resulting integrated antenna subtraction terms. It requires the introduction of 
fragmentation antenna functions~\cite{Gehrmann:2022cih,Gehrmann:2022pzd}, 
which are differential in the momentum of one final-state parton, which is associated with a final-state radiator.
The other radiator can be either in the final state (final-final fragmentation antenna functions) or 
in the initial state (initial-final fragmentation antenna functions).  These fragmentation antenna functions appear in 
those contributions to the antenna subtraction terms that involve the fragmenting parton $p$, indicated by 
a superscript `id.$p$'. In the following we briefly illustrate the structure of the antenna subtraction terms with identified final-state particles. We aim at a contained description of the key features of each subtraction component, rather than a detailed discussion of the whole antenna subtraction method, which is thoroughly documented in~\cite{Gehrmann-DeRidder:2005btv,Currie:2013vh,Gehrmann:2023dxm}. In particular, we highlight how fragmentation antenna functions are employed up to NNLO to subtract infrared divergences associated to the emission of unresolved radiation from the fragmenting parton. Clearly, for a generic process, standard antenna functions are also required to fully remove singularities originated from soft or collinear emission from other external legs.

\subsection{Final-final configurations}
At NLO, with fragmenting parton $k_p$ and non-identified partons $k_j,k_k$ 
all in the final state, we use three-parton tree-level fragmentation antenna function $X_3^{0,{\rm id.}p}$ to construct the real subtraction term:
\begin{eqnarray}
  \der\hsig_{{\rm NLO}}^{\S,{\rm id.}p} &\supseteq & 8\pi^2 {\cal N}_{B}
  \der\Phi_{n+1}(k_{1},\ldots,k_p,k_j,k_k,\ldots,k_{n+1};p_i,p_j)\, \frac{1}{S_{{n+1}}}\,X_3^{0,{\rm id.}p}(k_p,k_j,k_k)
  \nonumber \\ & \times&
  M^{0}_{n}(k_{1},\ldots,\tilde{k}_p,\tilde{K},\ldots,k_{n+1};p_i,p_j)\nn\\
   & \times&J(k_{1},\ldots,\tilde{k}_p,\tilde{K},\ldots,k_{n+1};p_i,p_j;\xi_1,\xi_2,z\eta )\,.
  \label{eq:SNLOxs}
\end{eqnarray}
The final-final phase space mapping~\cite{Gehrmann:2022pzd} is defined by
 the momentum fraction
$z$, used to define $\tilde{k}_p = k_p/z$, and by a recoil momentum
$\tilde{K}$:
\begin{equation}
  \label{eq:NLOmap} 
  z = \frac{s_{pj}+s_{pk}}{s_{pj}+s_{pk}+s_{jk}}\,, \qquad
  \tilde{K} = k_j + k_k - (1-z) \frac{k_p}{z}\,.
\end{equation}
At NNLO for final-final kinematics, the removal of double-unresolved divergences at the real-real level is done using final-final four-parton tree-level fragmentation antenna functions
$X_4^{0,{\rm id.}p}$:
 \begin{eqnarray}
  \der\hsig_{{\rm NNLO}}^{\S,{\rm id.}p} &\supseteq & (8\pi^2)^2 {\cal N}_{B}
  \der\Phi_{n+2}(k_{1},\ldots,k_p,k_j,k_k,k_l,\ldots,k_{n+1};p_i,p_j)\, \frac{1}{S_{{n+1}}}\,X_4^{0,{\rm id.}p}(k_p,k_j,k_k,k_l)
  \nonumber \\ & \times&
  M^{0}_{n}(k_{1},\ldots,\tilde{k}_p,\tilde{K},\ldots,k_{n+1};p_i,p_j)
  \nonumber \\ & \times&J(k_{1},\ldots,\tilde{k}_p,\tilde{K},\ldots,k_{n+1};p_i,p_j;\xi_1,\xi_2,z\eta  )\,,
  \label{eq:SNNLOxs}
\end{eqnarray}
now with 
\begin{equation}
  \label{eq:NNLOmap} 
  z = \frac{s_{pj}+s_{pk}+s_{pl}}{s_{pj}+s_{pk}+s_{jk}+s_{pl}+s_{jl}+s_{kl}} \,, \qquad
  \tilde{K} = k_j + k_k + k_l - (1-z) \frac{k_p}{z}\,.
\end{equation}
The removal of single-unresolved divergences at the real-virtual level is done analogously to \eqref{eq:SNLOxs} using final-final one-loop  fragmentation antenna functions 
$X_3^{1,{\rm id.}p}$:
\begin{eqnarray}
  \der\hsig_{{\rm NNLO}}^{\T,{\rm id.}p} &\supseteq & 8\pi^2 {\cal N}_{B}
  \der\Phi_{n+1}(k_{1},\ldots,k_p,k_j,k_k,\ldots,k_{n+1};p_i,p_j)\, \frac{1}{S_{{n+1}}}\,X_3^{1,{\rm id.}p}(k_p,k_j,k_k)
  \nonumber \\ & \times&
  M^{0}_{n}(k_{1},\ldots,\tilde{k}_p,\tilde{K},\ldots,k_{n+1};p_i,p_j)
  \nonumber \\ & \times&
  J(k_{1},\ldots,\tilde{k}_p,\tilde{K},\ldots,k_{n+1};p_i,p_j;\xi_1,\xi_2,z\eta )\,.
  \label{eq:TNLOxs}
\end{eqnarray}
The integration of the final-final fragmentation antenna functions is based on the phase space 
factorizations (with $q=k_p+k_j+k_k(+k_l)$ time-like):
\begin{eqnarray}
  \der\Phi_{n+1}(k_{1},\ldots,k_p,k_j,k_k,\ldots,k_{n+1};p_i,p_j) &=&
  \der\Phi_{n}(k_1,\dots,\tilde{k}_p,\tilde{K},\dots,k_{n+1};p_i,p_j)\nonumber\\
  &\times&\frac{q^2}{2\pi}\der\Phi_{2}(k_j,k_k;q-k_p)\,z^{1-2\e}\,\der z\,, \nonumber \\ 
    \der\Phi_{n+2}(k_{1},\ldots,k_p,k_j,k_k,k_l,\ldots,k_{n+2};p_i,p_j) &=&
  \der\Phi_{n}(k_1,\dots,\tilde{k}_p,\tilde{K},\dots,k_{n+2};p_i,p_j)\nonumber\\
  & \times&\frac{q^2}{2\pi}\der\Phi_{3}(k_j,k_k,k_l;q-k_p)\,z^{1-2\e}\,\der z.
\end{eqnarray}
These lead to the integrated antenna functions 
\begin{eqnarray}
{\cal X}_3^{0,{\rm id.}p}(z) & = & \frac{1}{C(\epsilon) }\int \der\Phi_2(k_j,k_k;q-k_p) \frac{q^2}{2\pi}
  \,z^{1-2\e}\,X_3^{0,{\rm id.}p}({k}_p,k_j,k_k)\,, \\
  {\cal X}_3^{1,{\rm id.}p}(z) & = & \frac{1}{C(\epsilon) }\int \der\Phi_2(k_j,k_k;q-k_p) \frac{q^2}{2\pi}
  \,z^{1-2\e}\,X_3^{1,{\rm id.}p}({k}_p,k_j,k_k)\,, \\
  {\cal X}_4^{0,{\rm id.}p}(z) & = & \frac{1}{C(\epsilon)^2 }\int \der\Phi_3(k_j,k_k,k_l;q-k_p) \frac{q^2}{2\pi}
  \,z^{1-2\e}\,X_4^{0,{\rm id.}p}({k}_p,k_j,k_k,k_l)\,,
\end{eqnarray}
with
\begin{equation}
C(\epsilon) = \frac{\left(4\pi e^{-\gamma_E}\right)^{\epsilon}}{8 \pi^2} \, ,
\end{equation}
which were all computed in~\cite{Gehrmann:2022pzd}.

Integrated fragmentation antenna functions are employed to subtract explicit infrared singularities from virtual corrections. It is possible to organize them within the so-called integrated dipoles~\cite{Currie:2013vh,Chen:2022clm,Chen:2022ktf,Gehrmann:2023dxm}. We discuss in detail the construction and the properties of integrated dipoles in the context of antenna subtraction with fragmentation in Section~\ref{sec:dipoles}. Here we simply summarize their generic structure at NLO and NNLO to present the typical expressions for virtual subtraction terms. 

One-loop integrated dipoles $J_2^{(1)}$ contain integrated fragmentation three-parton tree-level antenna functions and NLO mass factorization kernels $\Gamma^{(1)}$:
\begin{equation}\label{eq:1loopdip}
	J_2^{(1)}(p,a)=c_{\mathcal{X}^0_3}\mathcal{X}^{0,\text{id}.p}_3+c_{\Gamma^{(1)}}\Gamma^{(1)},
\end{equation}
where $c_{\mathcal{X}^{0}_3}$ and $c_{\Gamma^{(1)}}$ are constants which depend on the specific partonic and kinematical configuration. Parton $a$ is a non-identified hard radiator. 
Since both integrated fragmentation antenna functions and splitting kernels depend on $z$, the integrated dipoles will also depend on it, even if we keep this dependence implicit for ease of notation. One-loop integrated dipoles can be used to assemble the NLO virtual subtraction term, aimed at removing the explicit infrared singularities in one-loop matrix elements:
\begin{eqnarray}\label{eq:sigTNLO}
	\der\hsig_{{\rm NLO}}^{\T,{\rm id.}p} &\supseteq & {\cal N}_{B}
	\der\Phi_{n}(k_{1},\ldots,k_{p},k_{a},\ldots,k_{n};p_i,p_j)\, \frac{1}{S_{{n}}}J_2^{(1)}(p,a)
	\nonumber \\ & \times&
	M^{0}_{n}(k_{1},\ldots,k_p,k_a,\ldots,k_{n};p_i,p_j)\,
	\nonumber \\ & \times& J(k_{1},\ldots,k_p,k_a,\ldots,k_{n};p_i,p_j;\xi_1,\xi_2,z\eta )\, .
	\label{eq:TNLOxs2}
\end{eqnarray}
The specific form of the  integrated dipole $J^{(1)}_2$ depends on the partonic species of $p$ and $a$ as encoded by its arguments in the expression above. An analogous contribution is used at partonic multiplicity  $(n+1)$ to subtract the explicit poles of the real-virtual matrix element.

Two-loop integrated dipoles $J_2^{(2)}$ have a more complicated structure:
\begin{eqnarray}\label{eq:2loopdip}
	J_2^{(2)}(p,a)&=&c_{\mathcal{X}^0_4}\mathcal{X}^{0,\text{id}.p}_4+c_{\mathcal{X}^1_3}\mathcal{X}^{1,\text{id}.p}_3+c_{\mathcal{X}^0_3\mathcal{X}^0_3}\mathcal{X}^{0,\text{id}.p}_3\otimes\mathcal{X}^{0,\text{id}.p}_3+c_{\beta_0}\dfrac{\beta_0}{\e}\mathcal{X}^{0,\text{id}.p}_3\nn\\
	&+&c_{\overline{\Gamma}^{(2)}}\overline{\Gamma}^{(2)}+c_{\Gamma^{(1)}\Gamma^{(1)}}\Gamma^{(1)}\otimes\Gamma^{(1)}+c_{\Gamma^{(1)}\mathcal{X}_3^0}\Gamma^{(1)}\otimes\mathcal{X}_3^0,
\end{eqnarray}
where, as at one-loop, all constants depend on the partonic content and the kinematical configuration of the dipole. For a complete list of two-loop mass factorization kernels $\overline{\Gamma}^{(2)}$ see e.g.~\cite{Currie:2013vh} for the space-like expressions and Appendix~\ref{app:MFkern} for the time-like ones. Two-loop integrated dipoles are needed to construct the double-virtual subtraction terms. However, to properly remove the infrared singularities of two-loop matrix elements, one also has to consider the combination of one-loop integrated dipoles with one-loop matrix-elements and the convolution of two one-loop integrated dipoles~\cite{Currie:2013vh,Gehrmann:2023dxm}. The structures present in the double-virtual subtraction term are then:
\begin{eqnarray}\label{eq:sigUNNLO1}
	\der\hsig_{{\rm NNLO}}^{\U,{\rm id.}p} &\supseteq & {\cal N}_{B}
	\der\Phi_{n}(k_{1},\ldots,k_{p},k_{a},\ldots,k_{n};p_i,p_j)\, \frac{1}{S_{{n}}}J_2^{(2)}(p,a)
	\nonumber \\ & \times&
	M^{0}_{n}(k_{1},\ldots,k_p,k_a,\ldots,k_{n};p_i,p_j)\,\nn\\
	& \times&J(k_{1},\ldots,k_p,k_a,\ldots,k_{n};p_i,p_j;\xi_1,\xi_2,z\eta )\,
	\label{eq:UNLOxs2a}
\end{eqnarray}
\begin{eqnarray}\label{eq:sigUNNLO2}
	\der\hsig_{{\rm NNLO}}^{\U,{\rm id.}p} &\supseteq & {\cal N}_{B}
	\der\Phi_{n}(k_{1},\ldots,k_{p},k_{a},\ldots,k_{n};p_i,p_j)\, \frac{1}{S_{{n}}}J_2^{(1)}(p,a)
	\nonumber \\ & \times&
	M^{1}_{n}(k_{1},\ldots,k_p,k_a,\ldots,k_{n};p_i,p_j)\,\nn \\
	& \times&J(k_{1},\ldots,k_p,k_a,\ldots,k_{n};p_i,p_j;\xi_1,\xi_2,z\eta )
	\label{eq:UNLOxs2b},
\end{eqnarray}
and
\begin{eqnarray}\label{eq:sigUNNLO3}
	\der\hsig_{{\rm NNLO}}^{\U,{\rm id.}p} &\supseteq & {\cal N}_{B}
	\der\Phi_{n}(k_{1},\ldots,k_{a},k_{b},\ldots,k_{b},k_{c},\dots,k_{n};p_i,p_j)\, \frac{1}{S_{{n}}}J_2^{(1)}(p,a)\otimes J_2^{(1)}(b,c)
	\nonumber \\ & \times&
	M^{0}_{n}(k_{1},\ldots,k_p,k_a,\ldots,k_{b},k_{c},\ldots,k_{n};p_i,p_j)\,\nn\\
	& \times&J(k_{1},\ldots,k_p,k_a,\ldots,k_{b},k_{c},\ldots,k_{n};p_i,p_j;\xi_1,\xi_2,z\eta )\,,
	\label{eq:UNLOxs2c}
\end{eqnarray}
where $b$ and $c$ can also coincide with either $p$ or $a$.

\subsection{Initial-final configurations}
For configurations with momentum $p_i$ in the initial state, fragmenting parton momentum $k_p$ and unresolved momentum $k_k$ in the final state, the NLO real subtraction term is assembled using initial-final three-parton tree-level fragmentation antenna functions $X_{3,i}^{0,{\rm id.}p}$:
\begin{eqnarray}
  \der\hsig_{{\rm NLO}}^{\S,{\rm id.}p} &\supseteq & 8\pi^2 {\cal N}_{B}
  \der\Phi_{n+1}(k_{1},\ldots,k_p,k_k,\ldots,k_{n+1};p_i,p_j)\, \frac{1}{S_{{n+1}}}\,X_{3,i}^{0,{\rm id.}p}(k_p,k_k;p_i)
  \nonumber \\ & \times&
  M^{0}_{n}(k_{1},\ldots,,\tilde{k}_p,\ldots,k_{n+1};xp_i,p_j)
  \nonumber \\ & \times&
  J(k_{1},\ldots,\tilde{k}_p,\ldots,k_{n+1};xp_i,p_j;\xi_1/x,\xi_2,z\eta )\,.
  \label{eq:SNLOxs2}
\end{eqnarray}
In here, the phase space mapping is the initial-final one~\cite{Daleo:2006xa}, with final-state momentum fraction $z$ and initial-state momentum fraction $x$ defined as
\begin{equation} 
z= \frac{s_{ip}}{s_{ip}+s_{ik}}\equiv z_3 \,, \qquad x = \frac{s_{ip}+s_{ik}-s_{pk}} {s_{ip}+s_{ik}} \,.
\label{eq:IFmapNLO}
\end{equation}
At NNLO, we use initial-final four-parton fragmentation antenna functions $X_{4,i}^{0,{\rm id.}p}$ to remove double-unresolved singularities:
\begin{eqnarray}
  \der\hsig_{{\rm NNLO}}^{\S,{\rm id.}p} &\supseteq & (8\pi^2)^2 {\cal N}_{B}
  \der\Phi_{n+2}(k_{1},\ldots,k_p,k_k,k_l,\ldots,k_{n+2};p_i,p_j)\, \frac{1}{S_{{n+2}}}\,X_{4,i}^{0,{\rm id.}p}(k_p,k_k,k_l;p_i)
  \nonumber \\ & \times&
  M^{0}_{n}(k_{1},\ldots,,\tilde{k}_p,\ldots,k_{n+2};xp_i,p_j)
  \nonumber \\ & \times&
  J(k_{1},\ldots,\tilde{k}_p,\ldots,k_{n+2};xp_i,p_j;\xi_1/x,\xi_2,z\eta )\,,
  \label{eq:SNNLOxs2}
\end{eqnarray}
with the initial-final phase space mapping at NNLO, and $x$ and $z$ defined as:
\begin{equation} 
z= \frac{s_{ip}}{s_{ip}+s_{ik}+s_{il}}\equiv z_4 \,, \qquad x = \frac{s_{ip}+s_{ik}+s_{il}-s_{pk}-s_{pl}-s_{kl}} {s_{ip}+s_{ik}+s_{il}} \,.
\label{eq:IFmapNNLO}
\end{equation}
The removal of single-unresolved divergences from the real-virtual corrections is done analogously to~(\ref{eq:SNLOxs2}) using initial-final one-loop fragmentation antenna functions $X_{3,i}^{1,{\rm id.}p}$:
\begin{eqnarray}
	\der\hsig_{{\rm NNLO}}^{\T,{\rm id.}p} &\supseteq & 8\pi^2 {\cal N}_{B}
	\der\Phi_{n+1}(k_{1},\ldots,k_p,k_k,\ldots,k_{n+1};p_i,p_j)\, \frac{1}{S_{{n+1}}}\,X_{3,i}^{1,{\rm id.}p}(k_p,k_k;p_i)
	\nonumber \\ & \times&
	M^{0}_{n}(k_{1},\ldots,,\tilde{k}_p,\ldots,k_{n+1};xp_i,p_j)\,\nn\\
	&\times&J(k_{1},\ldots,\tilde{k}_p,\ldots,k_{n+1};xp_i,p_j;\xi_1/x,\xi_2,z\eta )\,.
	\label{eq:TNNLOxs2}
\end{eqnarray}

To integrate the initial-final fragmentation antenna functions, we employ the phase space mappings
(with $q=-p_i+k_p+k_k(+k_l)$ space-like and $Q^2=-q^2$): 
\begin{eqnarray}
  \der\Phi_{n+1}(k_{1},\ldots,k_p,k_k,\ldots,k_{n+1};p_i,p_j) &=&
  \der\Phi_{n}(k_1,\dots,\tilde{k}_p,\dots,k_{n+1};xp_i,p_j)\nonumber\\
 &&\hspace{-1cm}\times\frac{\der x}{x}\frac{Q^2}{2\pi}\der\Phi_{2}(k_p,k_k;q,p_i)
  \delta\left(z- z_3\right) \der z\,, \nonumber \\ 
    \der\Phi_{n+2}(k_{1},\ldots,k_p,k_k,k_l,\ldots,k_{n+2};p_i,p_j) &=&
  \der\Phi_{n}(k_1,\dots,\tilde{k}_p,\dots,k_{n+2};xp_i,p_j)\nonumber\\
 &&\hspace{-1cm}\times\frac{\der x}{x}\frac{Q^2}{2\pi}\der\Phi_{3}(k_p,k_k,k_l;q,p_i)
  \delta\left(z-z_4\right) \der z\,, 
\end{eqnarray}
resulting in the integrated antenna functions
\begin{eqnarray}
{\cal X}_{3,i}^{0,{\rm id.}p}(x,z) & = & \frac{1}{C(\epsilon) }\int \der\Phi_2(k_p,k_k;q,p_i) \frac{Q^2}{2\pi}
  \, X_{3,i}^{0,{\rm id.}p}({k}_p,k_k;p_i) \delta (z-z_3) \,, \\\label{eq:IFcalX31}
  {\cal X}_{3,i}^{1,{\rm id.}p}(x,z) & = & \frac{1}{C(\epsilon) }\int \der\Phi_2(k_p,k_k;q,p_i) \frac{Q^2}{2\pi}
  \,X_{3,i}^{1,{\rm id.}p}({k}_p,k_k;p_i) \delta (z-z_3) \,, \\ 
  {\cal X}_{4,i}^{0,{\rm id.}p}(x,z) & = & \frac{1}{C(\epsilon)^2 }\int \der\Phi_3(k_p,k_k,k_l;q,p_i) \frac{Q^2}{2\pi}
  \,X_{4,i}^{0,{\rm id.}p}(k_p,k_k,k_l;p_i) \delta (z-z_4) \,. \label{eq:IFcalX40}
\end{eqnarray}
The method for the analytic computation of these integrated initial-final fragmentation antenna functions has been 
developed in the context of extending the antenna subtraction method to photon fragmentation processes 
in~\cite{Gehrmann:2022cih}. In the case of the ${\cal X}_{3,i}^{0,{\rm id.}p}(x,z)$ and 
$  {\cal X}_{3,i}^{1,{\rm id.}p}(x,z)$, the $\delta (z-z_3)$ trivializes the $\der\Phi_2$ phase space integration, which 
subsequently amounts only to an expansion in distributions in $(1-z)$ and $(1-x)$. This has been performed for 
all three-parton  initial-final fragmentation antenna functions in~\cite{Gehrmann:2022cih}. The integration of 
the  four-parton initial-final fragmentation  antenna
functions is described in detail in Section~\ref{sec:intant} below, where the full set of  ${\cal X}_{4,i}^{0,{\rm id.}p}(x,z)$ 
is computed.   

As for the final-final case, the subtraction of explicit singularities at the virtual level occurs by means of one- and two-loop integrated dipoles. In this case, initial-final integrated fragmentation antenna functions are combined with initial- or final-state splitting kernels to construct the integrated dipoles, but their general expressions still follow~\eqref{eq:1loopdip} and~\eqref{eq:2loopdip}. Clearly, integrated dipoles in the initial-final configuration depend on both $z$ and $x$. We refer the reader to Section~\ref{sec:dipoles} for the explicit representation of final-final and initial-final integrated dipoles. The structure of the virtual subtraction terms is identical to the one in equations~\eqref{eq:sigTNLO} and~\eqref{eq:sigUNNLO1}--\eqref{eq:sigUNNLO3}. 

\section{Integration of  initial-final  fragmentation antenna functions }
\label{sec:intant}

The initial-final phase space for three-parton fragmentation antenna functions $\mathcal{X}^{0,{\rm id.} p}_{3,i}$ is fully constrained, such that no integration is required:
\begin{eqnarray}
\mathcal{X}^{0,{\rm id.} p}_{3,i}\left(x,z\right) &=& \frac{1}{C(\epsilon)}\int {\rm d} \Phi_2(k_p,k_k;q,p_i) \, X_{3,i}^{0,{\rm id.} p} \, \frac{Q^2}{2\pi} \, \delta\left( z -  \frac{s_{ip}}{s_{ip}+s_{ik}} \right) \nonumber  \\
 &= &\frac{Q^2}{2} \frac{e^{\gamma_E \epsilon}}{\Gamma(1-\epsilon)} \left(Q^2\right)^{-\epsilon} \mathcal{J}(x,z) \, X_{3,i}^{0,{\rm id.} p}(x,z) \, ,
\label{eq:intX30IFfrag}
\end{eqnarray}
with $q^2 = (p_i-k_j-k_k)^2 = -Q^2<0$ and the Jacobian factor is given by
\begin{equation}
\mathcal{J}(x,z) = (1-x)^{-\epsilon} x^{\epsilon} z^{-\epsilon} (1-z)^{-\epsilon} \, .
\label{eq:JacPhi2}
\end{equation}
After expressing the invariants in the antenna function in terms of $x$ and $z$, all terms of the form $(1-x)^{-1-\epsilon}$ and $(1-z)^{-1-\epsilon}$ are expanded in distributions, where we use the notation
\begin{equation}
\mathcal{D}_n(u) = \left[\frac{\ln^n (1-u) }{1-u} \right]_+ \, \, , \, n \in \mathbb{N}_0 \, .
\label{eq:Dndef}
\end{equation}
The same feature also holds true for the one-loop three-parton antenna functions $\mathcal{X}^{1,{\rm id.} p}_{3,i}$, where some additional care has to be taken 
to isolate the kinematical endpoint contributions in $x=1$ and $z=1$ from the one-loop box integrals. The integration of the initial-final fragmentation antenna functions  $\mathcal{X}^{0,{\rm id.} p}_{3,i}$ and $\mathcal{X}^{1,{\rm id.} p}_{3,i}$ is described in detail 
in~\cite{Gehrmann:2022cih}. Although the main focus of~\cite{Gehrmann:2022cih} was on photon fragmentation processes, the full set of 
$\mathcal{X}^{0,{\rm id.} p}_{3,i}$ and  $\mathcal{X}^{1,{\rm id.} p}_{3,i}$ required for generic hadron fragmentation was already computed there. 

The integration of four-parton  initial-final fragmentation antenna functions  $\mathcal{X}^{0,{\rm id.} p}_{4,i}$ is also described in~\cite{Gehrmann:2022cih}, where 
only a limited set of  antenna functions relevant to photon fragmentation has been computed. To complete the computation for all antenna functions relevant 
to hadron fragmentation processes $\mathcal{X}^{0,{\rm id.} p}_{4,i}$, technical extensions of the methods outlined in~\cite{Gehrmann:2022cih} are 
required. The antenna functions relevant to photon fragmentation processes were all 
integrable at $z=1$, since the photon must always be accompanied by a quark which can not become soft. In contrast, general hadron fragmentation processes 
can be accompanied by soft gluons only, thereby requiring an expansion of the relevant fragmentation antenna functions around $z=1$ (in addition to the expansion 
around $x=1$). Another technical aspect concerns the integration of the initial-final fragmentation antenna functions over $z$, which recovers the 
inclusive initial-final antenna functions~\cite{Daleo:2009yj}. These calculations help to determine several integration constants in the 
master integrals relevant for $\mathcal{X}^{0,{\rm id.} p}_{4,i}$ and they must now account for potentially singular behaviour in one or both endpoints 
$z=0$ and $z=1$.

The kinematics of the $X^0_4$ initial-final antenna functions is given by
\begin{equation}
q+p_i \rightarrow k_p + k_l + k_k \, ,
\end{equation}
with $p_i^2=k_p^2=k_l^2=k_k^2=0$ and  $q^2= -Q^2 <0$. Fully inclusive integration over the four-particle phase space yields the 
integrated $\mathcal{X}^0_4$ antenna functions~\cite{Daleo:2009yj}:
\begin{equation}
\mathcal{X}^0_{4,i}(x) = \frac{1}{C(\epsilon)^2} \int \text{d} \Phi_3(k_p,k_k,k_l;p_i,q) \frac{Q^2}{2 \pi} X^0_{4,i} \, ,
\label{eq:def_ifantenna_qcd}
\end{equation}
with $x= {Q^2}/({2p \cdot q})$.

For fragmentation antenna functions in initial-final kinematics, the integration remains differential in the final-state momentum fraction $z$ of the 
identified parton $p$
\begin{equation}
\mathcal{X}^{0, \, {\rm id.}  p}_{4,i}(x,z) = \frac{1}{C(\epsilon)^2} \int \text{d} \Phi_3(k_p,k_k,k_l;p_i,q) \, \delta\left(z -x \frac{(p_i+k_p)^2}{Q^2} \right) \frac{Q^2}{2\pi} X^{0, \, {\rm id.}  p}_{4,i} \, ,
\label{eq:def_intphotonicantenna}
\end{equation}
where as in (\ref{eq:intX30IFfrag}), the initial-state momentum $p$ is used as a reference momentum to define $z$:
\begin{equation}
z = x\frac{(k_p + p_i)^2}{Q^2} = \frac{s_{ip}}{s_{ip} + s_{ik} + s_{il}} \, .
\end{equation}
Using the reverse unitarity relation 
\begin{equation}
2 \pi i \delta(k^2) = \frac{1}{k^2+ i \epsilon} - \frac{1}{k^2-i \epsilon} \, ,
\end{equation}
the phase space integrals (\ref{eq:def_intphotonicantenna}) are rewritten as 
$2\to 2$  three-loop-integrals in  forward scattering kinematics with four cut propagators (three on-shell conditions and the definition of $z$). These 
are amenable to standard integral reduction techniques based on integration-by-parts (IBP) 
relations~\cite{Chetyrkin:1981qh} in the Laporta 
algorithm~\cite{Laporta:2000dsw}. The resulting integrals all contain four cut propagators and up to  three linearly independent ordinary propagators. After applying momentum conservation
$k_k = q + p_i - k_p - k_l$, the following set of denominator factors appears in the antenna functions: 
\begin{eqnarray}
D_1 &=& (q-k_p)^2 \, , \nonumber \\
D_2 &=& (p_i+q-k_p)^2 \, ,\nonumber  \\
D_3 &=& (p_i-k_l)^2 \, ,\nonumber  \\
D_4 &=& (q- k_l)^2 \, ,\nonumber  \\
D_5 &=& (p_i+q-k_l)^2 \, ,\nonumber  \\
D_6 &=& (q-k_p -k_l)^2\, ,\nonumber  \\
D_7 &=& (p_i-k_p-k_l)^2 \, , \nonumber \\
D_8 &=& (k_p + k_l)^2\, , \nonumber \\
D_9 &=& k_p^2\, ,\nonumber  \\
D_{10} &=& k_l^2 \, , \nonumber \\
D_{11} &=& (q+p_i-k_p-k_l)^2 \, , \nonumber \\
D_{12} &=& (p_i-k_p)^2 + Q^2 \frac{z}{x} \, ,
\end{eqnarray}
where the cut propagators are $D_9$ to $D_{12}$. Combining the cut propagators with any subset of three linearly independent ordinary propagators 
yields an integral family, for which an IBP reduction to master integrals 
can be performed. We use the \texttt{Reduze2}~\cite{vonManteuffel:2012np} code for this task. 

The master integrals are labelled by their propagators factors (omitting the cut propagators, which we require in each integral), for example:
\begin{equation}
I[-3,7] = \frac{Q^2(2\pi)^{-2d+3}}{x}  \int \text{d}^d k_{p} \, \text{d}^d k_l  \, \delta\left(D_9\right) \, \delta\left(D_{10}\right) \, \delta\left(D_{11}\right)  \delta\left(D_{12}\right) \frac{D_3}{D_7},
\end{equation}
where a negative sign on the propagator label indicates its occurrence in the numerator. We find 12 integral families and in total 21 master integrals which are summarised in Table~\ref{tabMI}. The integral family F derives from the $I[1,3,7]$ top-level integral which is reducible to known integrals from other families. 
\begin{table}[t]
\centering
{%
\begin{tabular}{c|c|c|c|c}
family                & master     & deepest pole    & at $x=1$     &     at $z=1$                \\ \hline
\multicolumn{1}{l|}{} & $I[0]$     & $\epsilon^0$    & $(1-x)^{1-2\epsilon}$  &  $(1-z)^{1-2\epsilon}$                        \\ \hline
\multirow{2}{*}{A}    & $I[5]$     & $\epsilon^{-1}$ & $(1-x)^{-2 \epsilon}$  &  $(1-z)^{1-2\epsilon}$                               \\
                      & $I[2,3,5]$ & $\epsilon^{-2}$ & $(1-x)^{-1-2\epsilon}$ &  $(1-z)^{-1-2\epsilon}$          \\ \hline
\multirow{4}{*}{B}    & $I[7]$     & $\epsilon^0$    & $(1-x)^{1-2\epsilon}$  &  $(1-z)^{1-2\epsilon}$                     \\
                      & $I[-2,7]$  & $\epsilon^0$    & $(1-x)^{1-2\epsilon}$  &  $(1-z)^{1-2\epsilon}$                     \\
                      & $I[-3,7]$  & $\epsilon^0$    & $(1-x)^{1-2\epsilon}$  &  $(1-z)^{1-2\epsilon}$                       \\
                      & $I[2,3,7]$ & $\epsilon^{-2}$ & $(1-x)^{-2\epsilon}$   &  $(1-z)^{-1-2\epsilon}$ \\ \hline
\multirow{2}{*}{C}    & $I[5,7]$   & $\epsilon^{-1}$ & $(1-x)^{-2 \epsilon}$  &  $(1-z)^{1-2\epsilon}$ \\
                      & $I[3,5,7]$ & $\epsilon^{-2}$ & $(1-x)^{- 2\epsilon}$  &  $(1-z)^{-2\epsilon}$ \\ \hline
\multirow{3}{*}{D}     & $I[1]$   & $\epsilon^0$ & $(1-x)^{-2 \epsilon}$  &  $(1-z)^{-2\epsilon}$ \\
                    & $I[1,4]$   & $\epsilon^0$  & $(1-x)^{-2 \epsilon}$  &  $(1-z)^{-2\epsilon}$ \\
                      & $I[1,3,4]$ & $\epsilon^{-1}$ & $(1-x)^{- 2\epsilon}$  &  $(1-z)^{-1-2\epsilon}$ \\ \hline
              {E}   & $I[1,3,5]$ & $\epsilon^{-2}$ & $(1-x)^{- 2\epsilon}$  &  $(1-z)^{-1-2\epsilon}$ \\ \hline
            {G}   & $I[1,3,8]$ & $\epsilon^{-2}$ & $(1-x)^{- 2\epsilon}$  &  $(1-z)^{-1-2\epsilon}$ \\ \hline
           {H}   & $I[1,4,5]$ & $\epsilon^{-1}$ & $(1-x)^{-1- 2\epsilon}$  &  $(1-z)^{-2\epsilon}$ \\ \hline
           {I}   & $I[2,4,5]$ & $\epsilon^{-2}$ & $(1-x)^{-1- 2\epsilon}$  &  $(1-z)^{-2\epsilon}$ \\ \hline
\multirow{2}{*}{J}    & $I[4,7]$   & $\epsilon^0$ & $(1-x)^{-2 \epsilon}$  &  $(1-z)^{-2\epsilon}$ \\
                      & $I[3,4,7]$ & $\epsilon^{-1}$ & $(1-x)^{- 2\epsilon}$  &  $(1-z)^{-2\epsilon}$ \\ \hline
       {K}   & $I[3,5,8]$ & $\epsilon^{-2}$ & $(1-x)^{-1- 2\epsilon}$  &  $(1-z)^{-2\epsilon}$ \\ \hline
       {L}   & $I[4,5,7]$ & $\epsilon^{-1}$ & $(1-x)^{-1- 2\epsilon}$  &  $(1-z)^{-2\epsilon}$ \\ \hline
       {M}   & $I[4,5,8]$ & $\epsilon^{-1}$ & $(1-x)^{-1- 2\epsilon}$  &  $(1-z)^{-2\epsilon}$ \\ \hline
\end{tabular}%
}
\caption{\label{tabMI} Summary of the double-real radiation master integrals.}
\end{table}
The master integrals are calculated using differential equations~\cite{Gehrmann:1999as}
 in the two kinematic variables $x$ and $z$. The boundary conditions are fixed by integrating the solution of the differential equations over $z$ and comparing the result with the inclusive master integrals calculated in~\cite{Daleo:2009yj}. This procedure is described 
in detail in~\cite{Gehrmann:2022cih}, where the master integrals of families A and B were computed already. 

The master integrals and the antenna functions potentially contain end-point singularities in $x=1$, $z=1$ and $z=0$. These are regulated by factors $(1-x)^{-2\epsilon}$, $z^{-n \epsilon}$ and $(1-z)^{-n\epsilon}$ with $n=1,2$ that need to be retained in exact form in solving the differential equations. Their $\epsilon$-expansion 
subsequently yields distributions around the endpoints. The 
singular behaviour at $z=0$, which is relevant to match the boundary conditions onto the inclusive master integrals, requires particular attention.

The phase space integral $I[0]$ can be derived by direct integration: 
\begin{equation}
I[0] (Q^2,x,z)= N_{\Gamma} \left(Q^2\right)^{1-2\epsilon} (1-x)^{1-2\epsilon} x^{-1+2\epsilon} z^{-\epsilon} (1-z)^{1-2\epsilon} \, , 
\label{eq:I0}
\end{equation}
with 
\begin{equation}
N_{\Gamma} = \frac{2^{-5+4\epsilon} \pi^{-3+2\epsilon}\, \Gamma^2(2-\epsilon)}{\Gamma^2\left(3- 2\epsilon\right)}\, .
\end{equation}

In the limit $z\to 0$, the identified particle momentum $k_p$ becomes soft, such that propagators $D_5 = (k_p+k_k)^2$ and $D_8= (k_p+k_l)^2$ become singular. 
The computation of the master integrals that contain both these propagators, $I[3,5,8]$ (family K) and $I[4,5,8]$ (family M), 
must account for this behaviour and retain the exact dependence on the dimensional regulator at least in the $z\to 0$ limit. For both integrals, a 
naive approach of solving the differential equations in $x$ and $z$ as a Laurent expansion in $\epsilon$ 
with symbolic boundary conditions at a regular point in $z$ yields integrals 
that contain at most $\epsilon^{-1}$, while the corresponding $z$-integrated inclusive initial-final master integrals with this combination of propagators 
diverge as $\epsilon^{-3}$. 

In the following, we provide a detailed description of the computation of $I[3,5,8]=I[358]$ in a closed form in $\epsilon$. The differential 
equations for this master integral contain only $I[0]$ and  
\begin{eqnarray}
I[5](Q^2,x,z) &=& N_{\Gamma} \left(\frac{1-2\epsilon}{\epsilon} \right)^2 \left(Q^2 \right)^{-2\epsilon} (1-x)^{-2\epsilon} x^{2\epsilon} \nonumber \\
&&\times \left( z^{-\epsilon}  {}_2F_1(\epsilon,2\epsilon,1+\epsilon;z) - z^{-2\epsilon} \frac{\Gamma\left(1-2\epsilon\right)\Gamma(1+\epsilon)}{\Gamma(1-\epsilon)} \right)\, .
\label{eq:I5}
\end{eqnarray}
as subtopologies, and the differential equations in $z$ and $x$ are fully separable. Moreover, its differential equation in $x$ is homogeneous, implying that the 
$x$-dependence of  $I[358]$ factorises fully. The differential equations read:
\begin{eqnarray}
\frac{\partial I[358](Q^2,x,z)}{\partial Q^2}&=&-\frac{2(1+\epsilon)}{Q^2}\, I[358](Q^2,x,z)\,, \nonumber \\
\frac{\partial I[358](Q^2,x,z)}{\partial x}&=&\left( \frac{1+2\epsilon}{1-x} + \frac{2+2\epsilon}{x}   \right)\, I[358](Q^2,x,z) \,, \nonumber \\
\frac{\partial I[358](Q^2,x,z)}{\partial z}&=& -\frac{1+2\epsilon}{z} I[358](Q^2,x,z) -\frac{2x^3(1-2\epsilon)^2(1+z)}{(Q^2)^3(1-x)^2\epsilon z^2 (1-z)^2}  I[0](Q^2,x,z)
\nonumber \\ && +\frac{2 x^2 \epsilon}{(Q^2)^2(1-x)z^2}  I[5](Q^2,x,z) \, .
\end{eqnarray}
From the above equations, the functional dependence of $I[358]$ on $Q^2$ and on $x$ can be read off. Moreover, the $z$-differential equation can be solved 
by means of an integrating factor $z^{1+2\epsilon}$. Introducing 
\begin{equation}
I[358](Q^2,x,z)=N_{\Gamma}\left(\frac{1-2\epsilon}{\epsilon}\right)^{2}(Q^2)^{-2-2\epsilon}(1-x)^{-1-2\epsilon}x^{2+2\epsilon}z^{-1-2\epsilon} I^{\prime}[358](z) \, ,
\end{equation}
and inserting (\ref{eq:I0}),(\ref{eq:I5}) yields
\begin{eqnarray}
\frac{\partial I^{\prime}[358](z)}{\partial z}&=& -\frac{4(1-2\e)^2}{\e} z^\e (1-z)^{-1-2\e}   - \frac{2(1-2\e)^2}{\e}z^{-1+\e} (1-z)^{-2\e}  
 \nonumber \\
&& +  \frac{2(1-2\e)^2}{\e} z^{-1+\e} \,_2F_1(\e,2\e;1+\e;z) \nonumber \\ &&
-\frac{2(1-2\e)^2}{\e}\, \frac{\Gamma(1 - 2\e)\Gamma(1 + \e)}{\Gamma(1 - \e)} \, z^{-1} \;.
\end{eqnarray}
This equation can be integrated in a straightforward manner in the form of a primitive:
\begin{eqnarray}
I^{\prime}[358](z) &=&  -\frac{4(1-2\e)^2}{\e(1+\e)} z^{1+\e} \,_2F_1(1+\e,1+2\e;2+\e;z) \nonumber \\
&&   - \frac{2(1-2\e)^2}{\e^2} z^{\e}   \,_2F_1(\e,2\e;1+\e;z) \nonumber \\
&& +  \frac{2(1-2\e)^2}{\e^2} z^{\e} \,_3F_2(\e,\e,2\e;1+\e,1+\e;z) \nonumber\\
&& -\frac{2(1-2\e)^2}{\e}\, \frac{\Gamma(1 - 2\e)\Gamma(1 + \e)}{\Gamma(1 - \e)} \ln (z) + C' \;.
\label{eq:I358prime}
\end{eqnarray}
where $C'$ is the constant of integration. In (\ref{eq:I358prime}), which is exact in $\e$, we note the simultaneous appearance of $z^\e$ and $\ln z$. This observation implies that the singular 
behaviour of the master
integral $I[358](Q^2,x,z)$ at $z\to 0$ can not be expressed by an ansatz containing 
a finite number $i$ of terms of the form $z^{-1-n\e}$ with integer $n\leq i$. Moreover, a naive $\e$-expansion of  (\ref{eq:I358prime}) shows that its most 
singular piece is only $1/\e$. The a priori unknown boundary constant $C'$ can be determined by computing the  
$z$-integral of $I[358](Q^2,x,z)$ and matching it onto the known inclusive result. The inclusive 
$I_{{\rm inc}}[358](Q^2,x)$ is itself not a master integral but can be expressed in terms of the inclusive phase space $I_{{\rm inc}}[0](Q^2,x)$:
\begin{equation} 
I_{{\rm inc}}[358](Q^2,x) = \frac{3(1-2\e)(4-6\e)(2-6\e)}{\e^3}\frac{x^3}{(Q^2)^3(1-x)^2} I[0](Q^2,x) \,  ,  \label{inclusiveI358red}
\end{equation}
with 
\begin{equation}
 I_{{\rm inc}}[0](Q^2,x)= N_{\Gamma}(Q^2)^{1-2\epsilon}(1-x)^{1-2\epsilon}x^{-1+2\epsilon}\frac{\Gamma(2-2\epsilon)\Gamma(1-\epsilon)}{\Gamma(3-3\epsilon)}\,  .\label{inclusiveI0}
\end{equation}
We note that (\ref{inclusiveI358red}) diverges as $1/\e^3$. Dividing (\ref{eq:I358prime}) by 
the integrating factor $z^{1+2\epsilon}$ 
and integrating over $z$, we obtain standard integrals yielding hypergeometric functions at unit argument as well as from the last two terms:
\begin{equation}
\int_0^1 {\rm d} z \, z^{-1-2\e} = -\frac{1}{2\e}\,, \qquad \int_0^1 {\rm d} z \,\ln(z) z^{-1-2\e} =- \frac{1}{4\e^2}\,,
\end{equation}
where in particular the occurrence of a double pole in the second integral is noteworthy. By matching onto the inclusive integral, we then obtain a closed form expression:
\begin{eqnarray}
I[358](Q^2,x,z)&=& N_{\Gamma}\left(\frac{1-2\epsilon}{\epsilon}\right)^{2}(Q^2)^{-2-2\epsilon}(1-x)^{-1-2\epsilon}x^{2+2\epsilon}z^{-1-2\epsilon}\nonumber \\
&&\times \bigg( -2(1-z)^{-2\epsilon}z^{\epsilon}+2z^{\epsilon} {}_3 F_2(\epsilon,\epsilon,2\epsilon,1+\epsilon,1+\epsilon,z) \nonumber \\
&& \qquad 
-\frac{2\epsilon \Gamma(1-2\epsilon)\Gamma(1+\epsilon)}{\Gamma(1-\epsilon)}
(\pi \cot(\pi \epsilon)+ \ln (z)) \bigg) \,  .
\end{eqnarray}

 The master integral $I[458]$ has considerably more subtopologies than $I[358]$ and its
differential equations in $z$ and $x$ do not separate. After  
extracting its dominant behaviour at $z=0$,
the computation for $I[458]$ follows in principle the same steps as for $I[358]$.
It is however much more cumbersome and 
consequently less intuitive to describe.

For their insertion into the integrated  antenna functions $\mathcal{X}^{0, \, {\rm id.}  p}_{4,i}(x,z)$,
 master integrals are calculated in terms of a Laurent expansion in $\epsilon$, after factoring the relevant regulating factors in $(1-x)$ and $(1-z)$ from each integral. 
The results for the integrals are collected in Appendix \ref{app:MI} and given in computer-readable format in an ancillary file. 
The regulating factors combine with potential endpoint singularities in the 
reduction coefficients of the integrated antenna functions to master integrals, and 
are subsequently expanded in distributions. The resulting integrated initial-final fragmentation antenna functions are included as an ancillary file. 

\subsection{List of integrated antenna functions}
In this section we list all integrated antenna functions in the initial-final kinematics with an identified parton in the final state, following the notation of \eqref{eq:IFcalX31} and \eqref{eq:IFcalX40}.  These relate to  the parent unintegrated antenna,  denoting  in the 
argument list  the parton crossed to the initial state with ` $\hat{}$ ' and  the identified parton with ` $\id{}$ '. The functional 
form of these
unintegrated antenna functions is identical for all kinematical crossings and does not change if a final-state parton 
is identified. 
To obtain the minimal set of antennae we exploit the symmetries of the unintegrated antennae.  For example, if the unintegrated antenna $X^0_4(i,j,k,l)$ is symmetric under the exchange of identical partons $i$ and $j$, then the two integrated antennae $\XTZe{X}{i}{j}$ and $\XTZe{X}{j}{i}$ are also identical. We avoid listing double-real antennae which only give a finite contribution since they will not be employed in the subtraction. 

The $\mathcal{X}^0_3$ antennae are listed in Table \ref{tab:calX30} along with their symmetries.  Due to the presence of convolutions in the integrated dipoles  -- namely in the subtraction terms -- involving $\mathcal{X}^0_3$ antennae and NLO splitting kernels, the latter must be known up to $\mathcal{O}(\eps^2)$. In \cite{Gehrmann:2022cih}, most of the antennae in Table \ref{tab:calX30} were already computed, with the exception of higher orders in $\eps$ for $\XTZd{A}{q}{g}$. Furthermore, only partial results were presented for the flavour-changing part of the gluon-initiated $\mathcal{D}^0_3$ antennae.  For identity-changing dipoles in particular, both flavour-preserving and flavour-changing contributions of the gluon-initiated $\mathcal{D}^0_3$ antenna must be included.  $\XTZd{D}{q}{g}$ and $\XTZd{D}{g}{g}$ therefore contain both flavour structures.  For $\XTZd{D}{q}{g}$ the flavour-changing contribution is finite.  Again, due to the presence of convolutions in the subtraction terms, also the finite $\XTZd{G}{q}{q}$ antenna is needed.

The symmetries of the $\mathcal{X}^1_3$ antennae are the same as the ones of the corresponding $\mathcal{X}^0_3$ due to the identical external momentum configuration.  $\mathcal{A}$, $\mathcal{E}$ and $\mathcal{G}$ types have leading ($\mathcal{X}^1_3$) and sub-leading ($\tilde{\mathcal{X}}^1_3$) colour structures, as well as closed quark loop contributions ($\hat{\mathcal{X}}^1_3$).  $\mathcal{D}$ and $\mathcal{F}$ types only have leading-colour and 
closed quark loop contributions. The $\mathcal{X}^1_3$ antennae are listed in Tables \ref{tab:calX31qq}--\ref{tab:calX31gg}.

In Tables \ref{tab:calX40qq}--\ref{tab:calX40gg} we list the leading-colour $\mathcal{X}^0_4$ and subleading-colour $\wt{\mathcal{X}}^0_4$ antennae with respective symmetries. The number of antennae within the same family is determined by the number of symmetries of the integrand.  For example $C$, $D$ and $\tilde{E}$ have only one symmetry and therefore we find $7$ antennae of $\mathcal{C}$, $\mathcal{D}$ and $\wt{\mathcal{E}}$ types respectively. Of the $\mathcal{C}$-type,
only four of these are divergent and thus retained in Table \ref{tab:calX40qq}.

We notice that some identity-changing antennae, despite presenting $\eps$-poles, are not employed in the subtraction. This is the case for all four parton tree-level and three-parton one-loop gluon-initiated $\mathcal{D}$-type antennae with an identified gluon. Indeed, such configuration of identity-changing limits $g\to q$ can be rendered by quark-antiquark $\mathcal{A}$-type antennae, due to the freedom in the choice of the spectator parton. Moreover,  $\mathcal{G}$- and $\mathcal{E}$-type antennae which have the secondary quark (antiquark) in the initial state and the secondary antiquark (quark) identified do not correspond to any physical unresolved configuration and are therefore not needed for subtraction at NNLO. Finally, the $H_4^0(\id{1}_q, 2_{\qb}, 3_{\qp}, \hat{4}_{\qbp})$ antenna function only encodes a double-collinear iterated configuration, which can be targeted with the combination of two NLO antennae. Analogous observations were made in the context of antenna subtraction without identified final-state particles if the identified particle is crossed into the initial state. The full set of final-final and initial-final fragmentation antenna functions is provided as ancillary files. 

\setlength{\tabcolsep}{10pt} 
\renewcommand{\arraystretch}{1.5} 
\begin{table}[h]
	\centering\small\scalebox{0.9}{
\begin{tabular}{c c c c}
  \toprule
  & Notation & Integral of & Integrand symm. \\
  \midrule
  & & Hard radiators: quark-quark \\
  \midrule
  & $\XTZd{A}{q}{g}$ & $A_3^0(\id{1}_q, \hat{3}_g,2_{\qb})$ &\multirow{3}{*}{$1 \leftrightarrow 2$}   \\
  & $\XTZd{A}{q}{q}$ & $A_3^0(\id{1}_q, 3_g, \hat{2}_{\qb})$ & \\ 
  & $\XTZd{A}{g}{q}$ & $A_3^0(1_q, \id{3}_g,\hat{2}_{\qb})$ &  \\ 
  \midrule
  & & Hard radiators: quark-gluon \\
  \midrule
  & $\XTZd{D}{q}{g}$ & $D_3^0(\id{1}_q, 3_g, \hat{2}_{g})$ &  \multirow{3}{*}{$2 \leftrightarrow 3$} \\ 
  & $\XTZd{D}{g}{q}$ & $D_3^0(\hat{1}_{\qb}, 3_g, \id{2}_{g})$ &  \\
  & $\XTZd{D}{g}{g}$ & $D_3^0(1_q, \hat{3}_g, \id{2}_{g})$ &  \\ 
  \cmidrule{2-4}  
  & $\XTZd{E}{q}{\qp}$ & $E_3^0(\id{1}_q, 3_{\qp}, \hat{2}_{\qbp})$ & \multirow{2}{*}{$2 \leftrightarrow 3$} \\
  & $\XTZd{E}{\qp}{q}$ & $E_3^0(\hat{1}_{\qb}, \id{3}_{\qp}, 2_{\qbp})$ & \\
  \midrule
  & & Hard radiators: gluon-gluon \\
  \midrule
  &\multirow{1}{*}{$\XTZd{F}{g}{g}$} & $F_3^0(\id{1}_g, 3_g, \hat{2}_{g})$ & $1 \leftrightarrow 2$ , $1 \leftrightarrow 3$ , $2 \leftrightarrow 3$ \\
  \cmidrule{2-4}  
  & $\XTZd{G}{g}{q}$ & $G_3^0(\id{1}_g, 3_{q}, \hat{2}_{\qb})$ &  \multirow{3}{*}{$2 \leftrightarrow 3$}   \\
  & $\XTZd{G}{q}{g}$ & $G_3^0(\hat{1}_g, \id{3}_{q}, 2_{\qb})$ & \\ 
  & $\XTZd{G}{q}{q}$ & $G_3^0(1_g, \id{3}_{q}, \hat{2}_{\qb})$ &  \\ 
  \bottomrule
\end{tabular}}
\caption{Integrated tree-level three-parton antenna functions.}
\label{tab:calX30}
\end{table}

\setlength{\tabcolsep}{10pt} 
\renewcommand{\arraystretch}{1.4} 
\begin{table}[h]
	\centering\small\scalebox{0.9}{
\begin{tabular}{c c c c}
  \toprule
  & Notation & Integral of & Integrand symm. \\
  \midrule
  & & Hard radiators: quark-quark \\
  \midrule
  & $\XTZf{A}{q}{g}$ & $A_3^1(\id{1}_q, \hat{3}_g,2_{\qb})$ &\multirow{9}{*}{$1 \leftrightarrow 2$}   \\
  & $\XTZf{A}{q}{q}$ & $A_3^1(\id{1}_q, 3_g, \hat{2}_{\qb})$ & \\ 
  & $\XTZf{A}{g}{q}$ & $A_3^1(1_q, \id{3}_g,\hat{2}_{\qb})$ &  \\ 
  \cmidrule{2-3} 
  & $\XTZf{\wt{A}}{q}{g}$ & $\wt{A}_3^1(\id{1}_q, \hat{3}_g,2_{\qb})$ &  \\
  & $\XTZf{\wt{A}}{q}{q}$ & $\wt{A}_3^1(\id{1}_q, 3_g, \hat{2}_{\qb})$ & \\ 
  & $\XTZf{\wt{A}}{g}{q}$ & $\wt{A}_3^1(1_q, \id{3}_g,\hat{2}_{\qb})$ &  \\ 
  \cmidrule{2-3} 
  & $\XTZf{\wh{A}}{q}{g}$ & $\wh{A}_3^1(\id{1}_q, \hat{3}_g,2_{\qb})$ &  \\
  & $\XTZf{\wh{A}}{q}{q}$ & $\wh{A}_3^1(\id{1}_q, 3_g, \hat{2}_{\qb})$ & \\ 
  & $\XTZf{\wh{A}}{g}{q}$ & $\wh{A}_3^1(1_q, \id{3}_g,\hat{2}_{\qb})$ &  \\ 
  \bottomrule
\end{tabular}}
\caption{Integrated quark-quark one-loop three-parton antenna functions.}
\label{tab:calX31qq}
\end{table}

\setlength{\tabcolsep}{10pt} 
\renewcommand{\arraystretch}{1.4} 
\begin{table}[h]
	\centering\small\scalebox{0.9}{
\begin{tabular}{c c c c}
  \toprule
  & Notation & Integral of & Integrand symm. \\
  \midrule
  & & Hard radiators: quark-gluon \\
  \midrule
  & $\XTZf{D}{q}{g}$ & $D_3^1(\id{1}_q, 3_g, \hat{2}_{g})$ &  \multirow{6}{*}{$2 \leftrightarrow 3$} \\ 
  & $\XTZf{D}{g}{q}$ & $D_3^1(\hat{1}_{\qb}, 3_g, \id{2}_{g})$ &  \\
  & $\XTZf{D}{g}{g}$ & $D_3^1(1_q, \hat{3}_g, \id{2}_{g})$ &  \\
  \cmidrule{2-3}  
  & $\XTZf{\wh{D}}{q}{g}$ & $\wh{D}_3^1(\id{1}_q, 3_g, \hat{2}_{g})$ & \\ 
  & $\XTZf{\wh{D}}{g}{q}$ & $\wh{D}_3^1(\hat{1}_{\qb}, 3_g, \id{2}_{g})$ &  \\
  & $\XTZf{\wh{D}}{g}{g}$ & $\wh{D}_3^1(1_q, \hat{3}_g, \id{2}_{g})$ &  \\
  \cmidrule{2-4}  
  & $\XTZf{E}{q}{\qp}$ & $E_3^1(\id{1}_q, 3_{\qp}, \hat{2}_{\qbp})$ & \multirow{9}{*}{$2 \leftrightarrow 3$} \\
  & $\XTZf{E}{\qp}{q}$ & $E_3^1(\hat{1}_{\qb}, \id{3}_{\qp}, 2_{\qbp})$ & \\
   & $\XTZf{E}{\qp}{\qp}$ & $E_3^1(1_{\qb}, \id{3}_{\qp}, \hat{2}_{\qbp})$ & \\
  \cmidrule{2-3} 
  & $\XTZf{\wt{E}}{q}{\qp}$ & $\wt{E}_3^1(\id{1}_q, 3_{\qp}, \hat{2}_{\qbp})$ & \\
  & $\XTZf{\wt{E}}{\qp}{q}$ & $\wt{E}_3^1(\hat{1}_{\qb}, \id{3}_{\qp}, 2_{\qbp})$ & \\
  & $\XTZf{\wt{E}}{\qp}{\qp}$ & $\wt{E}_3^1(1_{\qb}, \id{3}_{\qp}, \hat{2}_{\qbp})$ & \\
  \cmidrule{2-3} 
  & $\XTZf{\wh{E}}{q}{\qp}$ & $\wh{E}_3^1(\id{1}_q, 3_{\qp}, \hat{2}_{\qbp})$ & \\
  & $\XTZf{\wh{E}}{\qp}{q}$ & $\wh{E}_3^1(\hat{1}_{\qb}, \id{3}_{\qp}, 2_{\qbp})$ & \\
  & $\XTZf{\wh{E}}{\qp}{\qp}$ & $\wh{E}_3^1(1_{\qb}, \id{3}_{\qp}, \hat{2}_{\qbp})$ & \\
  \bottomrule
\end{tabular}}
\caption{Integrated quark-gluon one-loop three-parton antenna functions.}
\label{tab:calX31qg}
\end{table}

\setlength{\tabcolsep}{10pt} 
\renewcommand{\arraystretch}{1.4} 
\begin{table}[h]
	\centering\small\scalebox{0.9}{
\begin{tabular}{c c c c}
  \toprule
  & Notation & Integral of & Integrand symm. \\
  \midrule
  & & Hard radiators: gluon-gluon \\
  \midrule
  &$\XTZf{F}{g}{g}$ & $F_3^1(\id{1}_g, 3_g, \hat{2}_{g})$ & \multirow{2}{*}{ $1 \leftrightarrow 2$ , $1 \leftrightarrow 3$ , $2 \leftrightarrow 3$} \\
  \cmidrule{2-3} 
  &$\XTZf{\wh{F}}{g}{g}$ & $\wh{F}_3^1(\id{1}_g, 3_g, \hat{2}_{g})$ &  \\
  \cmidrule{2-4}  
  & $\XTZf{G}{g}{q}$ & $G_3^1(\id{1}_g, 3_{q}, \hat{2}_{\qb})$ &  \multirow{9}{*}{$2 \leftrightarrow 3$}   \\
  & $\XTZf{G}{q}{g}$ & $G_3^1(\hat{1}_g, \id{3}_{q}, 2_{\qb})$ & \\ 
  & $\XTZf{G}{q}{q}$ & $G_3^1(1_g, \id{3}_{q}, \hat{2}_{\qb})$ & \\ 
   \cmidrule{2-3}
  & $\XTZf{\wt{G}}{g}{q}$ & $\wt{G}_3^1(\id{1}_g, 3_{q}, \hat{2}_{\qb})$ &  \\
  & $\XTZf{\wt{G}}{q}{g}$ & $\wt{G}_3^1(\hat{1}_g, \id{3}_{q}, 2_{\qb})$ & \\ 
  & $\XTZf{\wt{G}}{q}{q}$ & $\wt{G}_3^1(1_g, \id{3}_{q}, \hat{2}_{\qb})$ & \\ 
   \cmidrule{2-3}
  & $\XTZf{\wh{G}}{g}{q}$ & $\wh{G}_3^1(\id{1}_g, 3_{q}, \hat{2}_{\qb})$ &  \\
  & $\XTZf{\wh{G}}{q}{g}$ & $\wh{G}_3^1(\hat{1}_g, \id{3}_{q}, 2_{\qb})$ & \\
  & $\XTZf{\wh{G}}{q}{q}$ & $\wh{G}_3^1(1_g, \id{3}_{q}, \hat{2}_{\qb})$ & \\  
  \bottomrule
\end{tabular}}
\caption{Integrated gluon-gluon one-loop three-parton antenna functions.}
\label{tab:calX31gg}
\end{table}

\begin{table}[h]
	\centering\small\scalebox{0.9}{
\begin{tabular}{c c c c}
  \toprule
  & Notation & Integral of & Integrand symm. \\
  \midrule
  & & Hard radiators: quark-quark \\
  \midrule
  & $\XTZe{A}{q}{q}$ & $A_4^0(\id{1}_q, 3_g, 4_g, \hat{2}_{\qb})$ &  \multirow{6}{*}{$1\leftrightarrow 2$ + $3\leftrightarrow 4$} \\ 
  & $\XTZe{A}{q}{g_3}$ & $A_4^0(\id{1}_q, \hat{3}_g, 4_g, 2_{\qb})$ & \\
  & $\XTZe{A}{q}{g_4}$ & $A_4^0(\id{1}_q, 3_g, \hat{4}_g,2_{\qb})$ & \\
  & $\XTZe{A}{g_3}{q}$ & $A_4^0(1_q, \id{3}_g, 4_g, \hat{2}_{\qb})$ & \\
  & $\XTZe{A}{g_4}{q}$ & $A_4^0(1_q, 3_g,\id{4}_g, \hat{2}_{\qb})$ & \\
  & $\XTZe{A}{g}{g}$ & $A_4^0(1_q, \id{3}_g, \hat{4}_g, 2_{\qb})$ & \\
  \cmidrule{2-4}
  & $\XTZe{\wt{A}}{q}{q}$ & $\wt{A}_4^0(\id{1}_q, 3_g, 4_g, \hat{2}_{\qb})$ &  \multirow{4}{*}{$1\leftrightarrow 2$ , $3\leftrightarrow 4$} \\ 
  & $\XTZe{\wt{A}}{q}{g}$ & $\wt{A}_4^0(\id{1}_q, \hat{3}_g, 4_g, 2_{\qb})$ & \\ 
  & $\XTZe{\wt{A}}{g}{q}$ & $\wt{A}_4^0(1_q, \id{3}_g, 4_g, \hat{2}_{\qb})$ & \\ 
  & $\XTZe{\wt{A}}{g}{g}$ & $\wt{A}_4^0(1_q, \id{3}_g, \hat{4}_g, 2_{\qb})$ & \\ 
  \cmidrule{2-4}
  & $\XTZe{B}{q}{q}$   & $B_4^0(\id{1}_q, 3_{\qp}, 4_{\qbp}, \hat{2}_{\qb})$ & \multirow{3}{*}{$1\leftrightarrow 2$ , $3\leftrightarrow 4$} \\ 
  & $\XTZe{B}{q}{\qp}$ & $B_4^0(\id{1}_q, 3_{\qp}, \hat{4}_{\qbp}, 2_{\qb})$ &  \\  
  & $\XTZe{B}{\qp}{q}$ & $B_4^0(1_q, \id{3}_{\qp}, 4_{\qbp}, \hat{2}_{\qb})$ &  \\  
  \cmidrule{2-4}
  & $\XTZe{C}{q_1}{q_2}$ & $C_4^0(\id{1}_q, 3_{q}, 4_{\qb}, \hat{2}_{\qb})$ & \multirow{4}{*}{$2\leftrightarrow 4$}  \\ 
  & $\XTZe{C}{q_1}{q_3}$ & $C_4^0(\id{1}_q, \hat{3}_{q}, 4_{\qb}, 2_{\qb})$ &  \\     
  & $\XTZe{C}{\qb_2}{\qb_1}$ & $C_4^0(\hat{1}_q, 3_{q}, 4_{\qb}, \id{2}_{\qb})$ &  \\     
  & $\XTZe{C}{q_3}{\qb_1}$ & $C_4^0(\hat{1}_q, \id{3}_{q}, 4_{\qb}, 2_{\qb})$ &  \\  
  \bottomrule
\end{tabular}}
\caption{Integrated quark-quark tree-level four-parton antenna functions. The notation $a\leftrightarrow b+c\leftrightarrow d$ indicates that the antenna is symmetric under \textit{simultaneous} exchange of $a$ with $b$ and $c$ with $d$.}
\label{tab:calX40qq}
\end{table}

\begin{table}[h]
	\centering\small\scalebox{0.9}{
\begin{tabular}{c c c c}
  \toprule
   & Notation & Integral of & Integrand symm. \\
   \midrule  
   & & Hard radiators : quark-gluon\\
  \midrule  
  & $\XTZe{D}{q}{g_2}$  & $D_4^0(\id{1}_q, \hat{2}_g, 3_g, 4_g)$ & \multirow{7}{*}{$2\leftrightarrow 4$} \\ 
  & $\XTZe{D}{q}{g_3}$  & $D_4^0(\id{1}_q,2_g, \hat{3}_g, 4_g)$ & \\  
  & $\XTZe{D}{g_2}{q}$  & $D_4^0(\hat{1}_{\qb}, \id{2}_g, 3_g, 4_g)$ & \\ 
  & $\XTZe{D}{g_2}{g_3}$  & $D_4^0(1_q, \id{2}_g, \hat{3}_g, 4_g)$ & \\ 
  & $\XTZe{D}{g_2}{g_4}$  & $D_4^0(1_q, \id{2}_g, 3_g, \hat{4}_g)$ & \\ 
  & $\XTZe{D}{g_3}{q}$  & $D_4^0(\hat{1}_{\qb}, 2_g, \id{3}_g, 4_g)$ & \\ 
  & $\XTZe{D}{g_3}{g_2}$  & $D_4^0(1_q, \hat{2}_g, \id{3}_g, 4_g)$ & \\ 
  \cmidrule{2-4}
  & $\XTZe{E}{q}{\qbp}$ & $E_4^0(\id{1}_q, \hat{2}_{\qp}, 3_{\qbp}, 4_g)$ & \multirow{12}{*}{No symm.} \\
  & $\XTZe{E}{q}{\qp}$ & $E_4^0(\id{1}_q,2_{\qp}, \hat{3}_{\qbp}, 4_g)$ & \\
  & $\XTZe{E}{q}{g}$ & $E_4^0(\id{1}_q, 2_{\qp}, 3_{\qbp}, \hat{4}_g)$ & \\
  & $\XTZe{E}{\qp}{q}$ & $E_4^0(\hat{1}_{\qb}, \id{2}_{\qp}, 3_{\qbp}, 4_g)$ & \\
  & $\XTZe{E}{\qp}{\qp}$ & $E_4^0(1_q, \id{2}_{\qp}, \hat{3}_{\qbp}, 4_g)$ & \\
  & $\XTZe{E}{\qp}{g}$ & $E_4^0(1_q, \id{2}_{\qp}, 3_{\qbp}, \hat{4}_g)$ & \\
  & $\XTZe{E}{\qbp}{q}$ & $E_4^0(\hat{1}_{\qb}, 2_{\qp}, \id{3}_{\qbp}, 4_g)$ & \\
  & $\XTZe{E}{\qbp}{\qbp}$ & $E_4^0(1_q, \hat{2}_{\qp},\id{3}_{\qbp}, 4_g)$ & \\
  & $\XTZe{E}{\qbp}{g}$ & $E_4^0(1_q, 2_{\qp}, \id{3}_{\qbp}, \hat{4}_g)$ & \\
  & $\XTZe{E}{g}{q}$ & $E_4^0(\hat{1}_{\qb}, 2_{\qp}, 3_{\qbp}, \id{4}_g)$ & \\
  & $\XTZe{E}{g}{\qbp}$ & $E_4^0(1_q, \hat{2}_{\qp}, 3_{\qbp}, \id{4}_g)$ & \\
  & $\XTZe{E}{g}{\qp}$ & $E_4^0(1_q, 2_{\qp}, \hat{3}_{\qbp}, \id{4}_g)$ & \\
  \cmidrule{2-4}
  & $\XTZe{\wt{E}}{q}{\qp}$ & $\wt{E}_4^0(\id{1}_q, 2_{\qp}, \hat{3}_{\qbp}, 4_g)$ & \multirow{7}{*}{$2\leftrightarrow 3$} \\
  & $\XTZe{\wt{E}}{q}{g}$ & $\wt{E}_4^0(\id{1}_q, 2_{\qp}, 3_{\qbp}, \hat{4}_g)$ & \\
  & $\XTZe{\wt{E}}{\qp}{q}$ & $\wt{E}_4^0(\hat{1}_{\qb}, \id{2}_{\qp}, 3_{\qbp}, 4_g)$ & \\
  & $\XTZe{\wt{E}}{\qp}{\qp}$ & $\wt{E}_4^0(1_q, \id{2}_{\qp}, \hat{3}_{\qbp}, 4_g)$ & \\
  & $\XTZe{\wt{E}}{\qp}{g}$ & $\wt{E}_4^0(1_q, \id{2}_{\qp}, 3_{\qbp}, \hat{4}_g)$ & \\
  & $\XTZe{\wt{E}}{g}{q}$ & $\wt{E}_4^0(\hat{1}_{\qb}, 2_{\qp}, 3_{\qbp}, \id{4}_g)$ & \\
  & $\XTZe{\wt{E}}{g}{\qp}$ & $\wt{E}_4^0(1_q, 2_{\qp}, \hat{3}_{\qbp}, \id{4}_g)$ & \\
  \bottomrule
\end{tabular}}
\caption{Integrated quark-gluon tree-level four-parton antenna functions. }
\label{tab:calX40qg}
\end{table}

\begin{table}[h]
	\centering\small\scalebox{0.9}{
\begin{tabular}{c c c c}
  \toprule
  & Notation & Integral of & Integrand symm. \\
  \midrule
  & & Hard radiators: gluon-gluon  & \\
  \midrule
  & $\XTZe{F}{g_1}{g_2}$ & $F_4^0(\id{1}_g, \hat{2}_g, 3_g,4_g)$ & \multirow{2}{*}{$1\leftrightarrow 3$ , $2\leftrightarrow 4$ , $1\leftrightarrow 2$ + $3\leftrightarrow 4$} \\
  & $\XTZe{F}{g_1}{g_3}$ & $F_4^0(\id{1}_g, 2_g, \hat{3}_g, 4_g)$ & \\
  \cmidrule{2-4}  
  & $\XTZe{G}{g}{g}$ & $G_4^0(\id{1}_g, 3_{q}, 4_{\qb}, \hat{2}_g)$ & \multirow{6}{*}{$1\leftrightarrow 2$ + $3\leftrightarrow 4$} \\ 
  & $\XTZe{G}{g_1}{q}$ & $G_4^0(\id{1}_g, 3_{q}, \hat{4}_{\qb}, 2_g)$ & \\ 
  & $\XTZe{G}{g_2}{q}$ & $G_4^0(1_g, 3_{q}, \hat{4}_{\qb}, \id{2}_g)$ & \\ 
  & $\XTZe{G}{q}{g_1}$ & $G_4^0(\hat{1}_g, \id{3}_{q}, 4_{\qb}, 2_g)$ & \\ 
  & $\XTZe{G}{q}{q}$ & $G_4^0(1_g, \id{3}_{q}, \hat{4}_{\qb}, 2_g)$ & \\
  & $\XTZe{G}{q}{g_2}$ & $G_4^0(1_g, \id{3}_{q}, 4_{\qb}, \hat{2}_g)$ & \\
  \cmidrule{2-4} 
  & $\XTZe{\wt{G}}{g}{q}$ & $\wt{G}_4^0(\id{1}_g, 3_{q}, \hat{4}_{\qb}, 2_g)$ & \multirow{4}{*}{$1\leftrightarrow 2$ , $3\leftrightarrow 4$} \\
  & $\XTZe{\wt{G}}{g}{g}$ & $\wt{G}_4^0(\hat{1}_g, 3_{q}, 4_{\qb}, \id{2}_g)$ &  \\
  & $\XTZe{\wt{G}}{q}{g}$ & $\wt{G}_4^0(\hat{1}_g, \id{3}_{q}, 4_{\qb}, 2_g)$ & \\  
  & $\XTZe{\wt{G}}{q}{q}$ & $\wt{G}_4^0(1_g, \id{3}_{q}, \hat{4}_{\qb}, 2_g)$ & \\    
  \cmidrule{2-4}  
  & $\XTZe{H}{q}{q}$ & $H_4^0(\id{1}_q, \hat{2}_{\qb}, 3_{\qp}, 4_{\qbp})$ & \multirow{2}{*}{$1\leftrightarrow 2$ , $3\leftrightarrow 4$ , $1\leftrightarrow 3$ + $2\leftrightarrow 4$} \\
  & $\XTZe{H}{q}{\qp}$ & $H_4^0(\id{1}_q, 2_{\qb}, 3_{\qp}, \hat{4}_{\qbp})$ & \\
  \bottomrule
\end{tabular}}
\caption{Integrated gluon-gluon tree-level four-parton antenna functions.}
\label{tab:calX40gg}
\end{table}
\FloatBarrier

\section{Mass factorization of initial- and final-state collinear singularities}
\label{sec:dipoles}

The integrated fragmentation antenna functions can be collected into so-called integrated dipoles, which allow for a natural organization of infrared singularities at the double-virtual level~\cite{Currie:2013vh,Chen:2022clm,Chen:2022ktf,Gehrmann:2023dxm}. It is noticeable that the assembly and the properties of integrated dipoles in the context on fragmentation antenna functions fully match the ones of standard antenna functions, suggesting that the overall antenna subtraction infrastructure in the presence of identified particles in the final state should not differ from the one without fragmentation. For this reason, the presentation of the integrated dipoles closely aligns with the discussion in~\cite{Gehrmann:2023dxm}, to which we refer for an in-depth illustration of characteristic structures which are summarized in the following. 

We distinguish between identity-preserving (IP) integrated dipoles, reproducing the infrared singularity structure of virtual corrections, and identity-changing (IC) integrated dipoles, addressing identity-changing initial and final-state collinear singularities. For fragmentation processes, we are interested in FF and IF integrated dipoles. To resort to a more compact notation, in the following we relabel $z$, the momentum fraction carried by the identified final-state parton, as $x_3$. Consistently, we label with $3$ the identified final-state parton in the list of external momenta, to distinguish from $1$ and $2$ used for initial-state partons. All integrated antenna functions depend therefore on the three momentum fractions $x_1$, $x_2$ and $x_3$ ($=z$), but here we keep this dependence implicit. We note that for FF fragmentation antenna functions, the dependence on $x_1$ and $x_2$ is trivially given by $\delta(1-x_1) \delta(1-x_2)$ and for IF fragmentation antenna function the $x_2$ dependence is trivially given by $ \delta(1-x_2)$. We also introduce the shorthand notation $\delta_i=\delta(1-x_i)$. In the presence of identity-changing configurations we use the subscripts $(a\to b)$ and $(a\gets b)$ to indicate a change of parton species from $a$ to $b$ respectively in the initial or final state.

\subsection{One-loop integrated dipoles}
We first consider one-loop integrated dipoles. They are given as a combination of integrated three-parton tree-level fragmentation antenna functions and NLO mass factorisation kernels. The general structure in the IP case reads:
\begin{eqnarray}
	\Jfull{1}(q,\bar{q})&=&\J{1}(q,\bar{q})\, , \\
	\Jfull{1}(i,g)&=&\J{1}(i,g)+\dfrac{N_f}{N_c}\Jh{1}(i,g),\quad i=q,g\,.
\end{eqnarray} 
The colour decomposition of the IC dipoles follows the one of splitting kernels at NLO:
\begin{eqnarray}
	\Jfullic{1}{q\to g}(g,i)&=&\left(\dfrac{N_c^2-1}{N_c}\right)\Jic{1}{q\to g}(g,i),\quad i=q,g\,, \\
	\Jfullic{1}{g\to q}(q,i)&=&\Jic{1}{g\to q}(q,i),\quad i=q,g\,\,,
\end{eqnarray}
and analogously for IC configurations on the final-state identified leg, simply obtained by inverting the direction of the arrow, for any kinematical configuration (FF or IF).
To properly adjust the spin-averaging factor for initial-state identity-changing collinear limits, we introduce~\cite{Currie:2013vh,Gehrmann:2023dxm}:
\begin{equation}
	\Sgtoq = \dfrac{2-2\e}{2}=1-\e \,,\quad\Sqtog=\dfrac{2}{2-2\e}=\dfrac{1}{1-\e} \, .
\end{equation}

In Tables \ref{tab:J21IPqqb}--\ref{tab:J21IPgg} we list the identity-preserving integrated dipoles and in Tables \ref{tab:J21ICqqb}--\ref{tab:J21ICgg} the identity-changing ones.

	\begingroup
	\renewcommand{\arraystretch}{1.8} 
	\begin{table}[t]
		\centering\small\scalebox{0.9}{
		\begin{tabular}{c|l}
			& Integrated dipoles
			\\ \hline
			
			\multirow{1}{*}{FF}  
			& $\J{1}(3_q,i_{\bar{q}})=\XTZ{A}{q}(s_{3i})-\Gammaone{qq}{x_3}\delta_1$  \\\cline{2-2}
			\hline
			
			\multirow{1}{*}{IF}  
			& $\J{1}(1_q,3_q)=\XTZd{A}{q}{q}(s_{13})-\Gammaone{qq}{x_1}\delta_3-\Gammaone{qq}{x_3}\delta_1$  \\\cline{2-2}
			\hline
		\end{tabular}}
		\caption{Identity-preserving quark-antiquark one-loop colour-stripped integrated dipoles.The subscripts indicate the identified final-state parton.}
		\label{tab:J21IPqqb}
	\end{table}
	\endgroup
	
	\begingroup
	\renewcommand{\arraystretch}{1.8} 
	\begin{table}[t]
		\centering\small\scalebox{0.9}{
		\begin{tabular}{c|l}
			& Integrated dipoles
			\\ \hline
			
			\multirow{2}{*}{FF$^{q}$} 
			& $\J{1}(3_q,i_g)=\frac{1}{2}\XTZ{D}{q}(s_{3i})-\Gammaone{qq}{x_3}\delta_1$  \\\cline{2-2}
			& $\Jh{1}(3_q,i_g)=\frac{1}{2}\XTZ{E}{q}(s_{3i})$  \\\cline{2-2}
			\hline
			
			\multirow{2}{*}{FF$^{g}$}
			& $\J{1}(3_g,i_q)=\XTZd{D}{g}{g\to g}(s_{3i})-\frac{1}{2}\Gammaone{gg}{x_3}\delta_1 $  \\\cline{2-2}
			& $\Jh{1}(3_g,i_q)=-\frac{1}{2}\GammaoneF{gg}{x_3}\delta_1$  \\\cline{2-2}
			\hline
			
			\multirow{2}{*}{IF$^g_q$}			
			& $\J{1}(1_q,3_g)=\XTZd{D}{g}{q}(s_{13})-\Gammaone{qq}{x_1}\delta_3-\frac{1}{2}\Gammaone{gg}{x_3}\delta_1$  \\\cline{2-2}
			& $\Jh{1}(1_q,3_g)=-\frac{1}{2}\GammaoneF{gg}{x_3}\delta_1$  \\\cline{2-2}
			\hline
			
			\multirow{2}{*}{IF$^q_g$}			
			& $\J{1}(3_q,1_g)=\XTZd{D}{q}{g}(s_{13})-\Gammaone{qq}{x_3}\delta_1-\frac{1}{2}\Gammaone{gg}{x_1}\delta_3$  \\\cline{2-2}
			& $\Jh{1}(3_q,1_g)=-\frac{1}{2}\GammaoneF{gg}{x_1}\delta_3$  \\\cline{2-2}
			\hline
		\end{tabular}}
		\caption{Identity-preserving quark-gluon and gluon-quark one-loop colour-stripped integrated dipoles. }
		\label{tab:J21IPqg}
	\end{table}
	\endgroup
	
	\begingroup
	\renewcommand{\arraystretch}{1.8} 
	\begin{table}[t]
		\centering\small\scalebox{0.9}{
		\begin{tabular}{c|l}
			& Integrated dipoles
			\\ \hline
			
			\multirow{2}{*}{FF}  
			& $\J{1}(3_g,i_g)=\frac{1}{2}\XTZ{F}{g}(s_{3i})-\frac{1}{2}\Gammaone{gg}{x_3}\delta_1$  \\\cline{2-2}
			& $\Jh{1}(3_g,i_g)=\frac{1}{2}\XTZ{G}{g}(s_{3i})-\frac{1}{2}\GammaoneF{gg}{x_3}\delta_1$  \\\cline{2-2}
			\hline
			
			\multirow{2}{*}{IF} 
			& $\J{1}(1_g,3_g)=\XTZd{F}{g}{g}(s_{13})-\frac{1}{2}\Gammaone{gg}{x_1}\delta_3-\frac{1}{2}\Gammaone{gg}{x_3}\delta_1$  \\\cline{2-2}
			& $\Jh{1}(1_g,3_g)=-\frac{1}{2}\GammaoneF{gg}{x_1}\delta_3-\frac{1}{2}\GammaoneF{gg}{x_3}\delta_1$  \\\cline{2-2}
			\hline
		\end{tabular}}
		\caption{Identity-preserving gluon-gluon one-loop colour-stripped integrated dipoles.}
		\label{tab:J21IPgg}
	\end{table}
	\endgroup
	\FloatBarrier
	
	\begingroup
	\renewcommand{\arraystretch}{1.8} 
	\begin{table}[t]
		\centering\small\scalebox{0.9}{
		\begin{tabular}{c|l}
			& Integrated dipoles
			\\ \hline
			
			\multirow{1}{*}{FF$^{g\gets q}$}  
			& $\Jic{1}{g\gets q}(3_q,i_{\bar{q}})=-\frac{1}{2}\XTZ{A}{g}(s_{3i})+\Gammaone{gq}{x_3}\delta_1$  \\\cline{2-2}
			\hline
			
			\multirow{1}{*}{IF$^q_{g\to q}$}  
			& $\Jic{1}{g\to q}(1_q,3_q)=\XTZd{A}{q}{g}(s_{13})-\Sgtoq\Gammaone{qg}{x_1}\delta_3$  \\\cline{2-2}
			\hline
			
			\multirow{1}{*}{IF$_q^{g\gets q}$}  
			& $\Jic{1}{g\gets q
			}(1_q,3_q)=-\XTZd{A}{g}{q}(s_{13})+\Gammaone{gq}{x_3}\delta_1$  \\\cline{2-2}
			\hline
		\end{tabular}}
		\caption{Identity-changing quark-antiquark one-loop colour-stripped integrated dipoles. The subscripts indicate different choices of identified final-state partons. }
		\label{tab:J21ICqqb}
	\end{table}
	\endgroup
	
	\begingroup
	\renewcommand{\arraystretch}{1.8} 
	\begin{table}[t]
		\centering\small\scalebox{0.9}{
		\begin{tabular}{c|l}
			& Integrated dipoles
			\\ \hline
			
			\multirow{1}{*}{FF$^{g\gets q}$} 
			& $\Jic{1}{g\gets q}(3_q,i_g)=-\XTZd{D}{g}{g\gets q}(s_{3i})+\Gammaone{gq}{x_3}\delta_1$  \\\cline{2-2}
			\hline
			
			\multirow{1}{*}{FF$^{q\gets g}$}
			& $\Jic{1}{q\gets g}(3_g,i_q)=-\XTZ{E}{\qp}(s_{3i})+\Gammaone{qg}{x_3}\delta_1$  \\\cline{2-2}
			\hline
			
			\multirow{1}{*}{IF$_g^{g\gets q}$}
			& $\Jic{1}{g\gets q}(1_g,3_q)=-\XTZd{D}{g}{g}(s_{13})+\Gammaone{gq}{x_3}\delta_1$  \\\cline{2-2}
			\hline
			
			\multirow{1}{*}{IF$^g_{g\to q}$}
			& $\Jic{1}{g\to q}(1_q,3_g)=-\XIFint{D}{3}{0}{g\to q}(s_{13})-\Sgtoq\Gammaone{qg}{x_1}\delta_3$  \\\cline{2-2}
			\hline
			
			\multirow{1}{*}{IF$_q^{q\gets g}$}
			& $\Jic{1}{q\gets g}(1_q,3_g)=-\XTZd{E}{\qp}{q}(s_{13})+\Gammaone{qg}{x_3}\delta_1$  \\\cline{2-2}
			\hline
			
			\multirow{1}{*}{IF$^q_{q\to g}$}
			& $\Jic{1}{q\to g}(1_g,3_q)=-\XTZd{E}{q}{\qp}(s_{13})-\Sqtog\Gammaone{gq}{x_1}\delta_3$  \\\cline{2-2}
			\hline
		\end{tabular}}
		\caption{Identity-changing quark-gluon one-loop colour-stripped integrated dipoles.}
		\label{tab:J21ICqg}
	\end{table}
	\endgroup
	
	\begingroup
	\renewcommand{\arraystretch}{1.8} 
	\begin{table}[t]
		\centering\small\scalebox{0.9}{
		\begin{tabular}{c|l}
			& Integrated dipoles
			\\ \hline
			
			\multirow{1}{*}{FF$^{q\gets g}$}  
			& $\Jic{1}{q\gets g}(3_g,i_g)=-\XTZ{G}{\qp}(s_{3i})+\Gammaone{qg}{x_3}\delta_1$  \\\cline{2-2}
			\hline
			
			\multirow{1}{*}{IF$^g_{q\to g}$} 
			& $\Jic{1}{q\to g}(1_g,3_g)=-\XTZd{G}{g}{\qp}(s_{13})-\Sqtog\Gammaone{gq}{x_1}\delta_3$  \\\cline{2-2}
			\hline
			
			\multirow{1}{*}{IF$_g^{q\gets g}$} 
			& $\Jic{1}{q\gets g}(3_g,1_g)=-\XTZd{G}{\qp}{g}(s_{13})+\Gammaone{qg}{x_3}\delta_1$  \\\cline{2-2}
			\hline
		\end{tabular}}
		\caption{Identity-changing gluon-gluon one-loop colour-stripped integrated dipoles.}
		\label{tab:J21ICgg}
	\end{table}
	\endgroup

In complete analogy to what happens for integrated dipoles without fragmentation~\cite{Gehrmann:2023dxm}, within IP one-loop integrated dipoles the mass factorization kernels absorb the poles of integrated antenna functions coming from initial-state PDF and  final-state FF collinear singularities. The residual infrared poles reproduce the ones of one-loop virtual corrections. In particular one can write~\cite{Gehrmann:2023dxm}:
\begin{equation}\label{J21relation}
	\poles\left[\Jfull{1}(i,j)\right]=\poles\left[\text{Re}\left(\ourIop{1}{i j}{\e,\mu_r^2}\right)\right],
\end{equation}
where the quantity on the right-hand-side is closely related to Catani's one-loop infrared insertion operators~\cite{Catani:1998bh} and is explicitly given in~\cite{Gehrmann:2023dxm}. The relation above holds for any kinematical and partonic configuration. On the other hand, IC one-loop integrated dipoles are free of poles, given that the singularities of integrated IC fragmentation antenna functions completely cancel against mass factorization kernels:
\begin{equation}\label{J21icrelation}
	\poles\left[\Jfullic{1}{a\to b}(i,j)\right]=\poles\left[\Jfullic{1}{a\gets b}(i,j)\right]=0,
\end{equation}
which again holds for any kinematical and partonic configuration.

\clearpage

\subsection{Two-loop integrated dipoles}

The ingredients needed to build the two-loop integrated dipoles in fragmentation are: four-parton tree-level integrated antenna functions,  three-parton one-loop antenna functions, convolutions of two three-partons tree-level integrated antenna functions, convolutions of three-parton tree-level integrated antenna functions with NLO mass factorisation kernels and NNLO mass factorisation kernels (see Appendix \ref{app:MFkern}). 

The general decomposition of IP two-loop integrated dipoles reads:
\begin{eqnarray}
	\Jfull{2}\left(q,\bar{q}\right)&=&\J{2}\left(q,\bar{q}\right)-\dfrac{1}{N_c^2}\Jt{2}\left(q,\bar{q}\right)+\dfrac{N_f}{N_c}\Jh{2}\left(q,\bar{q}\right)\\
	\Jfull{2}\left(g,i\right)&=&\J{2}\left(g,i\right)+\dfrac{N_f}{N_c}\,\Jh{2}\left(g,i\right)\nonumber \\
	&-&\dfrac{N_f}{N_c^3}\Jht{2}\left(g,i\right)+\dfrac{N_f^2}{N_c^2}\,\Jhh{2}\left(g,i\right),\quad i=g,q\,.
\end{eqnarray} 
As extensively explained in~\cite{Gehrmann:2023dxm}, the construction of two-loop integrated dipoles is significantly less straightforward than for their one-loop counterparts. In particular, the presence of spurious singularities in some of the NNLO antenna functions requires the introduction of so-called \textit{corrective}, \textit{auxiliary} and \textit{flip-flopping} integrated dipoles, to properly remove unwanted terms. Here we fully present these types of dipoles in the context of fragmentation antenna functions, but we direct the reader to~\cite{Gehrmann:2023dxm} for additional details on their definition.

The colour decomposition of identity-changing two-loop integrated dipoles follows the one of two-loop identity-changing splitting kernels:
\begin{eqnarray}
	\Jfullic{2}{q\to g}(g,i)&=&\left(\dfrac{N_c^2-1}{N_c}\right)\Big[\Big.N_c\Jic{2}{q\to g}(g,i)+\dfrac{1}{N_c}\Jtic{2}{q\to g}(g,i) \nn\\
	&&\hspace{5.5cm}+N_f\Jhic{2}{q\to g}(g,i)\Big.\Big]\,, \\
	\Jfullic{2}{g\to q}(q,i)&=&N_c\Jic{2}{g\to q}(q,i)+\dfrac{1}{N_c}\Jtic{2}{g\to q}(q,i)+N_f\Jhic{2}{g\to q}(q,i)\,, \\
	\Jfullic{2}{q\to \qb}(\qb,i)&=&\left(\dfrac{N_c^2-1}{N_c}\right)\left[\Jic{2}{q\to \qb}(\qb,i)+\dfrac{1}{N_c}\Jtic{2}{q\to \qb}(\qb,i)\right]\,, \\
	\Jfullic{2}{q\to q'}(q',i)&=&\left(\dfrac{N_c^2-1}{N_c}\right)\Jic{2}{q\to q'}(q',i)\,,
\end{eqnarray}
and analogously reverting the arrow for a change of identity happening on the final state. At NNLO we can also have a change of identity for both an initial- and a final-state parton.  These configurations are captured by the following integrated dipoles
\begin{eqnarray}
	\Jfullic{2}{g\to q,g\gets q}(q,\qb)&=&N_c\,\Jic{2}{g\to q,g\gets q}(q,\qb)+\dfrac{1}{N_c}\Jtic{2}{g\to q,g\gets q}(q,\qb)\,, \\
	\Jfullic{2}{q\to g,q\gets g}(g,g)&=&\left(\dfrac{N_c^2-1}{N_c}\right)\Jic{2}{q\to g,q\gets g}(g,g)\,, \\
	\Jfullic{2}{q'\to g,g\gets q}(g,q)&=&N_c\,\Jic{2}{q'\to g,g\gets q}(g,q)+\dfrac{1}{N_c}\Jtic{2}{q'\to g,g\gets q}(g,q)\,, 
\end{eqnarray}
and analogous ones where the order of identity changes is swapped. 
Finally,  a single parton can undergo a double change of identity (flip-flopping):
\begin{eqnarray}
	\Jfullic{2}{q\to g\to q}(q,i)&=&\left(\dfrac{N_c^2-1}{N_c}\right)\Jic{2}{q\to g\to q}(q,i)\, ,
\end{eqnarray}
for a flip-flop in the initial state
and analogously by inverting the direction of the arrows for a final-state flip-flop.

The IP quark-antiquark two-loop integrated dipoles are listed in Table~\ref{tab:J22IPqqb} and the IP quark-gluon two-loop colour stripped ones are given in Table~\ref{tab:J22IPqg}. The flip-flopping terms are given in Table~\ref{tab:J22qgff} while the corrective terms required to remove spurious identity-changing singularities from integrated identity-preserving gluon-initiated quark-gluon antenna functions are given in Table~\ref{tab:J22corr}. Auxiliary quark-antiquark two-loop integrated dipoles needed to remove spurious poles present in integrated quark-gluon antenna functions, which are not present in physical matrix elements, are listed in Table~\ref{tab:J22bar}. The identity-preserving gluon-gluon two-loop colour-stripped integrated dipoles are given in Table~\ref{tab:J22IPgg} and the required flip-flopping terms are given in Table~\ref{tab:J22ggff}. The IC quark-antiquark, quark-gluon and gluon-gluon two-loop integrated dipoles are listed in Tables~\ref{tab:J22ICqqbIF}--\ref{tab:J22IFICgg}.

As for the one-loop case, it is possible to relate the $\e$-poles of IP two-loop colour-stripped integrated dipoles to the infrared singularities of two-loop matrix elements~\cite{Gehrmann:2023dxm}:
\begin{eqnarray}
	\label{dipids1}&&\poles\left[N_c\,\Jfull{2}(q,\qb)-\dfrac{\beta_0}{\e}\Jfull{1}(q,\qb)\right]=\nn\\
	&&\hspace{5cm}\poles\left[\text{Re}\left(\ourIop{2}{q\qb}{\e,\mu_r^2}-\dfrac{\beta_0}{\e}\ourIop{1}{q\qb}{\e,\mu_r^2}\right)\right],\\
	\label{dipids2}&&\poles\left[N_c\,\Jfull{2}(g,g)-\dfrac{\beta_0}{\e}\Jfull{1}(g,g)\right]=\nn\\
	&&\hspace{5cm}\poles\left[\text{Re}\left(\ourIop{2}{gg}{\e,\mu_r^2}-\dfrac{\beta_0}{\e}\ourIop{1}{gg}{\e,\mu_r^2}\right)\right],\\
	\label{dipids3}&&\poles\Bigg[\Bigg.N_c\left(\Jfull{2}(q,g)+\Jfull{2}(g,\qb)-2\Jfullb{2}(q,\qb)\right)\nn\\&&\hspace{6cm}-\dfrac{\beta_0}{\e}\left(\Jfull{1}(q,g)+\Jfull{1}(g,\qb)\right)\Bigg.\Bigg]=\nn\\
	&&\poles\left[\text{Re}\left(\ourIop{2}{qg}{\e,\mu_r^2}+\ourIop{2}{g\qb}{\e,\mu_r^2}-\dfrac{\beta_0}{\e}\left(\ourIop{1}{gg}{\e,\mu_r^2}+\ourIop{1}{g\qb}{\e,\mu_r^2}\right)\right)\right],
\end{eqnarray}
where the quantities in right-hand-side are related to Catani's one- and two-loop infrared insertion operators~\cite{Catani:1998bh} and are given in~\cite{Gehrmann:2023dxm}. The relations above hold for any kinematical and partonic configuration. The IC two-loop dipoles satisfy
\begin{eqnarray}
	\poles\left[\Jfullic{2}{a\to b}(i,j)-\dfrac{\beta_0}{\eps}\Jfullic{1}{a\to b}(i,j)\right]&=&0,\\
	\poles\left[\Jfullic{2}{a\gets b}(i,j)-\dfrac{\beta_0}{\eps}\Jfullic{1}{a\gets b}(i,j)\right]&=&0 \, ,
\end{eqnarray}
where we notice that the one-loop integrated dipoles present in this formulae can vanish if the considered change of identity is not allowed at one-loop, for example $q\to q'$. The full set of final-final and initial-final fragmentation colour-stripped integrated dipoles is provided as ancillary files.

	\begingroup
	\renewcommand{\arraystretch}{1.8} 
	\begin{table}[h]
		\centering\small\scalebox{0.9}{
		\begin{tabular}{c|l}
			& Integrated dipoles
			\\ \hline
			
			\multirow{5}{*}{FF}  
			& $\J{2}\left(3_{q},i_{\bar{q}}\right)=\XFZ{A}{q}+\XTO{A}{q}+\boelite\QQslite{3i}\XTZ{A}{q}-\frac{1}{2}\left[\XTZ{A}{q}\otimes\XTZ{A}{q}\right]$ \\
			& $\phantom{\J{2}\left(3_{q},i_{\bar{q}}\right)}-\Gammatwo{qq}{x_3}\deltaone$ \\\cline{2-2}
			& $\Jt{2}\left(3_q,i_{\bar{q}}\right)=\frac{1}{2}\XFZ{\wt{A}}{q}+2\XFZ{C}{\qb}+\XFZ{C}{q_1}+\XTO{\wt{A}}{q}-\frac{1}{2}\left[\XTZ{A}{q}\otimes\XTZ{A}{q}\right]$ \\
			& $\phantom{\Jt{2}\left(3_q,i_{\bar{q}}\right)}+\Gammatwott{qq}{x_3}\deltaone$ \\\cline{2-2}
			& $\Jh{2}\left(3_q,i_{\bar{q}}\right)=\XFZ{B}{q}+\XTO{\wh{A}}{q}+\bFoelite\QQslite{3i}\XTZ{A}{q}-\GammatwoF{qq}{x_3}\deltaone$  \\\cline{2-2}
			\hline
			
			\multirow{6}{*}{IF}  
			& $\J{2}\left(1_{q},3_{q}\right)=\XTZe{A}{q}{q}+\XTZf{A}{q}{q}+\boelite\QQslite{13}\XTZd{A}{q}{q}-\frac{1}{2}\left[\XTZd{A}{q}{q}\otimes\XTZd{A}{q}{q}\right]$ \\ 
			&$\phantom{\J{2}\left(1_{q},3_{q}\right)}-\iGammatwo{qq}{x_1}\delta_3-\Gammatwo{qq}{x_3}\deltaone$ \\\cline{2-2}
			& $\Jt{2}\left(1_q,3_q\right)=\frac{1}{2}\XTZe{\wt{A}}{q}{q}+2\XTZe{C}{\qb_2}{\qb_1}+2\XTZe{C}{q_1}{q_4}+\XTZf{\wt{A}}{q}{q}-\frac{1}{2}\left[\XTZd{A}{q}{q}\otimes\XTZd{A}{q}{q}\right]$ \\
			& $\phantom{\Jt{2}\left(1_q,3_q\right)}+\iGammatwott{qq}{x_1}\delta_3+\Gammatwott{qq}{x_3}\deltaone$ \\\cline{2-2}
			& $\Jh{2}\left(1_q,3_{q}\right)=\XTZe{B}{q}{q}+\XTZf{\wh{A}}{q}{q}+\bFoelite\QQslite{13}\XTZd{A}{q}{q}$ \\
			& $\phantom{\Jh{2}\left(1_q,3_{q}\right)}-\iGammatwoF{qq}{x_1}\delta_3-\GammatwoF{qq}{x_3}\deltaone$  \\\cline{2-2}
			\hline
		\end{tabular}}
		\caption{Identity-preserving quark-antiquark two-loop colour-stripped integrated dipoles.The subscripts indicate different choices of initial-state partons.}
		\label{tab:J22IPqqb}
	\end{table}
	\endgroup

	\begingroup
	\renewcommand{\arraystretch}{1.8} 
	\begin{table}[h]
		\centering\small\scalebox{0.9}{
		\begin{tabular}{c|l}
			& Integrated dipoles
			\\ \hline
			
			\multirow{6}{*}{FF$^{q}$} 
			& $\J{2}\left(3_q,i_g\right)=\frac{1}{2}\XFZ{D}{q}+\frac{1}{2}\XTO{D}{q}+\frac{1}{2}\boelite\QQslite{3i}\XTZ{D}{q}-\frac{1}{4}\left[\XTZ{D}{q}\otimes\XTZ{D}{q}\right]$ \\
			& $\phantom{\J{2}\left(3_q,i_g\right)}-\Gammatwo{qq}{x_3}\deltaone$  \\\cline{2-2}
			& $\Jh{2}\left(3_q,i_g\right)=\XFZ{E}{q_1}+\frac{1}{2}\XTO{\wh{D}}{q}+\frac{1}{2}\XTO{E}{q}+\frac{1}{2}\bFoelite\QQslite{3i}\XTZ{D}{q}$ \\ 
			& $\phantom{\Jh{2}\left(3_q,i_g\right)}+\frac{1}{2}\boelite\QQslite{3i}\XTZ{E}{q}-\frac{1}{2}\left[\XTZ{E}{q}\otimes\XTZ{D}{q}\right]-\GammatwoF{qq}{x_3}\deltaone$  \\\cline{2-2}
			& $\Jht{2}\left(3_q,i_g\right)=\frac{1}{2}\XFZ{\wt{E}}{q_1}+\frac{1}{2}\XTO{\wt{E}}{q}$ \\\cline{2-2}
			& $\Jhh{2}\left(3_q,i_g\right)=\frac{1}{2}\XTO{\wh{E}}{q}+\frac{1}{2}\bFoelite\QQslite{3i}\XTZ{E}{q}-\frac{1}{4}\left[\XTZ{E}{q}\otimes\XTZ{E}{q}\right]$  \\\cline{2-2}
			\hline
			
			\multirow{6}{*}{FF$^{g}$}
			& $\J{2}\left(3_g,i_q\right)=\XFZ{D}{g_2}+\frac{1}{2}\XFZ{D}{g_3}+\XTO{D}{g}+\boelite\QQslite{3i}\XTZ{D}{g}$ \\ 
			& $\phantom{\J{2}\left(3_g,i_q\right)}-\left[\XTZd{D}{g}{g\gets g}\otimes\XTZd{D}{g}{g\gets g}\right]-\frac{1}{2}\Gammatwo{gg}{x_3}\deltaone
			+\Jic{2}{\mathrm{IC}\,\mathrm{corr}.}(3_q,i_{q})$ \\\cline{2-2}
			& $\Jh{2}\left(3_g,i_q\right)=\XFZ{E}{g}+\XTO{\wh{D}}{g}+\bFoelite\QQslite{3i}\XTZ{D}{g}$ \\ 
			& $\phantom{\Jh{2}\left(3_g,i_q\right)}-\frac{1}{2}\GammatwoF{gg}{x_3}\deltaone
			+\Jhic{2}{\mathrm{f/f}}\left(3_g,i_q\right)
			+\Jhic{2}{\mathrm{IC}\,\mathrm{corr}.}(3_q,i_{q})$ \\\cline{2-2}
			& $\Jht{2}\left(3_g,i_q\right)=\frac{1}{2}\XFZ{\wt{E}}{g}+\frac{1}{2}\GammatwoFt{gg}{x_3}\deltaone
			+\Jhtic{2}{\mathrm{f/f}}\left(3_g,i_q\right)$  \\\cline{2-2}
			& $\Jhh{2}\left(3_g,i_q\right)=-\frac{1}{2}\GammatwoFF{gg}{x_3}\deltaone$  \\\cline{2-2}
			\hline
			
			\multirow{6}{*}{IF$^{g}_q$}
			& $\J{2}\left(1_q,3_g\right)=\XTZe{D}{g_2}{q}+\frac{1}{2}\XTZe{D}{g_3}{q}+\XTZf{D}{g}{q}+\boelite\QQslite{13}\XTZd{D}{g}{q}$ \\ 
			& $\phantom{\J{2}\left(1_q,3_g\right)}-\left[\XTZd{D}{g}{q}\otimes\XTZd{D}{g}{q}\right]-\iGammatwoF{qq}{x_1}\delta_3-\frac{1}{2}\GammatwoF{gg}{x_3}\deltaone$ \\\cline{2-2}
			& $\Jh{2}\left(1_q,3_g\right)=\XTZe{E}{g}{q}+\frac{1}{2}\XTZf{\wh{D}}{g}{q}+\bFoelite\QQslite{13}\XTZd{D}{g}{q}-\iGammatwoF{qq}{x_1}\delta_3$ \\ 
			& $\phantom{\Jh{2}\left(1_q,3_g\right)}-\frac{1}{2}\GammatwoF{gg}{x_3}\deltaone
			+\Jhic{2}{\mathrm{f/f}}\left(1_q,3_g\right)$ \\\cline{2-2}
			& $\Jht{2}\left(1_q,3_g\right)=\frac{1}{2}\XTZe{\wt{E}}{g}{q}+\frac{1}{2}\GammatwoFt{gg}{x_3}\deltaone+\Jhtic{2}{\mathrm{f/f}}\left(1_q,3_g\right)$  \\\cline{2-2}
			& $\Jhh{2}\left(1_q,3_g\right)=-\frac{1}{2}\GammatwoFF{gg}{x_3}\deltaone$  \\\cline{2-2}
			\hline
			
			\multirow{6}{*}{IF$^{q}_g$}
			& $\J{2}\left(1_g,3_q\right)=\XTZe{D}{q}{g_2}+\frac{1}{2}\XTZe{D}{q}{g_3}+\XTZf{D}{q}{g}+\boelite\QQslite{13}\XTZd{D}{q}{g}$ \\ 
			& $\phantom{\J{2}\left(1_g,3_q\right)}-\left[\XTZd{D}{q}{g}\otimes\XTZd{D}{q}{g}\right]-\GammatwoF{qq}{x_3}\delta_1-\frac{1}{2}\GammatwoF{gg}{x_1}\delta_3$ \\\cline{2-2}
			& $\Jh{2}\left(1_g,3_q\right)=\XTZe{E}{q}{g}+\frac{1}{2}\XTZf{\wh{D}}{q}{g}+\bFoelite\QQslite{13}\XTZd{D}{q}{g}-\GammatwoF{qq}{x_3}\delta_1$ \\ 
			& $\phantom{\Jh{2}\left(3_q,1_g\right)}-\frac{1}{2}\iGammatwoF{gg}{x_1}\delta_3
			+\Jhic{2}{\mathrm{f/f}}\left(3_q,1_g\right)$ \\\cline{2-2}
			& $\Jht{2}\left(1_g,3_q\right)=\frac{1}{2}\XTZe{\wt{E}}{q}{g}+\frac{1}{2}\iGammatwoFt{gg}{x_1}\delta_3+\Jhtic{2}{\mathrm{f/f}}\left(3_q,1_g\right)$  \\\cline{2-2}
			& $\Jhh{2}\left(1_g,3_q\right)=-\frac{1}{2}\iGammatwoFF{gg}{x_1}\delta_3$  \\\cline{2-2}
			\hline
		\end{tabular}}
		\caption{Identity-preserving quark-gluon and gluon-quark two-loop colour-stripped integrated dipoles.}
		\label{tab:J22IPqg}
	\end{table}
	\endgroup

	\begingroup
	\renewcommand{\arraystretch}{1.8} 
	\begin{table}[h]
		\centering\small\scalebox{0.9}{
		\begin{tabular}{c|l}
			& Integrated dipoles
			\\ \hline
			
			\multirow{2}{*}{FF$^{g}$}  
			& $\Jhic{2}{\mathrm{f/f}}\left(3_g,i_q\right)=-\left[\Gammaoneconv{gq}{x_3}\otimes\XTZ{E}{q'}\right]+\frac{1}{2}\left[\Gammaoneconv{gq}{x_3}\otimes\Gammaoneconv{qg}{x_3}\right]\delta_1$ \\\cline{2-2}
			& $\Jhtic{2}{\mathrm{f/f}}\left(3_g,i_q\right)=-\left[\Gammaoneconv{gq}{x_3}\otimes\XTZ{E}{q'}\right]+\frac{1}{2}\left[\Gammaoneconv{gq}{x_3}\otimes\Gammaoneconv{qg}{x_3}\right]\delta_1$  \\\cline{2-2}
			\hline
			
			\multirow{2}{*}{IF$^{g}_q$} 
			& $\Jhic{2}{\mathrm{f/f}}\left(3_g,1_q\right)=-\left[\Gammaoneconv{gq}{x_3}\otimes\XTZd{E}{\qp}{q}\right]+\frac{1}{2}\left[\Gammaoneconv{qg}{x_3}\otimes\Gammaoneconv{gq}{x_3}\right]\delta_1$  \\\cline{2-2}
			& $\Jhtic{2}{\mathrm{f/f}}\left(3_g,1_q\right)=-\left[\Gammaoneconv{gq}{x_3}\otimes\XTZd{E}{\qp}{q}\right]+\frac{1}{2}\left[\Gammaoneconv{qg}{x_3}\otimes\Gammaoneconv{gq}{x_3}\right]\delta_1$  \\\cline{2-2}
			\hline
			
			\multirow{2}{*}{IF$^{q}_g$} 
			& $\Jhic{2}{\mathrm{f/f}}\left(1_g,3_q\right)=\Sgtoq\left[\Gammaoneconv{qg}{x_1}\otimes\XTZd{E}{q}{\qp}\right]+\frac{1}{2}\left[\Gammaoneconv{qg}{x_1}\otimes\Gammaoneconv{gq}{x_1}\right]\delta_3$  \\\cline{2-2}
			& $\Jhtic{2}{\mathrm{f/f}}\left(1_g,3_q\right)=\Sgtoq\left[\Gammaoneconv{qg}{x_1}\otimes\XTZd{E}{q}{\qp}\right]+\frac{1}{2}\left[\Gammaoneconv{qg}{x_1}\otimes\Gammaoneconv{gq}{x_1}\right]\delta_3$  \\\cline{2-2}
			\hline
		\end{tabular}}
		\caption{Flip-flopping contributions to identity-preserving quark-gluon two-loop integrated dipoles.}
		\label{tab:J22qgff}
	\end{table}
	\endgroup

	\begingroup
	\renewcommand{\arraystretch}{1.8} 
	\begin{table}[h]
		\centering\small\scalebox{0.9}{
		\begin{tabular}{c|l}
			& Integrated dipoles
			\\ \hline
			
			\multirow{8}{*}{FF}  
			& $\Jic{2}{\mathrm{IC}\,\mathrm{corr}}\left(3_q,i_{q}\right)=-\left[\XTZ{D}{q}\otimes\XTZd{D}{g}{g\gets q}\right]-\left[\Gammaoneconv{gg}{x_3}\otimes\XTZd{D}{g}{g\gets q}\right]$ \\
			& $\phantom{\Jic{2}{\mathrm{IC}\,\mathrm{corr}}\left(3_q,i_{\qb}\right)}+2\left[\Gammaoneconv{qq}{x_3}\otimes\XTZd{D}{g}{g\gets q}\right]-\XFZ{A}{g}-\frac{1}{2}\XFZ{\wt{A}}{g}-\frac{1}{2}\XTO{A}{g}$ \\
			& $\phantom{\Jic{2}{\mathrm{IC}\,\mathrm{corr}}\left(3_q,i_{\qb}\right)}-\frac{1}{2}\XTO{\wt{A}}{g}-\frac{1}{2}\boelite\left(\QQslite{3i}-1\right)\XTZ{A}{g}+\left[\XTZ{A}{q}\otimes\XTZ{A}{g}\right]$ \\
			& $\phantom{\Jic{2}{\mathrm{IC}\,\mathrm{corr}}\left(3_q,i_{\qb}\right)}+\frac{1}{2}\left[\Gammaoneconv{gg}{x_3}\otimes\XTZ{A}{g}\right]-\left[\Gammaoneconv{qq}{x_3}\otimes\XTZ{A}{g}\right]$ \\
			& $\phantom{\Jic{2}{\mathrm{IC}\,\mathrm{corr}}\left(3_q,i_{\qb}\right)}-\boelite\left(\XTZ{D}{g}-\XTZd{D}{g}{g\gets g}\right)$  \\\cline{2-2}
			& $\Jhic{2}{\mathrm{IC}\,\mathrm{corr}}\left(3_q,i_{q}\right)=-\left[\XTZ{E}{q}\otimes\XTZd{D}{g}{g\gets q}\right]-\left[\GammaoneFconv{gg}{x_3}\otimes\XTZd{D}{g}{g\gets q}\right]$ \\
			& $\phantom{\Jhic{2}{\mathrm{IC}\,\mathrm{corr}}\left(3_g,i_{\qb}\right)}-\frac{1}{2}\XTO{\wh{A}}{g}-\frac{1}{2}\bFoelite\left(\QQslite{3i}-1\right)\XTZ{A}{g}$ \\
			& $\phantom{\Jhic{2}{\mathrm{IC}\,\mathrm{corr}}\left(3_g,i_{\qb}\right)}+\frac{1}{2}\left[\GammaoneFconv{gg}{x_3}\otimes\XTZ{A}{g}\right]-\bFoelite\left(\XTZ{D}{g}-\XTZd{D}{g}{g\to g}\right)$  \\\cline{2-2}
			\hline
			
		\end{tabular}}
		\caption{Corrective terms required to remove spurious identity-changing singularities from integrated identity-preserving gluon-initiated quark-gluon antenna functions.}
		\label{tab:J22corr}
	\end{table}
	\endgroup

	\begingroup
	\renewcommand{\arraystretch}{1.8} 
	\begin{table}[h]
		\centering\small\scalebox{0.9}{
		\begin{tabular}{c|l}
			& Integrated dipoles
			\\ \hline
			
			\multirow{2}{*}{FF}  
			& $\Jb{2}(3_q,i_{\qb})=\frac{1}{2}\XFZ{\wt{A}}{q}+\XTO{\wt{A}}{q}-\frac{1}{2}\left[\XTZ{A}{q}\otimes\XTZ{A}{q}\right]$  \\\cline{2-2}
			& $\Jtb{2}(3_q,i_{\qb})=\Jt{2}(3_q,j_{\qb})$  \\\cline{2-2}
			\hline
			
			\multirow{2}{*}{IF}  
			& $\Jb{2}(1_q,3_q)=\frac{1}{2}\XTZe{\wt{A}}{q}{q}+\XTZf{\wt{A}}{q}{q}-\frac{1}{2}\left[\XTZd{A}{q}{q}\otimes\XTZd{A}{q}{q}\right]$  \\\cline{2-2}
			& $\Jtb{2}(1_q,3_q)=\Jt{2}(1_q,3_q)$  \\\cline{2-2}
			\hline
		\end{tabular}}
		\caption{Auxiliary quark-antiquark two-loop integrated dipoles needed to remove spurious poles present in integrated quark-gluon antenna functions, which are not present in physical matrix elements. $\Jt{2}(3_q,j_{\qb})$ is the FF fragmentation dipoles given in Table \ref{tab:J22IPqqb}.}
		\label{tab:J22bar}
	\end{table}
	\endgroup

	\begingroup
	\renewcommand{\arraystretch}{1.8} 
	\begin{table}[h]
		\centering\small\scalebox{0.9}{
		\begin{tabular}{c|l}
			& Integrated dipoles
			\\ \hline
			
			\multirow{8}{*}{FF}  
			& 
			$\J{2}\left(3_g,i_g\right)=\frac{1}{2}\XFZ{F}{g}+\frac{1}{2}\XTO{F}{g}+\frac{1}{2}\boelite\QQslite{3i}\XTZ{F}{g}
			-\frac{1}{4}\left[\XTZ{F}{g}\otimes\XTZ{F}{g}\right]$ \\ 
			& $\phantom{\J{2}\left(3_g,i_g\right)}-\frac{1}{2}\Gammatwo{gg}{x_3}\deltaone$  \\\cline{2-2}
			& $\Jh{2}\left(3_g,i_g\right)=\XFZ{G}{g}+\frac{1}{2}\XTO{\wh{F}}{g}+\frac{1}{2}\XTO{G}{g}+\frac{1}{2}\bFoelite\QQslite{3i}\XTZ{F}{g}
			$ \\ 
			& $\phantom{\Jh{2}\left(3_g,i_g\right)}+\frac{1}{2}\boelite\QQslite{3i}\XTZ{G}{g}-\frac{1}{2}\left[\XTZ{G}{g}\otimes\XTZ{F}{g}\right]-\frac{1}{2}\GammatwoF{gg}{x_3}\deltaone$ \\ 
			& $\phantom{\Jh{2}\left(3_g,i_g\right)}+\Jhic{2}{\mathrm{f/f}}\left(3_g,i_g\right)$  \\\cline{2-2}
			& $\Jht{2}\left(3_g,i_g\right)=\frac{1}{2}\XFZ{\wt{G}}{g}+\frac{1}{2}\XTO{\wt{G}}{g}+\frac{1}{2}\GammatwoFt{gg}{x_3}\deltaone
			+\Jhtic{2}{\mathrm{f/f}}\left(3_g,i_g\right)$  \\\cline{2-2}
			& $\Jhh{2}\left(3_g,i_g\right)=\frac{1}{2}\XTO{\wh{G}}{g}+\frac{1}{2}\bFoelite\QQslite{1i}\XTZ{G}{g}
			-\frac{1}{4}\left[\XTZ{G}{g}\otimes\XTZ{G}{g}\right]$ \\ 
			& $\phantom{\Jhh{2}\left(3_g,i_g\right)}-\frac{1}{2}\GammatwoFF{gg}{x_3}\deltaone$  \\\cline{2-2}
			\hline
			
			\multirow{6}{*}{IF} 
			& $\J{2}\left(1_g,3_g\right)=\XTZe{F}{g_1}{g_2}+\frac{1}{2}\XTZe{F}{g_1}{g_3}+\XTZf{F}{g}{g}+\boelite\QQslite{13}\XTZd{F}{g}{g}$ \\ 
			& $\phantom{\J{2}\left(1_g,3_g\right)}-\left[\XTZd{F}{g}{g}\otimes\XTZd{F}{g}{g}\right]-\frac{1}{2}\iGammatwo{gg}{x_1}\delta_3-\frac{1}{2}\Gammatwo{gg}{x_3}\deltaone$  \\\cline{2-2}
			& $\Jh{2}\left(1_g,3_g\right)=\XTZe{G}{g}{g}+\XTZf{\wh{F}}{g}{g}+\bFoelite\QQslite{13}\XTZd{F}{g}{g}
			-\frac{1}{2}\iGammatwoF{gg}{x_1}\delta_3$ \\ 
			& $\phantom{\Jh{2}\left(1_g,3_g\right)}-\frac{1}{2}\GammatwoF{gg}{x_3}\deltaone
			+\Jhic{2}{\mathrm{f/f}}\left(1_g,3_g\right)$  \\\cline{2-2}
			& $\Jht{2}\left(1_g,3_g\right)=\frac{1}{2}\XTZe{G}{g}{g}+\frac{1}{2}\iGammatwoFt{gg}{x_1}\delta_3+\frac{1}{2}\GammatwoFt{gg}{x_1}\delta_3
			+\Jhtic{2}{\mathrm{f/f}}\left(1_g,3_g\right)$  \\\cline{2-2}
			& $\Jhh{2}\left(1_g,3_g\right)=-\frac{1}{2}\iGammatwoFF{gg}{x_1}\delta_3-\frac{1}{2}\GammatwoFF{gg}{x_3}\deltaone$  \\\cline{2-2}
			\hline
		\end{tabular}}
		\caption{Identity-preserving gluon-gluon two-loop colour-stripped integrated dipoles.}
		\label{tab:J22IPgg}
	\end{table}
	\endgroup

	\begingroup
	\renewcommand{\arraystretch}{1.8} 
	\begin{table}[h]
		\centering\small\scalebox{0.9}{
		\begin{tabular}{c|l}
			& Integrated dipoles
			\\ \hline
			
			\multirow{2}{*}{FF}  
			& $\Jhic{2}{\mathrm{f/f}}\left(3_g,i_g\right)=-\left[\Gammaoneconv{gq}{x_3}\otimes\XTZ{G}{\qp}\right]+\frac{1}{2}\left[\Gammaoneconv{gq}{x_3}\otimes\Gammaoneconv{qg}{x_3}\right]\delta_	1$ \\\cline{2-2}
			& $\Jhtic{2}{\mathrm{f/f}}\left(3_g,i_g\right)=-\left[\Gammaoneconv{gq}{x_3}\otimes\XTZ{G}{\qp}\right]+\frac{1}{2}\left[\Gammaoneconv{gq}{x_3}\otimes\Gammaoneconv{qg}{x_3}\right]\delta_	1$  \\\cline{2-2}
			\hline
			
			\multirow{4}{*}{IF} 
			& $\Jhic{2}{\mathrm{f/f}}\left(1_g,2_g\right)=\Sgtoq\left[\Gammaoneconv{qg}{x_1}\otimes\XTZd{G}{g}{\qp}\right]+\frac{1}{2}\left[\Gammaoneconv{qg}{x_1}\otimes\Gammaoneconv{gq}{x_1}\right]\delta_	3$ \\
			& $\phantom{\Jhic{2}{\mathrm{f/f}}\left(1_g,3_g\right)}-\left[\Gammaoneconv{gq}{x_3}\otimes\XTZd{G}{q}{g}\right]+\frac{1}{2}\left[\Gammaoneconv{gq}{x_3}\otimes\Gammaoneconv{qg}{x_3}\right]\delta_	1$  \\\cline{2-2}
			& $\Jhtic{2}{\mathrm{f/f}}\left(1_g,2_g\right)=\Sgtoq\left[\Gammaoneconv{qg}{x_1}\otimes\XTZd{G}{g}{\qp}\right]+\frac{1}{2}\left[\Gammaoneconv{qg}{x_1}\otimes\Gammaoneconv{gq}{x_1}\right]\delta_	3$ \\
			& $\phantom{\Jhtic{2}{\mathrm{f/f}}\left(1_g,3_g\right)}-\left[\Gammaoneconv{gq}{x_3}\otimes\XTZd{G}{q}{g}\right]+\frac{1}{2}\left[\Gammaoneconv{gq}{x_3}\otimes\Gammaoneconv{qg}{x_3}\right]\delta_	1$ \\\cline{2-2}
			\hline
		\end{tabular}}
		\caption{Flip-flopping contributions to identity-preserving gluon-gluon two-loop integrated dipoles.}
		\label{tab:J22ggff}
	\end{table}
	\endgroup

	\begingroup
	\renewcommand{\arraystretch}{1.8} 
	\begin{table}[h]
		\centering\small\scalebox{0.9}{
		\begin{tabular}{c|l}
			& Integrated dipoles
			\\ \hline
			
			\multirow{8}{*}{FF$^{g\gets q}$}
			& $\Jic{2}{g\gets q}\left(3_q,i_{\qb}\right)=-\XFZ{A}{g}-\frac{1}{2}\XTO{A}{g}-\frac{1}{2}\boelite\QQslite{3i}\XTZ{A}{g}$\\
			& $\hspace{-1cm}\phantom{\Jic{2}{q\gets g}\left(3_q,i_{\qb}\right)}+\frac{1}{2}\left[\XTZ{A}{g}\otimes\XTZ{A}{q}\right]+\Gammatwo{gq}{x_3}\deltaone$\\
			& $\hspace{-1cm}\phantom{\Jic{2}{g\gets q}\left(3_q,i_{\qb}\right)}+\frac{1}{2}\left[\XTZ{A}{g}\otimes\Gammaoneconv{gg}{x_3}\right]-\frac{1}{2}\left[\Gammaoneconv{gq}{x_3}\otimes\Gammaoneconv{gg}{x_3}\right]\delta_1$\\
			& $\hspace{-1cm}\phantom{\Jic{2}{g\gets q}\left(3_q,i_{\qb}\right)}-\frac{1}{2}\left[\XTZ{A}{g}\otimes\Gammaoneconv{qq}{x_3}\right]+\frac{1}{2}\left[\Gammaoneconv{gq}{x_3}\otimes\Gammaoneconv{qq}{x_3}\right]\delta_1$\\
			& $\Jhic{2}{g\gets q}\left(3_q,i_{\qb}\right)=-\frac{1}{2}\XTO{\wh{A}}{g}-\frac{1}{2}\bFoelite\QQslite{3i}\XTZ{A}{g}+\GammatwoF{gq}{x_3}\deltaone$\\
			& $\hspace{-1cm}\phantom{\Jhic{2}{g\gets q}\left(3_q,i_{\qb}\right)}+\frac{1}{2}\left[\XTZ{A}{g}\otimes\GammaoneFconv{gg}{x_3}\right]-\frac{1}{2}\left[\Gammaoneconv{gq}{x_3}\otimes\GammaoneFconv{gg}{x_3}\right]\delta_1$\\
			& $\Jtic{2}{g\gets q}\left(3_q,i_{\qb}\right)=-\XFZ{\wt{A}}{g}-\XTO{\wt{A}}{g}+\left[\XTZ{A}{q}\otimes \XTZ{A}{g}\right]-\left[\Gammaoneconv{qq}{x_3}\otimes\XTZ{A}{g}\right]$\\
			& $\hspace{-1cm}\phantom{\Jtic{2}{g\gets q}\left(3_q,i_{\qb}\right)}+\left[\Gammaoneconv{gq}{x_3}\otimes\Gammaoneconv{qq}{x_3}\right]\delta_1-2\Gammatwot{gq}{x_3}\deltaone$  \\\cline{2-2}
			\hline
			
			\multirow{1}{*}{FF$^{q\gets \qb}$}
			& $\Jtic{2}{\qb\gets q}\left(3_{\qb},i_q\right)=\XFZ{C}{q_3}+\Gammatwot{q \qb}{x_3}\deltaone$  \\\cline{2-2}
			\hline
			
			\multirow{2}{*}{FF$^{q\gets g\gets q}$}
			& $\Jic{2}{q\gets g\gets q}\left(3_q,i_g\right)=\XFZ{B}{\qp}-\left[\Gammaoneconv{qg}{x_3}\otimes\XTZ{A}{g}\right]$\\
			& $\hspace{-1cm}\phantom{\Jic{2}{q\gets g\gets q}\left(3_q,i_g\right)}+\left[\Gammaoneconv{qg}{x_3}\otimes\Gammaoneconv{gq}{x_3}\right]\delta_1-2\Gammatwot{qq}{x_3}\deltaone$  \\\cline{2-2}
			\hline
			\end{tabular}}
			\caption{Final-final identity-changing quark-antiquark two-loop colour-stripped integrated dipoles.}
			\label{tab:J22ICqqbIF}
		\end{table}
	\endgroup
 
	\begingroup
	\renewcommand{\arraystretch}{1.8} 
	\begin{table}[h]
		\centering\small\scalebox{0.9}{
		\begin{tabular}{c|l}
			& Integrated dipoles
			\\ \hline
			
			\multirow{9}{*}{IF$^{q}_{g\to q}$}
			& $\Jic{2}{g\to q}\left(1_q,3_q\right)=-\XTZe{A}{q}{g_3}-\XTZe{A}{q}{g_4}-\XTZf{A}{q}{g}+\boelite\QQslite{13}\XTZd{A}{q}{g}$\\
			& $\hspace{-1cm}\phantom{\Jic{2}{g\to q}\left(1_q,3_q\right)}-\left[\XTZd{A}{q}{g}\otimes\XTZd{A}{q}{q}\right]-\Sgtoq\iGammatwo{qg}{x_1}\delta_3$\\
			& $\hspace{-1cm}\phantom{\Jic{2}{g\to q}\left(1_q,3_q\right)}-\left[\XTZd{A}{q}{g}\otimes\Gammaoneconv{gg}{x_1}\right]+\frac{1}{2}\Sgtoq\left[\Gammaoneconv{qg}{x_1}\otimes\Gammaoneconv{gg}{x_1}\right]\delta_3$\\
			& $\hspace{-1cm}\phantom{\Jic{2}{g\to q}\left(1_q,3_q\right)}+\left[\XTZd{A}{q}{g}\otimes\Gammaoneconv{qq}{x_1}\right]-\frac{1}{2}\Sgtoq\left[\Gammaoneconv{qg}{x_1}\otimes\Gammaoneconv{qq}{x_1}\right]\delta_3$\\
			& $\Jhic{2}{g\to q}\left(1_q,3_q\right)=-\frac{1}{2}\XTZf{\wh{A}}{q}{g}+\frac{1}{2}\bFoelite\QQslite{13}\XTZd{A}{q}{g}-\Sgtoq\iGammatwoF{qg}{x_1}\delta_3$\\
			& $\hspace{-1cm}\phantom{\Jic{2}{g\to q}\left(1_q,3_q\right)}-\left[\XTZd{A}{q}{g}\otimes\GammaoneFconv{gg}{x_1}\right]+\frac{1}{2}\Sgtoq\left[\Gammaoneconv{qg}{x_1}\otimes\GammaoneFconv{gg}{x_1}\right]\delta_3$\\
			& $\Jtic{2}{g\to q}\left(1_q,3_q\right)=-\XTZe{\wt{A}}{q}{g}-\XTZf{\wt{A}}{q}{g}$\\
			& $\hspace{-1cm}\phantom{\Jtic{2}{g\to q}\left(1_q,3_q\right)}+\left[\XTZd{A}{q}{q}\otimes \XTZe{A}{q}{g}\right]+\left[\Gammaoneconv{qq}{x_1}\otimes\XTZd{A}{q}{g}\right]$\\
			& $\hspace{-1cm}\phantom{\Jtic{2}{g\to q}\left(1_q,3_q\right)}-\frac{1}{2}\Sgtoq\left[\Gammaoneconv{qg}{x_1}\otimes\Gammaoneconv{qq}{x_1}\right]\delta_3+\Sgtoq\iGammatwot{qg}{x_1}\delta_3$  \\\cline{2-2}
			\hline
			
			\multirow{9}{*}{IF$_q^{g\gets q}$}
			& $\Jic{2}{g\gets q}\left(3_q,1_q\right)=-\XTZe{A}{g_3}{q}-\XTZe{A}{g_4}{q}-\XTZf{A}{g}{q}-\boelite\QQslite{13}\XTZd{A}{g}{q}$\\
			& $\hspace{-1cm}\phantom{\Jic{2}{q\gets g}\left(3_q,1_q\right)}+\left[\XTZd{A}{g}{q}\otimes\XTZd{A}{q}{q}\right]+\Gammatwo{gq}{x_3}\delta_1$\\
			& $\hspace{-1cm}\phantom{\Jic{2}{g\gets q}\left(3_q,1_q\right)}+\left[\XTZd{A}{g}{q}\otimes\Gammaoneconv{gg}{x_3}\right]-\frac{1}{2}\left[\Gammaoneconv{gq}{x_3}\otimes\Gammaoneconv{gg}{x_3}\right]\delta_1$\\
			& $\hspace{-1cm}\phantom{\Jic{2}{g\gets q}\left(3_q,1_q\right)}-\left[\XTZd{A}{g}{q}\otimes\Gammaoneconv{qq}{x_3}\right]+\frac{1}{2}\left[\Gammaoneconv{gq}{x_3}\otimes\Gammaoneconv{qq}{x_3}\right]\delta_1$\\
			& $\Jhic{2}{g\gets q}\left(3_q,1_q\right)=-\frac{1}{2}\XTZf{\wh{A}}{g}{q}-\frac{1}{2}\bFoelite\QQslite{13}\XTZd{A}{g}{q}+\GammatwoF{gq}{x_3}\delta_1$\\
			& $\hspace{-1cm}\phantom{\Jic{2}{g\gets q}\left(3_q,1_q\right)}+\left[\XTZd{A}{g}{q}\otimes\GammaoneFconv{gg}{x_3}\right]-\frac{1}{2}\left[\Gammaoneconv{gq}{x_3}\otimes\GammaoneFconv{gg}{x_3}\right]\delta_1$\\
			& $\Jtic{2}{g\gets q}\left(3_q,1_q\right)=-\XTZe{\wt{A}}{g}{q}-\XTZf{\wt{A}}{g}{q}$\\
			& $\hspace{-1cm}\phantom{\Jtic{2}{g\gets q}\left(3_q,1_q\right)}+\left[\XTZd{A}{q}{q}\otimes \XTZe{A}{g}{q}\right]-\left[\Gammaoneconv{qq}{x_3}\otimes\XTZd{A}{g}{q}\right]$\\
			& $\hspace{-1cm}\phantom{\Jtic{2}{g\gets q}\left(3_q,1_q\right)}+\frac{1}{2}\left[\Gammaoneconv{gq}{x_3}\otimes\Gammaoneconv{qq}{x_3}\right]\delta_1-\Gammatwot{gq}{x_3}\delta_1$  \\\cline{2-2}
			\hline
			
			\multirow{2}{*}{IF$^{q}_{\qp\to q}$}
			& $\Jic{2}{\qp\to q}\left(1_{\qp},3_q\right)=\XTZe{B}{q}{\qp}-\Sqtog\left[\Gammaoneconv{gq}{x_1}\otimes \XTZd{A}{q}{g}\right]$\\
			& $\hspace{-1cm}\phantom{\Jic{2}{\qp\to q}\left(1_{\qp},3_q\right)}+\frac{1}{2}\left[\Gammaoneconv{gq}{x_1}\otimes\Gammaoneconv{qg}{x_1}\right]\delta_3-\iGammatwo{q\qp}{x_1}\delta_3$  \\\cline{2-2}
			\hline
			
			\multirow{2}{*}{IF$_{q}^{q\gets \qp}$}
			& $\Jic{2}{q \gets \qp}\left(3_{\qp},1_q\right)=\XTZe{B}{\qp}{q}-\left[\Gammaoneconv{qg}{x_3}\otimes \XTZd{A}{g}{q}\right]$\\
			& $\hspace{-1cm}\phantom{\Jic{2}{q\gets \qp}\left(3_{\qp},1_q\right)}+\frac{1}{2}\left[\Gammaoneconv{qg}{x_3}\otimes\Gammaoneconv{gq}{x_3}\right]\delta_1-\Gammatwo{q\qp}{x_3}\delta_1$  \\\cline{2-2}
			\hline
			
			\multirow{1}{*}{IF$^{q}_{\qb\to q}$}
			& $\Jic{2}{\qb\to q}\left(1_{\qb},3_q\right)=\XTZe{C}{q_1}{\qb_3}+\iGammatwo{q\qb}{x_1}\delta_3$  \\\cline{2-2}
			\hline
			
			\multirow{1}{*}{IF$_{q}^{q\gets \qb}$}
			& $\Jic{2}{\qb\gets q}\left(3_{\qb},1_q\right)=\XTZe{C}{q_3}{\qb_1}+\Gammatwo{q\qb}{x_3}\delta_1$  \\\cline{2-2}
			\hline
			
			\multirow{4}{*}{IF$_{g\to q}^{g\gets q}$}
			& $\Jic{2}{g\to q,g\gets q}\left(1_q,3_{\qb}\right)=\XTZe{A}{g}{g}+\Sgtoq\left[\Gammaoneconv{qg}{x_1}\otimes \XTZd{A}{g}{q}\right]$\\
			& $\hspace{-1cm}\phantom{\Jic{2}{g\to q,\qb\gets g}\left(1_q,3_{\qb}\right)}+\left[\Gammaoneconv{gq}{x_3}\otimes \XTZd{A}{q}{g}\right]-2\Sgtoq\Gammaone{qg}{x_1}\Gammaone{gq}{x_3}$\\
			& $\Jtic{2}{g\to q,g\gets q}\left(1_q,3_{\qb}\right)=\XTZe{\wt{A}}{g}{g}+\Sgtoq\left[\Gammaoneconv{qg}{x_1}\otimes \XTZd{A}{g}{q}\right]$\\
			& $\hspace{-1cm}\phantom{\Jic{2}{g\to q,g\gets q}\left(1_q,3_{\qb}\right)}+\left[\Gammaoneconv{gq}{x_3}\otimes \XTZd{A}{q}{g}\right]-2\Sgtoq\Gammaone{qg}{x_1}\Gammaone{gq}{x_3}$ \\\cline{2-2}
			\hline
		\end{tabular}}
		\caption{Initial-final identity-changing quark-antiquark two-loop colour-stripped integrated dipoles.}
		\label{tab:J22ICqqbII}
	\end{table}
	\endgroup

	\begingroup
	\renewcommand{\arraystretch}{1.8} 
	\begin{table}[h]
		\centering\small\scalebox{0.9}{
		\begin{tabular}{c|l}
			& Integrated dipoles
			\\ \hline
			
			\multirow{8}{*}{FF$^{q\gets g}$}
			& $\Jic{2}{q\gets g}(3_g,i_{q})=-\XFZ{E}{q_2}-\XTO{E}{\qp}-\boelite\QQslite{3i}\XTZ{E}{\qp}$\\
			& $\hspace{-1cm}\phantom{\Jic{2}{q\gets g}(3_g,i_{q})}+2\left[\XTZd{D}{g}{g\gets g}\otimes\XTZ{E}{\qp}\right]+\left[\Gammaoneconv{qq}{x_3}\otimes\XTZ{E}{\qp}\right]$\\
			& $\hspace{-1cm}\phantom{\Jic{2}{q\gets g}(3_g,i_{q})}-\left[\Gammaoneconv{gg}{x_3}\otimes\XTZ{E}{\qp}\right]-\frac{1}{2}\left[\Gammaoneconv{qq}{x_3}\otimes\Gammaoneconv{qg}{x_3}\right]\delta_1$\\
			& $\hspace{-1cm}\phantom{\Jic{2}{q\gets g}(3_g,i_{q})}+\frac{1}{2}\left[\Gammaoneconv{gg}{x_3}\otimes\Gammaoneconv{qg}{x_3}\right]\delta_1+\Gammatwo{qg}{x_3}\deltaone$\\
			& $\Jhic{2}{q\gets g}(3_g,i_{q})=-\XTO{\wh{E}}{\qp}-\bFoelite\QQslite{3i}\XTZ{E}{\qp}-\left[\GammaoneFconv{gg}{x_3}\otimes\XTZ{E}{\qp}\right]$\\
			& $\hspace{-1cm}\phantom{\Jhic{2}{q\gets g}(3_g,i_{q})}+\frac{1}{2}\left[\GammaoneFconv{gg}{x_3}\otimes\Gammaoneconv{qg}{x_3}\right]\delta_1+\GammatwoF{qg}{x_3}\deltaone$\\
			& $\Jtic{2}{q\gets g}(3_g,i_{q})=-\XFZ{\wt{E}}{q_2}-\XTO{\wt{E}}{\qp}+\left[\Gammaoneconv{qq}{x_3}\otimes\XTZ{E}{\qp}\right]$\\
			& $\hspace{-1cm}\phantom{\Jtic{2}{q\gets g}(3_g,i_{q})}-\frac{1}{2}\left[\Gammaoneconv{qq}{x_3}\otimes\Gammaoneconv{qg}{x_3}\right]\delta_1-\Gammatwot{qg}{x_3}\deltaone$  \\\cline{2-2}
			\hline
		\end{tabular}}
		\caption{Identity-changing final-final quark-gluon two-loop colour-stripped integrated dipoles.}
		\label{tab:J22FFICqg}
	\end{table}
	\endgroup

	\begingroup
	\renewcommand{\arraystretch}{1.8} 
	\begin{table}[h]
		\centering\small\scalebox{0.9}{
		\begin{tabular}{c|l}
			& Integrated dipoles
			\\ \hline
			
			\multirow{10}{*}{IF$^{q}_{q\to g}$}
			& $\Jic{2}{q\to g}(1_g,3_q)=-\XTZe{E}{q}{\qp}-\XTZe{E}{q}{\qbp}-\XTZf{E}{q}{\qp}-\boelite\QQslite{13}\XTZd{E}{q}{\qp}$\\
			& $\hspace{-1cm}\phantom{\Jic{2}{q\to g}(1_g,3_q)}+2\left[\XTZd{D}{q}{g}\otimes\XTZd{E}{q}{\qp}\right]+\left[\Gammaoneconv{qq}{x_1}\otimes\XTZd{E}{q}{\qp}\right]$\\
			& $\hspace{-1cm}\phantom{\Jic{2}{q\to g}(1_g,3_q)}-\left[\Gammaoneconv{gg}{x_1}\otimes\XTZd{E}{q}{\qp}\right]+\frac{1}{2}\Sgtoq\left[\Gammaoneconv{qq}{x_1}\otimes\Gammaoneconv{gq}{x_1}\right]\delta_3$\\
			& $\hspace{-1cm}\phantom{\Jic{2}{q\to g}(1_g,3_q)}-\frac{1}{2}\Sqtog\left[\Gammaoneconv{gg}{x_1}\otimes\Gammaoneconv{gq}{x_1}\right]\delta_3$\\
			& $\hspace{-1cm}\phantom{\Jic{2}{q\to g}(1_g,3_q)}-\Sqtog\iGammatwo{gq}{x_1}\delta_3$\\ 
			& $\Jhic{2}{q\to g}(1_g,3_q)=-\XTZf{\wh{E}}{q}{\qp}-\bFoelite\QQslite{13}\XTZd{E}{q}{\qp}$\\
			& $\hspace{-1cm}\phantom{\Jic{2}{q\to g}(1_g,3_q)}-\left[\GammaoneFconv{gg}{x_1}\otimes\XTZd{E}{q}{\qp}\right]$\\
			& $\hspace{-1cm}\phantom{\Jic{2}{q\to g}(1_g,3_q)}-\frac{1}{2}\Sqtog\left[\GammaoneFconv{gg}{x_1}\otimes\Gammaoneconv{gq}{x_1}\right]\delta_3-\Sqtog\iGammatwoF{gq}{x_1}\delta_3$\\
			& $\Jtic{2}{q\to g}(1_g,3_q)=-\XTZe{\wt{E}}{q}{\qp}-\XTZf{E}{q}{\qp}+\left[\Gammaoneconv{qq}{x_1}\otimes\XTZd{E}{q}{\qp}\right]$\\
			& $\hspace{-1cm}\phantom{\Jtic{2}{q\to g}(1_g,3_q)}+\frac{1}{2}\Sqtog\left[\Gammaoneconv{qq}{x_1}\otimes\Gammaoneconv{gq}{x_1}\right]\delta_3+\Sqtog\iGammatwot{gq}{x_1}\delta_3$  \\\cline{2-2}
			\hline
			
			\multirow{10}{*}{IF$_{q}^{q\gets g}$}
			& $\Jic{2}{q\gets g}(1_q,3_g)=-\XTZe{E}{\qp}{q}-\XTZe{E}{\qbp}{q}-\XTZf{E}{\qp}{q}-\boelite\QQslite{13}\XTZd{E}{\qp}{q}$\\
			& $\hspace{-1cm}\phantom{\Jic{2}{q\gets g}(3_g,1_q)}+2\left[\XTZd{D}{g}{q}\otimes\XTZd{E}{\qp}{q}\right]+\left[\Gammaoneconv{qq}{x_3}\otimes\XTZd{E}{\qp}{q}\right]$\\
			& $\hspace{-1cm}\phantom{\Jic{2}{q\gets g}(3_g,1_q)}-\left[\Gammaoneconv{gg}{x_3}\otimes\XTZd{E}{\qp}{q}\right]-\frac{1}{2}\left[\Gammaoneconv{qq}{x_3}\otimes\Gammaoneconv{qg}{x_3}\right]\delta_1$\\
			& $\hspace{-1cm}\phantom{\Jic{2}{q\gets g}(3_g,1_q)}+\frac{1}{2}\left[\Gammaoneconv{gg}{x_3}\otimes\Gammaoneconv{qg}{x_3}\right]\delta_1$\\
			& $\hspace{-1cm}\phantom{\Jic{2}{q\gets g}(3_g,1_q)}+\iGammatwo{qg}{x_3}\delta_1$\\ 
			& $\Jhic{2}{q\gets g}(1_q,3_g)=-\XTZf{\wh{E}}{\qp}{q}-\bFoelite\QQslite{13}\XTZd{E}{\qp}{q}$\\
			& $\hspace{-1cm}\phantom{\Jic{2}{q\gets g}(3_g,1_q)}-\left[\GammaoneFconv{gg}{x_3}\otimes\XTZd{E}{\qp}{q}\right]$\\
			& $\hspace{-1cm}\phantom{\Jic{2}{q\gets g}(3_g,1_q)}+\frac{1}{2}\left[\GammaoneFconv{gg}{x_3}\otimes\Gammaoneconv{qg}{x_3}\right]\delta_1+\GammatwoF{qg}{x_3}\delta_1$\\
			& $\Jtic{2}{q\gets g}(1_q,3_g)=-\XTZe{\wt{E}}{\qp}{q}-\XTZf{E}{\qp}{q}+\left[\Gammaoneconv{qq}{x_3}\otimes\XTZd{E}{\qp}{q}\right]$\\
			& $\hspace{-1cm}\phantom{\Jtic{2}{q\gets g}(3_g,1_q)}-\frac{1}{2}\left[\Gammaoneconv{qq}{x_3}\otimes\Gammaoneconv{qg}{x_3}\right]\delta_1-\iGammatwot{qg}{x_3}\delta_1$  \\\cline{2-2}
			\hline
			
			\multirow{4}{*}{IF$_{\qp \to g}^{g\gets q}$}
			& $\Jic{2}{\qp \to g,g\gets q}(1_g,3_q)=2\XTZe{E}{g}{\qp}+2\XTZe{E}{g}{\qbp}-2\Sqtog\left[\Gammaoneconv{gq}{x_1}\otimes \XTZd{D}{g}{g}\right]$\\
			& $\hspace{-1cm}\phantom{\Jic{2}{\qp \to g,g\gets q}(1_g,3_q)}-\left[\Gammaoneconv{gq}{x_3}\otimes \XTZd{E}{q}{\qp}\right]-2\Sqtog\left[\Gammaoneconv{gq}{x_1}\otimes\Gammaone{gq}{x_3}\right]$\\
			& $\hspace{-1cm}\phantom{\Jic{2}{\qp \to g,g\gets q}(1_g,3_q)}-\mathcal{B}^0_{4,\qp}+\left[\mathcal{E}^0_{3,\qp}\otimes \mathcal{A}^0_{3,g}\right]$\\
			& $\hspace{-1cm}\phantom{\Jic{2}{\qp \to g,g\gets q}(1_g,3_q)}+2\Sgtoq\left[\Gammaoneconv{qg}{x_1}\otimes \mathcal{E}^0_{3,\qp}\right]+2\left[\Gammaoneconv{gq}{x_1}\otimes\Gammaone{qg}{x_1}\right]\delta_3$\\
			\hline
			
			\multirow{4}{*}{IF$^{\qp\gets g}_{g\to q}$}
			& $\Jic{2}{g\to q,\qp\gets g}(3_g,1_q)=2\XTZe{E}{\qp}{g}+2\XTZe{E}{\qbp}{g}-2\left[\Gammaoneconv{qg}{x_3}\otimes \XTZd{D}{g}{g}\right]$\\
			& $\hspace{-1cm}\phantom{\Jic{2}{g\to q,\qp\gets g}(3_g,1_q)}+2\Sgtoq\left[\Gammaoneconv{qg}{x_1}\otimes \XTZd{E}{\qp}{q}\right]-2\Sqtog\left[\Gammaoneconv{qg}{x_3}\otimes\Gammaone{qg}{x_1}\right]$\\
			& $\hspace{-1cm}\phantom{\Jic{2}{g\to q,\qp\gets g}(3_g,1_q)}-\mathcal{B}^{0,\mathrm{id.}\qp}_{4}+\left[\mathcal{E}^{0,\mathrm{id.}\qp}_{3}\otimes \mathcal{A}^{0,\mathrm{id.}g}_{3}\right]$\\
			& $\hspace{-1cm}\phantom{\Jic{2}{g\to q,\qp\gets g}(3_g,1_q)}-2\left[\Gammaoneconv{gq}{x_3}\otimes \mathcal{E}^{0,\mathrm{id.}\qp}_{3}\right]+2\left[\Gammaoneconv{qg}{x_3}\otimes\Gammaone{gq}{x_3}\right]\delta_1$\\
			\hline
		\end{tabular}}
		\caption{Identity-changing initial-final quark-gluon two-loop colour-stripped integrated dipoles.}
		\label{tab:J22IFICqg}
	\end{table}
	\endgroup

	\begingroup
	\renewcommand{\arraystretch}{1.8} 
	\begin{table}[h]
		\centering\small\scalebox{0.9}{
		\begin{tabular}{c|l}
			& Integrated dipoles
			\\ \hline
			
			\multirow{10}{*}{FF$^{q\gets g}$}
			& $\Jic{2}{q\gets g}\left(3_g,i_g\right)=-\XFZ{G}{q}-\XTO{G}{q}-\boelite\QQslite{3i}\XTZ{G}{\qp}$\\
			& $\hspace{-1cm}\phantom{\Jic{2}{q\gets g}\left(3_g,i_g\right)}+\left[\XTZ{G}{\qp}\otimes\XTZ{F}{g}\right]+\Gammatwo{qg}{x_3}\deltaone$\\
			& $\hspace{-1cm}\phantom{\Jic{2}{q\gets g}\left(3_g,i_g\right)}+\left[\Gammaoneconv{qq}{x_3}\otimes\XTZ{G}{\qp}\right]-\left[\Gammaoneconv{gg}{x_3}\otimes\XTZ{G}{\qp}\right]$\\
			& $\hspace{-1cm}\phantom{\Jic{2}{q\gets g}\left(3_g,i_g\right)}-\frac{1}{2}\left[\Gammaoneconv{qq}{x_3}\otimes\Gammaoneconv{qg}{x_3}\right]+\frac{1}{2}\left[\Gammaoneconv{gg}{x_3}\otimes\Gammaoneconv{qg}{x_3}\right]\delta_1$\\
			& $\Jhic{2}{q\gets g}(3_g,i_g)=\XFZ{H}{q}-\XTO{\wh{G}}{q}-\bFoelite\QQslite{3i}\XTZ{G}{\qp}$\\
			& $\hspace{-1cm}\phantom{\Jhic{2}{q\gets g}(3_g,i_g)}-\left[\Gammaoneconv{qg}{x_3}\otimes\XTZ{G}{g}\right]-\frac{1}{2}\left[\Gammaoneconv{qg}{x_3}\otimes\GammaoneF{gg}{x_3}\right]\delta_1$\\
			& $\hspace{-1cm}\phantom{\Jhic{2}{q\gets g}(3_g,i_g)}+\GammatwoF{qg}{x_3}\deltaone$\\
			& $\Jtic{2}{q\gets g}\left(3_g,i_g\right)=-\frac{1}{2}\XFZ{\wt{G}}{q}-\XTO{\wt{G}}{q}+\left[\XTZ{G}{\qp}\otimes\Gammaoneconv{qq}{x_3}\right]$\\
			& $\hspace{-1cm}\phantom{\Jtic{2}{q\gets g}\left(3_g,i_g\right)}-\frac{1}{2}\left[\Gammaoneconv{qq}{x_3}\otimes\Gammaoneconv{qg}{x_3}\right]\delta_1-\Gammatwot{qg}{x_3}\deltaone$  \\\cline{2-2}
			\hline
		
		\end{tabular}}
		\caption{Identity-changing final-final gluon-gluon two-loop colour-stripped integrated dipoles.}
		\label{tab:J22FFICgg}
	\end{table}
	\endgroup

	\begingroup
	\renewcommand{\arraystretch}{1.8} 
	\begin{table}[h]
		\centering\small\scalebox{0.9}{
		\begin{tabular}{c|l}
			& Integrated dipoles
			\\ \hline
			
			\multirow{10}{*}{IF$^{g}_{q\to g}$}
			& $\Jic{2}{q\to g}\left(1_g,3_g\right)=-\XTZe{G}{g_1}{q}-\XTZe{G}{g_2}{q}-\XTZf{G}{g}{\qp}-\boelite\QQslite{13}\XTZd{G}{g}{\qp}$\\
			& $\hspace{-1cm}\phantom{\Jic{2}{q\to g}\left(1_g,3_g\right)}+2\left[\XTZd{G}{g}{\qp}\otimes\XTZd{F}{g}{g}\right]-\Sqtog\iGammatwo{gq}{x_1}\delta_3$\\
			& $\hspace{-1cm}\phantom{\Jic{2}{q\to g}\left(1_g,3_g\right)}+\left[\Gammaoneconv{gq}{x_3}\otimes\XTZd{G}{q}{q}\right]+\left[\Gammaoneconv{qq}{x_1}\otimes\XTZd{G}{g}{\qp}\right]-\left[\Gammaoneconv{gg}{x_1}\otimes\XTZd{G}{g}{\qp}\right]$\\
			& $\hspace{-1cm}\phantom{\Jic{2}{q\to g}\left(1_g,3_g\right)}+\frac{1}{2}\Sqtog\left[\Gammaoneconv{qq}{x_1}\otimes\Gammaoneconv{gq}{x_1}\right]\delta_3$\\
			& $\hspace{-1cm}\phantom{\Jic{2}{q\to g}\left(1_g,3_g\right)}-\frac{1}{2}\Sqtog\left[\Gammaoneconv{gg}{x_1}\otimes\Gammaoneconv{gq}{x_1}\right]\delta_3$\\
			& $\Jhic{2}{q\to g}\left(1_g,3_g\right)=-\XTZf{\wh{G}}{g}{\qp}-\bFoelite\QQslite{13}\XTZd{G}{g}{\qp}-\Sqtog\iGammatwoF{gq}{x_1}\delta_3$\\
			& $\hspace{-1cm}\phantom{\Jhic{2}{q\to g}\left(1_g,3_g\right)}-\left[\XTZd{G}{g}{\qp}\otimes\GammaoneFconv{gg}{x_1}\right]-\frac{1}{2}\Sqtog\left[\Gammaoneconv{gq}{x_1}\otimes\GammaoneFconv{gg}{x_1}\right]\delta_3$\\
			& $\Jtic{2}{q\to g}\left(1_g,3_g\right)=-\XTZe{G}{g}{q}-\XTZf{G}{g}{\qp}+\left[\XTZd{G}{g}{\qp}\otimes\Gammaoneconv{qq}{x_1}\right]$\\
			& $\hspace{-1cm}\phantom{\Jtic{2}{q\to g}\left(1_g,3_g\right)}+\frac{1}{2}\Sqtog\left[\Gammaoneconv{qq}{x_1}\otimes\Gammaoneconv{gq}{x_1}\right]\delta_3+\Sqtog\iGammatwot{gq}{x_1}\delta_3$ \\
			&$\hspace{-1cm}\phantom{\Jtic{2}{q\to g}\left(1_g,3_g\right)}-\left[\XTZd{G}{q}{q}\otimes\Gammaoneconv{gq}{x_3}\right]$  \\\cline{2-2}
			\hline
			
			\multirow{10}{*}{IF$_{g}^{q\gets g}$}
			& $\Jic{2}{q\gets g}\left(3_g,1_g\right)=-\XTZe{G}{q}{g_1}-\XTZe{G}{q}{g_2}-\XTZf{G}{\qp}{g}-\boelite\QQslite{13}\XTZd{G}{q}{g}$\\
			& $\hspace{-1cm}\phantom{\Jic{2}{q\gets g}\left(3_g,1_g\right)}+2\left[\XTZd{G}{q}{g}\otimes\XTZd{F}{g}{g}\right]+\Gammatwo{qg}{x_3}\delta_1$\\
			& $\hspace{-1cm}\phantom{\Jic{2}{q\gets g}\left(3_g,1_g\right)}+\Sgtoq\left[\Gammaoneconv{qg}{x_1}\otimes\XTZd{G}{q}{q}\right]+\left[\Gammaoneconv{qq}{x_3}\otimes\XTZd{G}{q}{g}\right]-\left[\Gammaoneconv{gg}{x_3}\otimes\XTZd{G}{q}{g}\right]$\\
			& $\hspace{-1cm}\phantom{\Jic{2}{q\gets g}\left(3_g,1_g\right)}-\frac{1}{2}\left[\Gammaoneconv{qq}{x_3}\otimes\Gammaoneconv{qg}{x_3}\right]\delta_1$\\
			& $\hspace{-1cm}\phantom{\Jic{2}{q\gets g}\left(3_g,1_g\right)}+\frac{1}{2}\left[\Gammaoneconv{gg}{x_3}\otimes\Gammaoneconv{qg}{x_3}\right]\delta_1$\\
			& $\Jhic{2}{q\gets g}\left(3_g,1_g\right)=-\XTZf{\wh{G}}{\qp}{g}-\bFoelite\QQslite{13}\XTZd{G}{q}{g}+\GammatwoF{qg}{x_3}\delta_1$\\
			& $\hspace{-1cm}\phantom{\Jhic{2}{q\gets g}\left(3_g,1_g\right)}-\left[\XTZd{G}{\qp}{g}\otimes\GammaoneFconv{gg}{x_3}\right]+\frac{1}{2}\left[\Gammaoneconv{qg}{x_3}\otimes\GammaoneFconv{gg}{x_3}\right]\delta_1$\\
			& $\Jtic{2}{q\gets g}\left(3_g,1_g\right)=-\XTZe{G}{q}{g}-\XTZf{G}{\qp}{g}+\left[\XTZd{G}{q}{g}\otimes\Gammaoneconv{qq}{x_3}\right]$\\
			& $\hspace{-1cm}\phantom{\Jtic{2}{q\gets g}\left(3_g,1_g\right)}-\frac{1}{2}\left[\Gammaoneconv{qq}{x_3}\otimes\Gammaoneconv{qg}{x_3}\right]\delta_1-\Gammatwot{qg}{x_3}\delta_1$ \\
			&$\hspace{-1cm}\phantom{\Jtic{2}{q\gets g}\left(3_g,1_g\right)}+\Sgtoq\left[\XTZd{G}{q}{q}\otimes\Gammaoneconv{qg}{x_1}\right]$  \\\cline{2-2}
			\hline
			
			\multirow{2}{*}{IF$_{q \to g}^{q\gets g}$}
			& $\Jic{2}{q\to g,q\gets g}\left(1_g,3_g\right)=\XTZe{H}{q}{\qp}+\Sqtog\left[\Gammaoneconv{gq}{x_1}\otimes \XTZd{G}{q}{g}\right]$\\
			& $\hspace{-1cm}\phantom{\Jic{2}{q\to g,q\gets g}(1_g,3_g)}-\left[\Gammaoneconv{qg}{x_3}\otimes \XTZd{G}{g}{\qp}\right]-\Sqtog\Gammaone{gq}{x_1}\Gammaone{qg}{x_3}$  \\\cline{2-2}
			\hline
		\end{tabular}}
		\caption{Identity-changing initial-final gluon-gluon two-loop colour-stripped integrated dipoles.}
		\label{tab:J22IFICgg}
	\end{table}
	\endgroup
	\FloatBarrier


\section{Conclusions}
\label{sec:conc}

Processes with identified final-state hadrons play an increasingly important role in precision studies at the LHC. To compute NNLO QCD corrections to these observables requires an extension of existing NNLO subtraction methods, in order to keep 
track of the momentum fraction of the final-state fragmenting parton. In the antenna subtraction method, this is accomplished 
through fragmentation antenna functions, which were derived previously at NNLO only for final-final kinematics~\cite{Gehrmann:2022pzd}, relevant  to $e^+e^-$ colliders. In the present work, we complete the set of fragmentation antenna functions for hadron colliders  at NNLO by computing them in initial-final kinematics. To enable the 
efficient and systematic construction of the relevant antenna subtraction terms, we combine the integrated 
fragmentation antenna functions in all kinematical settings with the respective mass factorization counterterms
into dipole operators~\cite{Currie:2013vh,Gehrmann:2023dxm}. These operators  mirror the infrared singularity  structure 
of the corresponding virtual loop amplitudes. 

Our results enable the extension of antenna subtraction at NNLO to important hadron collider processes such as 
 single-inclusive 
hadron production~\cite{Aversa:1988vb} or vector-boson-plus-hadron production~\cite{Caletti:2024xaw}, which were previously accessible only to NLO QCD. Likewise, hadron production processes inside jets at hadron-hadron or 
lepton-hadron colliders can now be computed to NNLO. 

The kinematical setting of the   integrated initial-final fragmentation antenna functions 
is identical to semi-inclusive hadron production in deep inelastic scattering (SIDIS). We could thus employ the same integration techniques as described in~\cite{Gehrmann:2022cih} and in Section~\ref{sec:intant} to compute the 
NNLO QCD coefficient functions 
for unpolarized and polarized  SIDIS, which we already presented 
elsewhere~\cite{Bonino:2024qbh,Bonino:2024wgg}. 

The ability to derive precise predictions for identified hadron cross sections has important phenomenological implications. It will 
for the first time enable global NNLO determinations of fragmentation functions for a variety of hadron species, thereby 
bringing the description of fragmentation processes to a new level of quantitative accuracy. Specific final-state hadron 
observables also allow to determine important  aspects of the structure of the colliding hadrons, such as the flavour decomposition of 
the quark sea. The identification of heavy-flavoured hadrons is moreover a common approach to jet flavour tagging, which can now be mirrored in a direct manner in theory calculations, thus allowing a potential alternative to 
the application of a flavoured jet algorithm~\cite{Caletti:2022hnc,Czakon:2022wam,Gauld:2022lem,Caola:2023wpj}.

\acknowledgments
This work has received funding from the Swiss National Science Foundation (SNF)
under contract 200020-204200 and from the European Research Council (ERC) under
the European Union's Horizon 2020 research and innovation programme grant
agreement 101019620 (ERC Advanced Grant TOPUP). MM is supported by a Royal Society Newton International Fellowship (NIF/R1/232539).

\begin{appendix}
\section{Master integrals for double-real radiation}
\label{app:MI}
In the following, we document all master integrals relevant for the integrated double-real radiation fragmentation 
antenna functions in initial-final kinematics $\mathcal{X}^{0,\,\mathrm{id.} j}_{4,i}(x,z)$. The Laurent expansions of these 
 integrals are required for generic $(x,z)$ up to transcendental weight 2, while they are required for generic $z$ at $x=1$ and 
 for generic $x$ at $z=1$ up to transcendental weight 3 and at the soft endpoint $x=z=1$ up to weight 4. 

\subsection{Normalization of integrals}
The full set of master integrals reads:
\begin{eqnarray}
I[0] (Q^2,x,z)&=& N_{\Gamma} \left(Q^2\right)^{1-2\epsilon} (1-x)^{1-2\epsilon} (1-z)^{1-2\epsilon}x^{-1} r_0(x,z) \, ,\nonumber  \\
I[1](Q^2,x,z) &=& N_{\Gamma} (Q^2)^{-2 \e} (1-x)^{-2 \e} (1-z)^{-2 \e} \; r_1(x,z)\, ,\nonumber \\
I[5](Q^2,x,z) &=& N_{\Gamma}   (Q^2)^{-2 \e} (1-x)^{-2 \e} (1-z)^{1-2 \e} \; r_5(x,z) \, ,\nonumber  \\
I[7](Q^2,x,z) &=& N_{\Gamma}   (Q^2)^{-2 \e} (1-x)^{1-2 \e} (1-z)^{1-2 \e}  \;   r_7(x,z) \, ,\nonumber \\
I[-2,7](Q^2,x,z) &=& N_{\Gamma}   (Q^2)^{1-2 \e} (1-x)^{2-2 \e} (1-z)^{1-2 \e}  \; r_{m27}(x,z)\, ,   \nonumber \\
I[-3,7](Q^2,x,z) &=& N_{\Gamma}   (Q^2)^{1-2 \e} (1-x)^{1-2 \e} (1-z)^{1-2 \e} \;  r_{m37}(x,z) \, ,\nonumber \\
I[1,4](Q^2,x,z) &=& N_{\Gamma}    (Q^2)^{-2 \e-1} (1-x)^{-2 \e} (1-z)^{-2 \e} \;  r_{14}(x,z) \, ,\nonumber\\
I[4,7](Q^2,x,z) &=& N_{\Gamma}      (Q^2)^{-2 \e-1} (1-x)^{-2 \e} (1-z)^{-2 \e}\;  r_{47}(x,z) \, , \nonumber\\
I[5,7](Q^2,x,z) &=& N_{\Gamma}     (Q^2)^{-2 \e-1} (1-x)^{-2 \e} (1-z)^{1-2 \e}\;  r_{57}(x,z)  \, ,\nonumber\\
I[1,3,4](Q^2,x,z) &=& N_{\Gamma}    (Q^2)^{-2 \e-2} (1-x)^{-2 \e} (1-z)^{-2 \e-1} 2x^2 (1+x)^{-1} \; r_{134}(x,z)  \, , \nonumber\\
I[1,3,5](Q^2,x,z) &=& N_{\Gamma}     (Q^2)^{-2 \e-2} (1-x)^{-2 \e} (1-z)^{-2 \e-1}x\;  r_{135}(x,z) \, , \nonumber\\
I[1,3,8](Q^2,x,z) &=& N_{\Gamma}    (Q^2)^{-2 \e-2} (1-x)^{-2 \e} (1-z)^{-2 \e-1}x^2z^{-1} \; r_{138}(x,z)   \, , \nonumber \\
I[1,4,5](Q^2,x,z) &=& N_{\Gamma}      (Q^2)^{-2 \e-2} (1-x)^{-2 \e-1} (1-z)^{-2 \e}x^2z^{-1} \;  r_{145}(x,z)  \, , \nonumber \\
I[2,3,5](Q^2,x,z) &=& N_{\Gamma}      (Q^2)^{-2 \e-2} (1-x)^{-2 \e-1} (1-z)^{-2 \e-1}x^2 \;  r_{235}(x,z) \, ,\nonumber \\
I[2,3,7](Q^2,x,z) &=& N_{\Gamma}      (Q^2)^{-2 \e-2} (1-x)^{-2 \e} (1-z)^{-2 \e-1}x^2z^{-1}\;  r_{237}(x,z) \, , \nonumber\\
I[2,4,5](Q^2,x,z) &=& N_{\Gamma}        (Q^2)^{-2 \e-2} (1-x)^{-2 \e-1} (1-z)^{-2 \e}x^2z^{-1}\;  r_{245}(x,z)  \, , \nonumber\\
I[3,4,7](Q^2,x,z) &=& N_{\Gamma}    (Q^2)^{-2 \e-2} (1-x)^{-2 \e} (1-z)^{-2 \e}2x^2 (1+x)^{-1}z^{-1}  \; r_{347}(x,z)  \, , \nonumber\\
I[3,5,7](Q^2,x,z) &=& N_{\Gamma}    (Q^2)^{-2 \e-2} (1-x)^{-2 \e} (1-z)^{-2 \e}xz^{-1} \; r_{357}(x,z)   \, ,\nonumber \\
I[3,5,8](Q^2,x,z) &=& N_{\Gamma}     (Q^2)^{-2 \e-2} (1-x)^{-2 \e-1} (1-z)^{-2 \e}x^2z^{-1}\; r_{358}(x,z)  \, , \nonumber \\
I[4,5,7](Q^2,x,z) &=& N_{\Gamma}      (Q^2)^{-2 \e-2} (1-x)^{-2 \e-1} (1-z)^{-2 \e}x^2z^{-1}\; r_{457}(x,z) \, , \nonumber \\
I[4,5,8](Q^2,x,z) &=& N_{\Gamma}      (Q^2)^{-2 \e-2} (1-x)^{-2 \e-1} (1-z)^{-2 \e}x^2z^{-1} \;  r_{458}(x,z) \, .
\end{eqnarray}
The dimensionless functions $r_i(x,z)$ can be expressed in Laurent expansions in $\e$. The prefactors are chosen to properly isolate the 
dominant behaviour in the $x=1$ and $z=1$ endpoints, and to account for common rational prefactors where appropriate. 

The $r_i(x,z)$ are documented below, truncated in $\e$ to the required orders for the respective kinematical regions.

\subsection{Hard region}
The master integrals in the hard region ($x<1$, $z<1$) read:
\begin{eqnarray}
r_{0}(x,y) &=& 1+\e(2\ln(x)-\ln(z))+\e^2\left(2\ln^2(x)-2 \ln(x)\ln(z)+\frac{\ln^2(z)}{2}\right)+{\cal O}(\e^3)\,,\nonumber \\
r_{1}(x,z) &=& \ln(x) 
+ \e \left(2 \text{Li}_2(x)+\ln (x) (2 \ln (1-x)-\ln (z)-2)+\frac{3 \ln ^2(x)}{2}-\frac{\pi ^2}{3}\right) \nonumber \\ &&+ {\cal O}(\e^2)\,, \nonumber \\
r_{5}(x,z) &=&
\frac{\ln (z)}{\e (1-z)} -
\frac{-12 \text{Li}_2(z)-12 \ln (z) (\ln (x)+\ln (1-z)-2)+9 \ln ^2(z)+2 \pi ^2}{6 (1-z)} \nonumber \\ &&
+{\cal O}(\e)\,,\nonumber \\
r_{7}(x,z) &=& -\frac{\ln(x)\ln(z)}{(1-x)(1-z)}+{\cal O}(\e)\,,
 \nonumber \\
 r_{m27}(x,z) &=& \frac{(1-x) (2-2 z+(z+1) \ln (z))+\ln (x) ((x+1) (1-z)+(x+z) \ln (z))}{(1-x)^2 x (1-z)} \nonumber \\ &&+{\cal O}(\e)\,,
 \nonumber \\
 r_{m37}(x,z) &=& \frac{\ln (x) (z-z \ln (z)-1)}{(1-x) x (1-z)}+{\cal O}(\e)\,,
 \nonumber \\
r_{14}(x,z) &=& \frac{2x}{u}  \Bigg(-\text{Li}_2\left(\frac{1-u-x}{2}\right)+\text{Li}_2\left(\frac{1+u-x}{2} \right)-\text{Li}_2\left(\frac{u+x-1}{2 x}\right)
 \nonumber \\ &&+\text{Li}_2\left(-\frac{u-x+1}{2 x}\right)+\ln (x) \ln (1-u+x)-\ln (x) \ln (1+u+x)\Bigg)+{\cal O}(\e)\,,
\nonumber \\
r_{47}(x,z) &=& -2 \sqrt{\frac{x}{z}} \Bigg( \text{Ti}_2\left( \sqrt{xz}\right)-\text{Ti}_2\left(-\sqrt{xz}\right)-\text{Ti}_2\left( \sqrt{\frac{x}{z}}\right)+\text{Ti}_2\left(- \sqrt{\frac{x}{z}}\right) \nonumber \\
&&
-\ln \left(xz\right) \arctan\left( \sqrt{xz}\right)+\ln \left(\frac{x}{z}\right) \arctan\left(\sqrt{\frac{x}{z}}\right)\Bigg) +{\cal O}(\e)\,,\nonumber \\
r_{57}(x,z) &=& \frac{x}{(1-z)v}\Bigg(\frac{2\left( \ln(v-z+1) - \ln (v+z-1) \right) }{\e} + 
8 \text{Li}_2\left(\frac{1-z}{v}\right)\nonumber \\ &&
-4 \text{Li}_2\left(\frac{1}{2} (-v-z+1)\right)+4 \text{Li}_2\left(\frac{v+z-1}{v-z+1}\right)+8 \text{Li}_2\left(\frac{v}{z-1}\right)\nonumber \\ &&
+4 \text{Li}_2\left(\frac{-v+z+1}{2 z}\right)+4 \text{Li}_2\left(-\frac{v-z+1}{2 z}\right)-4 \text{Li}_2\left(\frac{1}{2} (-v+z+1)\right)\nonumber \\ &&
+3 \ln ^2(v-z+1)- \ln ^2(v+z-1)+2 \ln (x) \ln (v-z+1)\nonumber \\ &&
+16 \ln (2) \ln (v+z-1)-4 \ln (1- z) \ln (v-z+1)-8 \ln (v-z+1)\nonumber \\ &&
-2 \ln (z) \ln (v-z+1)-2 \ln (v-z+1) \ln (v+z-1)+6 \ln (z) \ln (v+z-1)\nonumber \\ &&
+8 \ln (v+z-1)+4 \ln (v-z+1) \ln (v+z+1)-4 \ln (z) \ln (v+z+1)\nonumber \\ &&
-4 \ln (v+z-1) \ln (v+z+1)-8 \ln (v) \ln (1- z)+4 \ln (1- z) \ln (v+z-1)\nonumber \\ &&
-2\ln(x)\ln(v+z-1)+4 \ln ^2(v)+4 \ln ^2(1-z)-8 \ln(2)\ln(x)\nonumber \\ &&
-8 \ln (2) \ln (z)-4 \ln (z)\ln(x)+\frac{2 \pi ^2}{3}-16 \ln ^2(2)
\Bigg) +{\cal O}(\e)\,,
\nonumber \\
r_{134}(x,z) &=& -\frac{\ln(x)}{\e} +2 \text{Li}_2(-x)-2 \text{Li}_2(x)-2 \ln^2(x)-2 \ln (1-x) \ln (x) \nonumber \\ &&
+2 \ln (x+1) \ln (x)+4 \ln (x)+\frac{\pi ^2}{2}+{\cal O}(\e)\,,\nonumber \\
r_{135}(x,z) &=& \frac{1}{\e^2} - \frac{4}{\e} +
2 \text{Li}_2\left(1-\frac{x}{z}\right)-4 \text{Li}_2(x)-4 \ln (1-x) \ln (x)+\frac{2\pi ^2}{3}+4+{\cal O}(\e)\,,\nonumber \\
r_{138}(x,z) &=& \frac{1}{\e^2} + \frac{2  (2 \ln (x)-\ln (z)-2)}{\e}
-2 \text{Li}_2\left(1-\frac{x}{z}\right)+4 \text{Li}_2(x)+5 \ln ^2(x)-16 \ln (x) \nonumber \\ &&
-6 \ln (x) \ln (z) +4 \ln (1-x) \ln (x)+ \ln ^2(z)+8 \ln (z)-\frac{2 \pi ^2}{3}+4+{\cal O}(\e)\,,\nonumber \\
r_{145}(x,z) &=&\frac{\ln(x)}{\e} + 2 \text{Li}_2\left(1-\frac{x}{z}\right)-2 \text{Li}_2(z)-\frac{1}{2} 5 \ln ^2(x)+4 \ln (x)+\ln ^2(z)\nonumber \\ &&-2 \ln (1-z) \ln (z)+\frac{\pi ^2}{3}+{\cal O}(\e)\,,\nonumber \\
r_{235}(x,z) &=&-\frac{3}{\e^2}+\frac{2 (-3 \ln (x)+\ln (z)+6)}{\e}+4 \ln (x) \ln (z)-6 \ln ^2(x)+24 \ln (x) \nonumber \\ &&-8 \ln (z)+\frac{\pi ^2}{3}-12+{\cal O}(\e)\,,\nonumber \\
r_{237}(x,z) &=&r_{138}(x,z)\,,\nonumber \\
r_{245}(x,z) &=&-\frac{1}{\e^2}  + \frac{ -3 \ln (x)+2 \ln (z)+4}{\e} +
2 \text{Li}_2\left(1-\frac{x}{z}\right)+2 \text{Li}_2(z)+4 \ln (x) \ln (z)\nonumber \\ &&
-\frac{7}{2}  \ln ^2(x)+12 \ln (x)-\ln ^2(z)+2 \ln (1-z) \ln (z)-8 \ln (z)\nonumber \\ &&-\frac{\pi ^2}{3}-4  +{\cal O}(\e)\,,
\nonumber \\
r_{347}(x,z) &=&-\frac{\ln (x)}{\e} + 
-\text{Li}_2\left(-\frac{x}{z}\right)-\text{Li}_2(-x z)+2 \text{Li}_2(-x)-2 \text{Li}_2(x)-\ln (x) \ln \left(\frac{x+z}{z}\right)
\nonumber \\ &&
+\ln (x) \ln (z)-\ln (x) \ln (1+x z)+\ln (z) \ln \left(\frac{x+z}{z}\right)-\ln (z) \ln (1+x z)
\nonumber \\ &&
-\frac{3}{2}  \ln ^2(x)-2 \ln (1-x) \ln (x)+2 \ln (1+x) \ln (x)+4 \ln (x)+\frac{\ln ^2(z)}{2}
\nonumber \\ &&
+\frac{\pi ^2}{3}+{\cal O}(\e)\,,
 \nonumber \\
r_{357}(x,z) &=& \frac{(1-z)^2}{2x} r_{57}(x,z) + \frac{1}{\e^2} + \frac{\ln (x)-\ln (z)-4}{\e} 
-2 \text{Li}_2\left(\frac{1}{2} (-v-z+1)\right)\nonumber \\ &&
+2 \text{Li}_2\left(\frac{-v+z+1}{2 z}\right)-2 \text{Li}_2\left(-\frac{v-z+1}{2 z}\right)+2 \text{Li}_2\left(\frac{1}{2} (-v+z+1)\right)\nonumber \\ &&
+\frac{3}{2} \ln ^2(v-z+1)+\frac{3}{2} \ln ^2(v+z-1)+4 \ln (2) \ln (v+z+1)\nonumber \\ &&
-4 \ln (v-z+1)-\ln (v-z+1) \ln (v+z-1)-2 \ln (z) \ln (v+z-1)\nonumber \\ &&-4 \ln (v+z-1)
-2 \ln (v-z+1) \ln (v+z+1)+2 \ln (z) \ln (v+z+1)\nonumber \\ &&-2 \ln (v+z-1) \ln (v+z+1)+8 \ln (z)-\frac{\pi ^2}{3}\nonumber \\ &&
+4-2 \ln ^2(2)+8 \ln (2)
+{\cal O}(\e)\,,
\nonumber \\
r_{358}(x,z) &=&-\frac{2}{\e^2} +\frac{2 (-2 \ln (x)+\ln (z)+4)}{\e}   +  4 \ln (x) \ln (z)-4 \ln^2(x)+16 \ln (x)\nonumber \\ &&
-8 \ln (z)-8   +{\cal O}(\e)\,, \nonumber \\
r_{457}(x,z) &=& \frac{1}{\e}\left[ -\ln(z) + \frac{1+z}{v} \left( \ln(4xz) -2\ln(1-z+v) \right) \right]  \nonumber \\
&& + 
2 \text{Li}_2\left(\frac{-v-z+1}{2}\right)+2 \text{Li}_2\left(\frac{-v+z+1}{2 z}\right)-2 \text{Li}_2\left(-\frac{v-z+1}{2 z}\right)\nonumber \\  &&
-2 \text{Li}_2\left(\frac{-v+z+1}{2} \right)+2 \text{Li}_2\left(-\frac{x}{z}\right)-2 \text{Li}_2(-x z)-2 \text{Li}_2(z)\nonumber \\  &&
+2 \ln (x) \ln (v+z+1)+4 \ln (2) \ln (v-z+1)+4 \ln (2) \ln (v+z+1)\nonumber \\  &&
+2 \ln (z) \ln (v-z+1)-4 \ln (v-z+1) \ln (v+z+1)+4 \ln (z) \ln (v+z+1)\nonumber \\  &&
-3\ln (x) \ln (z)+2 \ln (x) \ln (x+z)-2 \ln (z) \ln (x+z)-2 \ln (x) \ln (x z+1)\nonumber \\  &&
-2 \ln (z) \ln (x z+1)-2 \ln (2) \ln (x)+\frac{3\ln ^2(z)}{2}-6 \ln (2) \ln (z)\nonumber \\  &&-2 \ln (1-z) \ln (z)+4 \ln (z)+\frac{\pi ^2}{3}-4 \ln ^2(2)
\nonumber \\  &&+ \frac{1+z}{v} \Bigg[
-4 \text{Li}_2\left(\frac{1-z}{v}\right)+2 \text{Li}_2\left(\frac{-v-z+1}{2} \right)-2 \text{Li}_2\left(\frac{v+z-1}{v-z+1}\right)\nonumber \\  &&
-4 \text{Li}_2\left(\frac{v}{z-1}\right)-2 \text{Li}_2\left(\frac{-v+z+1}{2 z}\right)-2 \text{Li}_2\left(-\frac{v-z+1}{2 z}\right)\nonumber \\  &&
+2 \text{Li}_2\left(\frac{-v+z+1}{2} \right)
-2 \ln (x) \ln (v-z+1)+2 \ln (x) \ln (v+z+1)\nonumber \\  &&
-2 \ln ^2(v-z+1)+8 \ln (2) \ln (v-z+1)+4 \ln (2) \ln (v+z+1)\nonumber \\  &&
+4 \ln (v) \ln (1-z)+4 \ln (1-z) \ln (v-z+1)+8 \ln (v-z+1)\nonumber \\  &&
+4 \ln (z) \ln (v-z+1)-4 \ln (v-z+1) \ln (v+z+1)+4 \ln (z) \ln (v+z+1)\nonumber \\  &&
-2 \ln ^2(v)-2 \ln (x) \ln (1-z)+\ln (x) \ln (z)+\frac{3 \ln ^2(x)}{2}-4 \ln (x)-2 \ln ^2(1-z)\nonumber \\  &&
-\frac{5 \ln ^2(z)}{2}-4 \ln (2) \ln (1-z)-8 \ln (2) \ln (z)-2 \ln (1-z) \ln (z)-4 \ln (z)\nonumber \\  &&
-\frac{\pi ^2}{3}-6 \ln ^2(2)-8 \ln (2)
\Bigg]+{\cal O}(\e)\,,
 \nonumber \\ 
r_{458}(x,z) &=&- r_{457}(x,z) -\frac{2 \ln (z)}{\e}+2 \text{Li}_2\left(-\frac{x}{z}\right)-2 \text{Li}_2(-x z)-4 \text{Li}_2(z)-4 \ln (x) \ln (z)\nonumber \\ &&
-2 \ln (z) \ln (x+z)-2 \ln (z) \ln (x z+1)+2 \ln (x) \ln (x+z)\nonumber \\ &&-2 \ln (x) \ln (x z+1)
+5 \ln ^2(z)-4 \ln (1-z) \ln (z)+8 \ln (z)+\frac{2 \pi ^2}{3}+{\cal O}(\e)\,,
 \nonumber \\
\end{eqnarray}
where we introduced the abbreviations
\begin{equation}
u = u(x,z) = \sqrt{(1+x)^2-4xz} \,, \qquad v = v(x,z) = \sqrt{(1-z)^2+4xz}\;.
\end{equation}
In $r_{47}(x,z)$, the inverse tangent integral function 
\begin{equation}
 \text{Ti}_2 (y) = \int_0^y \frac{\arctan x}{x} {\rm d} x
\end{equation}
appears. It is not commonly encountered in higher-order perturbative calculations and is  
related to the dilogarithm for purely imaginary argument as~\cite{lewin:1981}:
\begin{equation}
 \text{Li}_2 (iy) = \frac{1}{4}  \text{Li}_2 (-y^2) + i  \text{Ti}_2 (y) 
 \end{equation} 

\subsection{Initial state endpoint region}
The master integrals in the initial-state endpoint region ($x=1$, $z<1$) become:
\begin{eqnarray}
r_{0}(1,z) &=& 1-\e\ln(z)+\e^2\frac{\ln^2(z)}{2}+{\cal O}(\e^3)\,,\nonumber \\
r_{1}(1,z) &=& {\cal O}(\e^3) \,, \nonumber\\
r_{5}(1,z) &=& \frac{\ln (z)}{\e (1-z)} + \frac{12 \text{Li}_2(z)-9 \ln ^2(z)+12 (\ln (1-z)-2) \ln (z)-2 \pi ^2}{6 (1-z)}\nonumber \\ &&
+ \frac{\e}{6(1-z)}
 \Big(
 -48 \text{Li}_2(z)-24 \text{Li}_3(1-z)-12 \text{Li}_3(z)-12 \text{Li}_2(z) \ln (z)\nonumber \\ &&
 +24 \text{Li}_2(1-z) \ln (1-z)+24 \text{Li}_2(z) \ln (1-z)+7 \ln ^3(z)-18 \ln (1-z) \ln ^2(z)\nonumber \\ &&
 +36 \ln ^2(z)+24 \ln ^2(1-z) \ln (z)+4 \pi ^2 \ln (z)-48 \ln (1-z) \ln (z)+24 \ln (z)\nonumber \\ &&-4 \pi ^2 \ln (1-z)+12 \zeta (3)+8 \pi ^2
 \Big)
+{\cal O}(\e^2) \,, \nonumber\\
r_{7}(1,z) &=& \frac{\ln (z)}{1-z} + \frac{\e}{(1-z)}\Big( 2 \text{Li}_2(z)-\frac{3}{2}  \ln ^2(z)+2 (\ln (1-z)-1) \ln(z)-\frac{\pi ^2}{3}\Big) 
+{\cal O}(\e^2) \,, \nonumber\\
r_{m27}(1,z) &=& \frac{\ln (z)}{2}+ 
\frac{\e}{12} \left(12 \text{Li}_2(z)-9 \ln ^2(z)+12 \ln (1-z) \ln (z)-12 \ln (z)-2 \pi ^2\right)\nonumber \\ &&+{\cal O}(\e^2) \,, \nonumber\\
r_{m37}(1,z) &=&\frac{1-z+ z\ln (z)}{1-z} + \frac{\e}{1-z} \bigg( 2 z \text{Li}_2(z)-\frac{\pi ^2 z}{3}-\frac{1}{2} 3 z \ln ^2(z)-z \ln (z)\nonumber \\ &&
+2 z \ln (1-z) \ln (z)-\ln (z)\bigg) +{\cal O}(\e^2) \,,\nonumber \\
r_{14}(1,z) &=&{\cal O}(\e^2) \,, \nonumber\\
r_{47}(1,z) &=&{\cal O}(\e^2) \,, \nonumber\\
r_{57}(1,z) &=&-\frac{2 \ln (z)}{\e(1-z^2)} + \frac{1}{1-z^2}\bigg(
4 \ln ^2(1-z)+8 \ln (z)+4 \ln (1-z) \ln (z)+3 \ln ^2(z)\nonumber \\ &&
-8 \ln (1-z) \ln (z+1)-8 \ln (z) \ln (z+1)+4 \ln ^2(z+1)-8 \text{Li}_2(-z)\nonumber \\ &&
+4 \text{Li}_2(z)+8 \text{Li}_2\left(\frac{z+1}{z-1}\right)+8 \text{Li}_2\left(\frac{1-z}{z+1}\right) \bigg)\nonumber \\
&& + \frac{\e}{1-z^2}\Bigg(
16 \text{Li}_2(z)+8 \text{Li}_3(1-z)+4 \text{Li}_3(z)+4 \text{Li}_2(z) \ln (z)\nonumber \\ &&
-32 \text{Li}_2\left(\frac{z+1}{z-1}\right) \ln (z+1)+32 \text{Li}_2(-z) \ln (z+1)-32 \text{Li}_2(z) \ln (z+1)\nonumber \\ &&
-32 \text{Li}_2\left(\frac{1-z}{z+1}\right) \ln (z+1)-\frac{1}{3} 7 \ln ^3(z)-16 \ln ^3(z+1)+6 \ln (1-z) \ln ^2(z)\nonumber \\ &&
-12 \ln ^2(z)+32 \ln ^2(z+1) \ln (z)+32 \ln (1-z) \ln ^2(z+1)\nonumber \\ &&
-16 \ln ^2(1-z) \ln (z+1)-\frac{4}{3} \pi ^2 \ln (z)+16 \ln (1-z) \ln (z)\nonumber \\ &&
-32 \ln (1-z) \ln (z+1) \ln (z)-8 \ln (z)+\frac{8}{3} \pi ^2 \ln (z+1)-4 \zeta (3)-\frac{8 \pi ^2}{3}\Bigg)
\nonumber \\ &&
+{\cal O}(\e^2) \,,\nonumber \\
r_{134}(1,z) &=&{\cal O}(\e^2) \,, \nonumber\\
r_{135}(1,z) &=& \frac{1}{\e^2} - \frac{4}{\e} + 2 \text{Li}_2(z)-\ln ^2(z)+2 \ln (1-z) \ln (z)-\frac{\pi ^2}{3}+4\nonumber \\ &&
 + \e\Big(-8 \text{Li}_2(z)-4 \text{Li}_3(1-z)-2 \text{Li}_3(z)-2 \text{Li}_2(z) \ln (z)\nonumber \\ &&+4 \text{Li}_2(1-z) \ln (1-z)
 +4 \text{Li}_2(z) \ln (1-z)+\ln ^3(z)-3 \ln (1-z) \ln ^2(z)\nonumber \\ &&+4 \ln ^2(z)
 +4 \ln ^2(1-z) \ln (z)+\frac{2}{3} \pi ^2 \ln (z)-8 \ln (1-z) \ln (z)\nonumber \\ &&-\frac{2}{3} \pi ^2 \ln (1-z)
 +2 \zeta (3)+\frac{4 \pi ^2}{3}\Big)
+{\cal O}(\e^2) \,, \nonumber\\
r_{138}(1,z) &=& \frac{1}{\e^2}   -\frac{2 (\ln (z)+2)}{\e} -2 \text{Li}_2(z)+2 \ln ^2(z)-2 \ln (1-z) \ln (z)+8 \ln (z)+\frac{\pi ^2}{3}+4 \nonumber \\ &&
+\e\Big(   8 \text{Li}_2(z)+4 \text{Li}_3(1-z)+2 \text{Li}_3(z)+2 \text{Li}_2(z) \ln (z)-4 \text{Li}_2(1-z) \ln (1-z)\nonumber \\ &&
-4 \text{Li}_2(z) \ln (1-z)-\frac{1}{3} 4 \ln ^3(z)+3 \ln (1-z) \ln ^2(z)-8 \ln ^2(z)\nonumber \\ &&
-4 \ln ^2(1-z) \ln (z)-\frac{2}{3} \pi ^2 \ln (z)+8 \ln (1-z) \ln (z)-8 \ln (z)\nonumber \\ &&
+\frac{2}{3} \pi ^2 \ln (1-z)-2 \zeta (3)-\frac{4 \pi ^2}{3}  \Big)
+{\cal O}(\e^2) \,, \nonumber\\
r_{145}(1,z) &=&{\cal O}(\e^2) \,, \nonumber\\
r_{235}(1,z) &=& -  \frac{3}{\e^2} +\frac{2 (\ln (z)+6)}{\e}-8 \ln (z)+\frac{\pi ^2}{3}-12
+\e\Big(8 \text{Li}_3(z)-4 \text{Li}_2(z) \ln (z) \nonumber \\ &&
-\frac{2}{3}  \ln ^3(z)-\frac{2}{3} \pi ^2 \ln (z)+8 \ln (z)+2 \zeta (3)-\frac{4 \pi ^2}{3}\Big)
+ {\cal O}(\e^2) \,, \nonumber\\
r_{237}(1,z) &=&r_{138}(1,z)\,,\nonumber \\
r_{245}(1,z) &=& - \frac{1}{\e^2}   +\frac{2 (\ln (z)+2)}{\e} 
+4 \text{Li}_2(z)-2 \ln ^2(z)+4 \ln (1-z) \ln (z)-8 \ln (z)\nonumber \\ &&
-\frac{2 \pi ^2}{3}-4
+\e \Big(-16 \text{Li}_2(z)-12 \text{Li}_3(1-z)-4 \text{Li}_3(z)-4 \text{Li}_2(z) \ln (z)\nonumber \\ &&+12 \text{Li}_2(1-z) \ln (1-z)
+12 \text{Li}_2(z) \ln (1-z)+\frac{4 \ln ^3(z)}{3} -6 \ln (1-z) \ln ^2(z)\nonumber \\ &&+8 \ln ^2(z)
+12 \ln ^2(1-z) \ln (z)+\frac{4}{3} \pi ^2 \ln (z)-16 \ln (1-z) \ln (z)\nonumber \\ &&+8 \ln (z)
-2 \pi ^2 \ln (1-z)+4 \zeta (3)+\frac{8 \pi ^2}{3} \Big) 
+ {\cal O}(\e^2) \,, \nonumber\\
r_{347}(1,z) &=&{\cal O}(\e^2) \,, \nonumber\\
r_{357}(1,z) &=& \frac{1}{\eps^2} +  \frac{1}{\eps} \left(-4 -\frac{2\ln (z)}{1+z}  \right) +4+4 \ln (z) + \frac{\ln^2(z)}{2} 
\nonumber \\ &&
+ \frac{1-z}{1+z} \bigg(-2 \text{Li}_2\left(z^2\right)+4 \text{Li}_2\left(\frac{z+1}{z-1}\right)+6 \text{Li}_2(z)+4 \text{Li}_2\left(\frac{1-z}{z+1}\right)\nonumber \\ &&
+2 \ln ^2(1-z)+\frac{3 \ln ^2(z)}{2}+2 \ln ^2(z+1)+2 \ln (z) \ln (1-z)\nonumber \\ &&-4 \ln (z+1) \ln (1-z)+4 \ln (z)-4 \ln (z) \ln (z+1) 
\bigg)
\nonumber \\ &&
+\e \Bigg[ -\frac{1}{6} \ln ^3(z)-2 \ln ^2(z)-4 \ln (z)
+\frac{1-z}{1+z}\bigg( 
4 \text{Li}_2\left(z^2\right) \ln (z)+8 \text{Li}_2(z)\nonumber \\ &&
+4 \text{Li}_3(1-z)+2 \text{Li}_3(z)-8 \text{Li}_2\left(\frac{z+1}{z-1}\right) \ln (z)-14 \text{Li}_2(z) \ln (z)\nonumber \\ &&
-8 \text{Li}_2\left(\frac{1-z}{z+1}\right) \ln (z)-\frac{7}{6} \ln ^3(z)-5 \ln (1-z) \ln ^2(z)+8 \ln (z+1) \ln ^2(z)\nonumber \\ &&
-6 \ln ^2(z)-4 \ln ^2(1-z) \ln (z)-4 \ln ^2(z+1) \ln (z)+8 \ln (1-z) \ln (z)\nonumber \\ &&+8 \ln (1-z) \ln (z+1) \ln (z)-4 \ln (z)-2 \zeta (3)-\frac{4 \pi ^2}{3}
\bigg)\Bigg]  +{\cal O}(\e^2) \,,
 \nonumber\\
r_{358}(1,z) &=&-\frac{2}{\e^2} +\frac{2 \ln (z)+8}{\e} -8 - 8 \ln(z)
+\e \Big(4 \text{Li}_3(z)-\frac{4}{3}  \ln ^3(z)\nonumber \\ &&
+2 \ln (1-z) \ln ^2(z)-\frac{2}{3} \pi ^2 \ln (z)+8 \ln (z)-4 \zeta (3)
\Big)
+{\cal O}(\e^2) \,, \nonumber\\
r_{457}(1,z) &=&  {\cal O}(\e^2) \,, \nonumber\\
r_{458}(1,z) &=& -\frac{2\ln (z)}{\e} 
-2 \text{Li}_2\left(z^2\right)-4 \ln \left(1-z^2\right) \ln (z)+4 \ln ^2(z)+8 \ln (z)+\frac{\pi ^2}{3}
\nonumber \\ 
&&  + \e \Bigg[ 
8 \text{Li}_2\left(z^2\right)+2 \text{Li}_2\left(z^2\right) \log (z)-8 \text{Li}_3\left(\frac{1-z}{2}\right)-8 \text{Li}_3\left(\frac{z+1}{2}\right) \nonumber \\ &&
+16 \text{Li}_3(1-z)+8 \text{Li}_3\left(\frac{2 z}{z-1}\right)-4 \text{Li}_3(-z)+8 \text{Li}_3(z)-8 \text{Li}_3\left(\frac{z}{z+1}\right)\nonumber \\ &&
+8 \text{Li}_3\left(\frac{2 z}{z+1}\right)-4 \log ^2(2) \log \left(1-z^2\right)+16 \log (z) \log \left(1-z^2\right)\nonumber \\ &&
-\frac{4}{3} \log ^3(1-z)-4 \log ^3(z)+4 \log (2) \log ^2(1-z)+4 \log (2) \log ^2(z+1)\nonumber \\ &&
+8 \log (1-z) \log ^2(z)-16 \log ^2(z)+4 \log ^2(1-z) \log (z)+6 \log ^2(z) \log (z+1)\nonumber \\ &&
-\frac{4}{3} \pi ^2 \log (1-z)-\frac{4}{3} \pi ^2 \log (z)-8 \log (z)+2 \pi ^2 \log (z+1)\nonumber \\ &&
-8 \log (1-z) \log (z) \log (z+1)-4 \zeta (3)-\frac{4 \pi ^2}{3}+\frac{8 \log ^3(2)}{3}-\frac{4}{3} \pi ^2 \log (2)
\Bigg]\nonumber \\ &&
 +{\cal O}(\e^2) \,.
\end{eqnarray}

\subsection{Final state endpoint region}
The master integrals in the final-state endpoint region ($x<1$, $z=1$) become:
\begin{eqnarray}
r_{0}(x,1) &=& 1+\e\left(2\ln(x)\right)+\e^2\left(2\ln^2(x)\right)+{\cal O}(\e^3)\,,\nonumber \\
r_{1}(x,1) &=&
\ln (x) + \e \left(2 \text{Li}_2(x)+\frac{3 \ln ^2(x)}{2}+2 (\ln (1-x)-1) \ln (x)-\frac{\pi ^2}{3}\right) \nonumber \\ &&
+\e^2\bigg( -4 \text{Li}_2(x)-4 \text{Li}_3(1-x)-2 \text{Li}_3(x)+4 \text{Li}_2(x) \ln (x)\nonumber \\ &&+4 \text{Li}_2(1-x) \ln (1-x)
+4 \text{Li}_2(x) \ln (1-x)+\frac{7 \ln ^3(x)}{6}\nonumber \\ &&+3 \ln (1-x) \ln ^2(x)-3 \ln ^2(x)
+4 \ln ^2(1-x) \ln (x)-\frac{1}{3} \pi ^2 \ln (x)\nonumber \\ &&-4 \ln (1-x) \ln (x)-\frac{2}{3} \pi ^2 \ln (1-x)+2 \zeta (3)+\frac{2 \pi ^2}{3}
\bigg) +{\cal O}(\e^3) \,, \nonumber\\
r_{5}(x,1) &=&
-\frac{1}{\e}-2 (\ln (x)-1)-2 \e \left(\ln ^2(x)-2 \ln (x)\right)+{\cal O}(\e^2) \,, \nonumber\\
r_{7}(x,1) &=&
\frac{\ln (x)}{1-x} + \frac{\e \left(12 \text{Li}_2(x)+9 \ln ^2(x)+12 \ln (1-x) \ln (x)-12 \ln (x)-2 \pi ^2\right)}{6 (1-x)}\nonumber \\ &&+{\cal O}(\e^2) \,, \nonumber\\
r_{m27}(x,1) &=&{\cal O}(\e^2) \,, \nonumber\\
r_{m37}(x,1) &=&{\cal O}(\e^2) \,, \nonumber\\
r_{14}(x,1) &=&\frac{x \ln ^2(x)}{1-x}
+\e\,\frac{ x \ln (x)}{3 (1-x)}  \Big(12 \text{Li}_2(x)+3 \ln ^2(x)+12 \ln (1-x) \ln (x)\nonumber \\ &&
-12 \ln (x)-2 \pi ^2\Big)
+{\cal O}(\e^2) \,, \nonumber\\
r_{47}(x,1) &=&{\cal O}(\e^2) \,, \nonumber\\
r_{57}(x,1) &=& \frac{1}{\e}+\ln (x)-2 + \e\bigg(-2 \text{Li}_2(x)+\frac{1}{2} \ln (x) (-4 \ln (1-x)+\ln (x)-4)\nonumber \\ &&
+\frac{\pi ^2}{3}
\bigg)
+{\cal O}(\e^2) \,, \nonumber\\
r_{134}(x,1) &=& -\frac{\ln (x)}{\e}  + 2 \text{Li}_2(-x)-2 \text{Li}_2(x)-2 \ln ^2(x)-2 \ln (1-x) \ln (x)\nonumber \\ &&
+2 \ln (x+1) \ln (x)
+4 \ln (x)+\frac{\pi ^2}{2}
-\e \bigg(8 \text{Li}_2(-x)-8 \text{Li}_2(x)-4 \text{Li}_3\left(\frac{1-x}{2}\right)\nonumber \\ &&
-4 \text{Li}_3\left(\frac{x+1}{2}\right)+4 \text{Li}_3\left(\frac{2 x}{x-1}\right)+4 \text{Li}_3\left(\frac{2 x}{x+1}\right)-4 \text{Li}_2(-x) \ln (x)\nonumber \\ &&
+4 \text{Li}_2(x) \ln (x)-\frac{2}{3} \ln ^3(1-x)+\frac{5 \ln ^3(x)}{3}-\frac{2}{3} \ln ^3(x+1)-2 \ln ^2(2) \ln (1-x)\nonumber \\ &&
-2 \ln ^2(2) \ln (x+1)+2 \ln (2) \ln ^2(1-x)+2 \ln (2) \ln ^2(x+1)\nonumber \\ &&
+4 \ln (1-x) \ln ^2(x)-8 \ln ^2(x)+2 \ln (x) \ln ^2(x+1)+2 \ln ^2(1-x) \ln (x)\nonumber \\ &&
-4 \ln ^2(x) \ln (x+1)-\frac{2}{3} \pi ^2 \ln (1-x)-\pi ^2 \ln (x)-8 \ln (1-x) \ln (x)\nonumber \\ &&
+4 \ln (x)+\frac{4}{3} \pi ^2 \ln (x+1)-4 \ln (1-x) \ln (x) \ln (x+1)+8 \ln (x) \ln (x+1)\nonumber \\ &&
+2 \pi ^2+\frac{4 \ln ^3(2)}{3}-\frac{1}{6} \pi ^2 \ln (16)
\bigg)
+{\cal O}(\e^2) \,, \nonumber\\
r_{135}(x,1) &=& \frac{1}{\e^2} - \frac{4}{\e} +4 +\pi ^2-6 \ln (1-x) \ln (x)-6 \text{Li}_2(x) \nonumber \\ &&
+\e\bigg(
24 \text{Li}_2(x)+16 \text{Li}_3(1-x)+14 \text{Li}_3(x)-16 \text{Li}_2(1-x) \ln (1-x)\nonumber \\ &&
-16 \text{Li}_2(x) \ln (1-x)-14 \text{Li}_2(x) \ln (x)-16 \ln (x) \ln ^2(1-x)\nonumber \\ &&
-7 \ln ^2(x) \ln (1-x)+\frac{8}{3} \pi ^2 \ln (1-x)+24 \ln (x) \ln (1-x)\nonumber \\ &&
-14 \zeta (3)-4 \pi ^2
\bigg) 
+{\cal O}(\e^2) \,, \nonumber\\
r_{138}(x,1) &=& \frac{1}{\e^2}+\frac{4(\ln(x)-1)}{\e} + 6\text{Li}_2(x)+5\ln^2(x)+6\ln(1-x)\ln(x)-16\ln(x) \nn \\&&
+4-\pi^2+\e\bigg(-2\text{Li}_3(x)-16\text{Li}_3(1-x)+10\text{Li}_2(x)\ln(x)\nn\\&&
+16\text{Li}_2(x)\ln(1-x)-24\text{Li}_2(x)+16\text{Li}_2(1-x)\ln(1-x)+\frac{11}{3}\ln^3(x) \nn \\ &&
+9\ln(1-x)\ln^2(x)-20\ln^2(x)+16\ln^2(1-x)\ln(x)-24\ln(1-x)\ln(x)\nn\\&&
-\frac{4\pi^2}{3}\ln(x)+16\ln(x)-\frac{8\pi^2}{3}\ln(1-x)+4\pi^2+2\zeta_3\bigg)+\mathcal{O}(\e^2) \nn \\
r_{145}(x,1) &=& 
\frac{\ln (x)}{\e}+2 \text{Li}_2(x)+\frac{5 \ln ^2(x)}{2}+2 \ln (1-x) \ln (x)-4 \ln (x)-\frac{\pi ^2}{3}\nonumber \\ &&
+\e\bigg(-8 \text{Li}_2(x)-4 \text{Li}_3(1-x)-2 \text{Li}_3(x)+8 \text{Li}_2(x) \ln (x)+\frac{13 \ln ^3(x)}{6}\nonumber \\ &&
+7 \ln (1-x) \ln ^2(x)-10 \ln ^2(x)-\pi ^2 \ln (x)-8 \ln (1-x) \ln (x)\nonumber \\ &&+4 \ln (x)
+2 \zeta (3)+\frac{4 \pi ^2}{3}\bigg) 
+{\cal O}(\e^2) \,, \nonumber\\
r_{235}(x,1) &=&-\frac{3}{\e^2}+\frac{12-6 \ln (x)}{\e}-6 \ln ^2(x)+24 \ln (x)+\frac{\pi ^2}{3}-12\nonumber \\ &&
+\e\left(-4 \ln ^3(x)+24 \ln ^2(x)+\frac{2}{3} \pi ^2 \ln (x)-24 \ln (x)+10 \zeta (3)-\frac{4 \pi ^2}{3}\right)
+{\cal O}(\e^2) \,, \nonumber\\ 
r_{237}(x,1) &=&r_{138}(x,1)\,,\nonumber \\
r_{245}(x,1) &=&
-\frac{1}{\e^2}+\frac{4-3 \ln (x)}{\e}-2 \text{Li}_2(x)-\frac{7}{2}  \ln ^2(x)-2 \ln (1-x) \ln (x)+12 \ln (x)\nonumber \\ &&
+\frac{\pi ^2}{3}-4
+\e\bigg(8 \text{Li}_2(x)+4 \text{Li}_3(1-x)+2 \text{Li}_3(x)-4 \text{Li}_2(x) \ln (x)\nonumber \\ &&
-4 \text{Li}_2(1-x) \ln (1-x)-4 \text{Li}_2(x) \ln (1-x)-\frac{5}{2} \ln ^3(x)\nonumber \\ &&
-3 \ln (1-x) \ln ^2(x)+14 \ln ^2(x)-4 \ln ^2(1-x) \ln (x)+\frac{1}{3} \pi ^2 \ln (x)\nonumber \\ &&
+8 \ln (1-x) \ln (x)-12 \ln (x)+\frac{2}{3} \pi ^2 \ln (1-x)-2 \zeta (3)
-\frac{4 \pi ^2}{3}
\bigg)
+{\cal O}(\e^2) \,, \nonumber\\
r_{347}(x,1) &=& -\frac{\ln (x)}{\e}-2 \text{Li}_2(x)-\frac{1}{2} 3 \ln ^2(x)-2 \ln (1-x) \ln (x)+4 \ln (x)+\frac{\pi ^2}{3}\nonumber \\ &&
+\e\bigg(8 \text{Li}_2(x)+4 \text{Li}_3(1-x)+2 \text{Li}_3(x)-4 \text{Li}_2(x) \ln (x)-\frac{7}{6}  \ln ^3(x)\nonumber \\ &&
-3 \ln (1-x) \ln ^2(x)+6 \ln ^2(x)+\frac{1}{3} \pi ^2 \ln (x)+8 \ln (1-x) \ln (x)\nonumber \\ &&-4 \ln (x)-2 \zeta (3)-\frac{4 \pi ^2}{3}
\bigg)
+{\cal O}(\e^2) \,, \nonumber\\
r_{357}(x,1) &=& \frac{1}{\e^2}+\frac{\ln (x)-4}{\e}-2 \text{Li}_2(x)+\frac{\ln ^2(x)}{2}-2 \ln \left(1-x\right) \ln (x)-4 \ln (x)+\frac{\pi ^2}{3}+4\nonumber \\ &&
+\e\bigg(
8 \text{Li}_2(x)+4 \text{Li}_3(1-x)+2 \text{Li}_3(x)-4 \text{Li}_2(x) \ln (x)+\frac{\ln ^3(x)}{6}\nonumber \\ &&
-3 \ln (1-x) \ln ^2(x)-2 \ln ^2(x)+\frac{1}{3} \pi ^2 \ln (x)+8 \ln (1-x) \ln (x)\nonumber \\ &&
+4 \ln (x)-2 \zeta (3)-\frac{4 \pi ^2}{3}
\bigg)
 \nonumber\\
r_{358}(x,1) &=&-\frac{2}{\e^2}+\frac{8-4 \ln (x)}{\e}-4 \ln ^2(x)+16 \ln (x)-8 \nonumber \\ &&
+\e\left(-\frac{8}{3}  \ln ^3(x)+16 \ln ^2(x)-16 \ln (x)
\right)
+{\cal O}(\e^2) \,, \nonumber\\
r_{457}(x,1) &=&{\cal O}(\e^2) \,, \nonumber\\
r_{458}(x,1) &=& {\cal O}(\e^2) \,.
\end{eqnarray}
\subsection{Soft region}

In the soft endpoint ($x=1,z=1$) the  master integrals become:
\begin{eqnarray}
r_{0}(1,1) &=& 1\,, \nonumber \\
r_{1}(1,1) &=&{\cal O}(\e^4)\,, \nonumber \\
r_{5}(1,1) &=&-\frac{1}{\e} +2 + {\cal O}(\e^3)\,, \nonumber \\
r_{7}(1,1) &=& -1 + {\cal O}(\e^3)\,, \nonumber \\
r_{m27}(1,1) &=& {\cal O}(\e^3)\,, \nonumber \\
r_{m37}(1,1) &=& {\cal O}(\e^3)\,, \nonumber \\
r_{14}(1,1) &=& {\cal O}(\e^3)\,, \nonumber \\
r_{47}(1,1) &=& {\cal O}(\e^3)\,, \nonumber \\
r_{57}(1,1) &=&\frac{1}{\e} -2 + {\cal O}(\e^3)\,, \nonumber \\
r_{134}(1,1) &=& {\cal O}(\e^3)\,, \nonumber \\
r_{135}(1,1) &=& \frac{1}{\e^2} - \frac{4}{\e} +4 + {\cal O}(\e^3)\,, \nonumber \\
r_{138}(1,1) &=& \frac{1}{\e^2} - \frac{4}{\e} +4 + {\cal O}(\e^3)\,, \nonumber \\
r_{145}(1,1) &=& {\cal O}(\e^3)\,, \nonumber \\
r_{235}(1,1) &=& -\frac{3}{\e^2}+\frac{12}{\e}+\frac{1}{3} \left(\pi ^2-36\right)+\frac{2}{3} \e \left(15 \zeta (3)-2 \pi ^2\right)
 \nonumber \\ &&
+\frac{1}{90} \e^2 \left(-3600 \zeta (3)+31 \pi ^4+120 \pi ^2\right)+{\cal O}(\e^3)\,, \nonumber \\
r_{237}(1,1) &=& \frac{1}{\e^2} - \frac{4}{\e} +4 + {\cal O}(\e^3)\,, \nonumber \\
r_{245}(1,1) &=& -\frac{1}{\e^2} + \frac{4}{\e} -4 + {\cal O}(\e^3)\,, \nonumber \\
r_{347}(1,1) &=& {\cal O}(\e^3)\,, \nonumber \\
r_{357}(1,1) &=& \frac{1}{\e^2} - \frac{4}{\e} +4 + {\cal O}(\e^3)\,, \nonumber \\
r_{358}(1,1) &=& -\frac{2}{\e^2} + \frac{8}{\e} -8 + {\cal O}(\e^3)\,, \nonumber \\
r_{457}(1,1) &=& {\cal O}(\e^3)\,, \nonumber \\
r_{458}(1,1) &=& {\cal O}(\e^3)\,, 
\end{eqnarray}

\section{NNLO time-like mass factorisation kernels}\label{app:MFkern}
While the NLO mass factorisation kernels are identical for time-like and space-like kinematics, the NNLO ones are sensitive on whether the identified particle is in the initial or final state. In the following we list the NNLO time-like mass factorisation kernels. 
The reduced time-like two-loop mass factorization kernel is defined similarly to the space-like case~\cite{Currie:2013vh}:
	\begin{equation}
		\Gammatwo{ab;cd}{z,x_1,x_2}=\Gammatwo{ca,\text{full}}{z}\delta_{db}
		\delta(1-x_1)\delta(1-x_2)+\Gammatwo{db,\text{full}}{z}\delta_{ca}\delta(1-x_1)\delta(1-x_2)\, ,
	\end{equation}
	where $\Gammatwo{ca,\text{full}}{z}$ is directly related to the LO and NLO time-like Altarelli-Parisi spitting kernels~\cite{Altarelli:1977zs,Curci:1980uw,Furmanski:1980cm,Rijken:1996ns,Rijken:1996vr}
	\begin{equation}\label{Gamma2}
		\Gammatwo{ab,\text{full}}{z}=-\dfrac{1}{2\e}\left(P^1_{ab}(z)+\dfrac{\beta_0}{\e}P^0_{ab}(z)\right) \, .
	\end{equation}
	We can decompose $\Gammatwo{ca,\text{full}}{z}$ into colour layers as discussed in~\cite{Currie:2013vh}
	\begingroup
	\allowdisplaybreaks
	\begin{eqnarray}
		{\overline{\Gamma}}_{qq}^{(2)\text{id.}}(z)&=&\bigg(\frac{N_c^{2}-1}{N_c}\bigg)\bigg[N_c\overline{\Gamma}_{qq}^{(2)\text{id.}}(z)+\wt{\overline{\Gamma}}_{qq}^{(2)\text{id.}}(z)+\frac{1}{N_c}\wt{\wt{\overline{\Gamma}}}_{qq}^{(2)\text{id.}}(z)+N_f\wh{\overline{\Gamma}}_{qq}^{(2)\text{id.}}(z)\bigg]\, , \\
		{\overline{\Gamma}}_{q\qb}^{(2)\text{id.}}(z)&=&\bigg(\frac{N_c^{2}-1}{N_c}\bigg)\bigg[\overline{\Gamma}_{q\qb}^{(2)\text{id.}}(z)+\frac{1}{N_c}\wt{\overline{\Gamma}}_{q\qb}^{(2)\text{id.}}(z)\bigg]\,,\\
		{\overline{\Gamma}}_{qq'}^{(2)\text{id.}}(z)&=&\bigg(\frac{N_c^{2}-1}{N_c}\bigg)\ \overline{\Gamma}_{qq'}^{(2)\text{id.}}(z)\, ,\\
		{\overline{\Gamma}}_{q\qb'}^{(2)\text{id.}}(z)&=&\bigg(\frac{N_c^{2}-1}{N_c}\bigg)\ \overline{\Gamma}_{q\qb'}^{(2)\text{id.}}(z)\,,\\
		{\overline{\Gamma}}_{gq}^{(2)\text{id.}}(z)&=&\bigg(\frac{N_c^{2}-1}{N_c}\bigg)\bigg[N_c\overline{\Gamma}_{gq}^{(2)\text{id.}}(z)+\frac{1}{N_c}\wt{\overline{\Gamma}}_{gq}^{(2)\text{id.}}(z)+N_f\wh{\overline{\Gamma}}_{gq}^{(2)\text{id.}}(z)\bigg]\, ,\\
		{\overline{\Gamma}}_{qg}^{(2)\text{id.}}(z)&=&N_c\overline{\Gamma}_{qg}^{(2)\text{id.}}(z)+\frac{1}{N_c}\wt{\overline{\Gamma}}_{qg}^{(2)\text{id.}}(z)+N_f\wh{\overline{\Gamma}}_{qg}^{(2)\text{id.}}(z)\,,\\
		{\overline{\Gamma}}_{gg}^{(2)\text{id.}}(z)&=&N_c^{2}\overline{\Gamma}_{gg}^{(2)\text{id.}}(z)+N_cN_f\wh{\overline{\Gamma}}_{gg}^{(2)\text{id.}}(z)+\frac{N_f}{N_c}\wh{\wt{\overline{\Gamma}}}_{gg}^{(2)\text{id.}}(z)+N_f^{2}\wh{\wh{\overline{\Gamma}}}_{gg,}^{(2)\text{id.}}(z)\, .
	\end{eqnarray}
	\endgroup
	We report for completeness the explicit form of each colour stripped function
\begin{eqnarray}
\overline{\Gamma}_{qq}^{(2)\text{id.}}(z)&=&\frac{1}{\eps^{2}}\bigg[-\frac{11}{12}p^0_{qq}(z)\bigg]\nn \\
&+&\frac{1}{\eps}\bigg[\left(\frac{\pi ^2}{12}-\frac{67}{36}+\frac{1}{4} \log ^2(z)-\frac{31}{24} \log (z)-\frac{1}{2} \log (1-z) \log (z)\right) {\cal{D}}_{0}(z) \nn \\
   &-&\left(\frac{43}{192}+\frac{13 \pi ^2}{144}\right)\delta(1-z) -\frac{\pi ^2 z}{24}+\frac{71
   z}{36}+\left(-\frac{3 z}{16}-\frac{3}{16}\right) \log ^2(z) \nn \\
   &+&\left(\frac{7 z}{12}+\frac{5}{6}\right) \log (z) +\left(\frac{z}{4}+\frac{1}{4}\right) \log (1-z) \log (z)-\frac{\pi
   ^2}{24}-\frac{1}{9} \bigg]\, ,\\
\wt{\overline{\Gamma}}_{qq}^{(2)\text{id.}}(z)&=&\frac{1}{\eps}\bigg[-\frac{7 z^2}{9}+\left(\frac{z^2}{3}+\frac{9 z}{8}+\frac{5}{8}\right) \log
   (z)-\frac{z}{2} \nn \\
   &+&\frac{5}{18 z}+\left(-\frac{z}{8}-\frac{1}{8}\right) \log
   ^2(z)+1\bigg]\, ,\\
\wt{\wt{\overline{\Gamma}}}_{qq}^{(2)\text{id.}}(z)&=&\frac{1}{\eps}\bigg[\left(-\frac{1}{2} \log ^2(z)+\frac{3}{8} \log (z)+\frac{1}{2} \log (1-z) \log (z)\right) {\cal{D}}_{0}(z) \nn \\
&+&\left(\frac{3
   \zeta(3)}{4}+\frac{3}{64}-\frac{\pi ^2}{16}\right) \delta(1-z) +\frac{5 z}{8}+\left(\frac{5 z}{16}+\frac{5}{16}\right) \log ^2(z)\nn \\
   &+&\left(-\frac{3
   z}{8}-\frac{5}{8}\right) \log (z)-\left(\frac{z}{4}+\frac{1}{4}\right) \log (1-z) \log (z)-\frac{5}{8}\bigg]\, ,\\
\wh{\overline{\Gamma}}_{qq}^{(2)\text{id.}}(z)&=&\frac{1}{\eps^{2}}\bigg[\frac{1}{6}{\cal{D}}_{0}(z)+\frac{\delta(1-z)}{8}-\frac{z}{12}-\frac{1}{12}\bigg]\nn\\
&+&\frac{1}{\eps}\bigg[\left(\frac{5}{18}+\frac{1}{6} \log (z) \right) {\cal{D}}_{0}(z)+\left(\frac{\pi ^2}{36}+\frac{1}{48}\right)\delta(1-z)\nn \\
&-&\frac{11 z}{36}+\left(-\frac{z}{12}-\frac{1}{12}\right) \log (z)+\frac{1}{36}\bigg]\, ,\\
\overline{\Gamma}_{q\b{q}}^{(2)\text{id.}}(z)&=&\wt{\overline{\Gamma}}_{qq}^{(2)\text{id.}}(z)\, ,\\
\wt{\overline{\Gamma}}_{q\b{q}}^{(2)\text{id.}}(z)&=&\frac{1}{\eps}\bigg[ \left(\frac{1}{2}+\frac{\pi^2}{24}\right)(1-z)-\frac{\pi^2}{12(1+z)}\nn \\
&+&\left(\log(1+z)\log(z)+\text{Li}_2(z)\right)\left(\frac{1}{2}-\frac{1}{2}z-\frac{1}{1-z}\right)+\log(z)\frac{1}{4}\left(1+z\right)\nn \\
   &+&\log^2(z)\left(-\frac{1}{8}+\frac{1}{8}z+\frac{1}{4}\frac{1}{1-z}\right)\bigg]\,,\\
\overline{\Gamma}_{q\qp}^{(2)\text{id.}}(z)&=&\wt{\overline{\Gamma}}_{qq}^{(2)\text{id.}}(z)\,,\\
\overline{\Gamma}_{q\qbp}^{(2)\text{id.}}(z)&=&\wt{\overline{\Gamma}}_{qq}^{(2)\text{id.}}(z)\,,\\
\overline{\Gamma}_{gq}^{(2)\text{id.}}(z)&=&\frac{1}{\eps^{2}}\bigg[-\frac{11}{12}p^0_{gq}(z)\bigg]\nn\\
&+&\frac{1}{\eps}\bigg[\left(-\frac{z}{2}-\frac{1}{z}-1\right)
   \text{Li}_2(-z)+\left(-z-\frac{2}{z}+2\right) \text{Li}_2(z)+\frac{11
   z^2}{9}\nn \\
   &+&\left(-\frac{2 z^2}{3}-\frac{37 z}{16}+\frac{3}{2 z}-1\right) \log
   (z)-\frac{5 z}{16}-\frac{17}{36 z}+\left(\frac{z}{8}+\frac{1}{4
   z}-\frac{1}{4}\right) \log ^2(1-z) \nn \\
   &+&\left(\frac{13
   z}{16}+\frac{1}{z}+\frac{3}{8}\right) \log ^2(z)+\frac{1}{4} z \log
   (1-z)+\left(-z-\frac{2}{z}+2\right) \log (1-z) \log(z) \nn \\
   &+&\left(-\frac{z}{2}-\frac{1}{z}-1\right) \log (z) \log (z+1)-\frac{\pi
   ^2}{6}-\frac{19}{16}\bigg]\, ,\\
\wt{\overline{\Gamma}}_{gq}^{(2)\text{id.}}(z)&=&\frac{1}{\eps}\bigg[\left(-z-\frac{2}{z}+2\right) \text{Li}_2(z)+\frac{9z}{16}+\left(\frac{z}{8}+\frac{1}{4 z}-\frac{1}{4}\right) \log
   ^2(1-z)\nn \\
   &+&\left(\frac{1}{8}-\frac{z}{16}\right) \log ^2(z)+\frac{1}{4} z \log(1-z)-\left(\frac{z}{2}+\frac{1}{z}-1\right) \log (z) \log(1-z) \nn \\
   &+&\left(\frac{z}{16}-1\right) \log (z)-\frac{1}{16}\bigg]\, ,\\
\wh{\overline{\Gamma}}_{gq}^{(2)\text{id.}}(z)&=&\frac{1}{\eps^{2}}\bigg[\frac{1}{6}p^0_{gq}(z)\bigg]\, ,\\
\overline{\Gamma}_{qg}^{(2)\text{id.}}(z)&=&\frac{1}{\eps^{2}}\bigg[-\frac{11}{12}p^0_{qg}(z)\bigg]\nn\\
&+&\frac{1}{\eps}\bigg[\left(z^2+z+\frac{1}{2}\right) \text{Li}_2(-z)+\left(2 z^2-2 z+1\right)
   \text{Li}_2(z)+\frac{\pi ^2 z^2}{12}-\frac{133 z^2}{36} \nn \\
   &+&\left(\frac{1}{4} z^2-\frac{1}{4} z +\frac{1}{8}\right) \log ^2\left(z-z^2\right)+\left(-\frac{z^2}{2}+\frac{z}{2}-\frac{1}{4}\right) \log ^2(1-z)\nn \\
   &+&\left(-\frac{7 z^2}{12}+\frac{7
   z}{12}-\frac{13}{24}\right) \log (1-z)+\left(\frac{z^2}{12}+\frac{31
   z}{12}+\frac{31}{48}\right) \log (z) \nn \\
   &+&\left(z^2+z+\frac{1}{2}\right) \log (z)
   \log (z+1)+\frac{\pi ^2 z}{12}+\frac{173 z}{144}+\frac{5}{9 z} \nn \\
   &+&\left(-\frac{11z}{8}-\frac{5}{16}\right) \log ^2(z)+\frac{\pi ^2}{24}+\frac{7}{18}\bigg]\, , \\
\wt{\overline{\Gamma}}_{qg}^{(2)\text{id.}}(z)&=&\frac{1}{\eps}\bigg[\left(2 z^2-2 z+1\right) \text{Li}_2(z)-\frac{\pi ^2 z^2}{12}-\frac{5
   z^2}{4} \nn \\
   &+&\left(-\frac{1}{4} z^2 +\frac{1}{4} z -\frac{1}{8}\right)\log ^2\left(z-z^2\right)\nn \\
   &-&\left(\frac{z^2}{4}-\frac{z}{4}+\frac{3}{8}\right) \log
   (1-z)+\left(\frac{z^2}{4}+\frac{z}{4}-\frac{5}{16}\right) \log (z) \nn \\
   &-&\left(-2 z^2+2 z-1\right) \log (1-z) \log (z) \nn \\
   &+&\frac{\pi ^2 z}{12} +\frac{23 z}{16}+\left(\frac{1}{16}-\frac{z}{8}\right) \log ^2(z)-\frac{\pi
   ^2}{24}-\frac{3}{4}\bigg]\, ,\\
   \wh{\overline{\Gamma}}_{qg}^{(2)\text{id.}}(z)&=&\frac{1}{\eps ^{2}}\bigg[\frac{1}{6}p^0_{qg}(z)\bigg] \nn \\
   &+&\frac{1}{\eps}\bigg[\frac{2 z^2}{9}+\left(\frac{z^2}{3}-\frac{z}{3}+\frac{1}{6}\right) (\log(1-z)+\log(z))-\frac{2 z}{9}+\frac{5}{18}\bigg]\, ,\\
\overline{\Gamma}_{gg}^{(2)\text{id.}}(z)&=&\frac{1}{\eps^{2}}\bigg[-\frac{11}{12}p^0_{gg}(z)\bigg] \nn \\
   &+&\frac{1}{\eps}\bigg[ \left(\frac{\pi ^2}{6}-\frac{67}{18}+\frac{3}{2} \log ^2(z)-\frac{11}{3} \log (z)+2\log (1-z) \log (z) \right){\cal{D}}_0(z)\nn \\
   &-&\left(\frac{3 \zeta_3}{2}+\frac{4}{3} \right) \delta(1-z) 
   +\left(-2 z^2-2 z+\frac{2}{z+1}-\frac{2}{z}-4\right) \text{Li}_2(-z)-\frac{1}{3} \pi ^2 z^2\nn \\
   &+&\left(-z^2+4 z-\frac{1}{2 (z+1)}+\frac{2}{z}\right) \log ^2(z) +\left(\frac{11 z^2}{3}+\frac{z}{2}+\frac{11}{3 z}+\frac{11}{2}\right)
   \log (z)\nn \\
   &+&\left(2 z^2-2 z-\frac{2}{z}+4\right) \log (1-z) \log (z)  \nn \\
   &+&\left(-2 z^2-2 z+\frac{2}{z+1}-\frac{2}{z}-4\right) \log (z) \log (z+1)  \nn \\
   &+&\frac{109 z}{36}+\frac{\pi ^2}{6 (z+1)}-\frac{2 \pi ^2}{3}+\frac{25}{36}\bigg]
   \, ,\\
\wh{\overline{\Gamma}}_{gg}^{(2)}(z)&=&\frac{1}{\eps^{2}}\bigg[\frac{1}{6} p^0_{gg}(z)-\frac{11}{12} p^0_{gg,F}(z)\bigg]\nn\\
&+&\frac{1}{\eps}\bigg[\left(\frac{5}{9} +\frac{2}{3} \log (z) \right) {\cal{D}}_0(z)+z^2 +\left(-\frac{4 z^2}{3}-\frac{3 z}{4}-\frac{9}{4}\right)  \log (z)\nn \\
     &-&\frac{4 z}{9}+\frac{11 \delta(1-z)}{24}+\left(-\frac{z}{4}-\frac{1}{4}\right) \log
   ^2(z)-\frac{10}{9}\bigg]\, ,\\
\wh{\wt{\overline{\Gamma}}}_{gg}^{(2)}(z)&=&\frac{1}{\eps}\bigg[-\frac{\delta(1-z)}{8}-\frac{41 z^2}{18}+\left(\frac{2 z^2}{3}+\frac{7
   z}{4}+\frac{2}{3 z}+\frac{5}{4}\right) \log (z) \nn \\
   &+&\frac{3 z}{2}+\frac{23}{18
   z}+\left(\frac{z}{4}+\frac{1}{4}\right) \log ^2(z)-\frac{1}{2}\bigg]\, ,\nn\\
\wh{\wh{\overline{\Gamma}}}_{gg}^{(2)}(z)&=&\frac{1}{\eps^{2}}\bigg[-\frac{\delta(1-z)}{18}\bigg]\, ,
\end{eqnarray}
where
\begin{eqnarray}
{p}_{qq}^{0}(z)&=&{\cal{D}}_{0}(z)-\frac{(1+z)}{2}+\frac{3}{4}\delta(1-z)\label{eq:p0}\, , \nn\\
{p}_{gq}^{0}(z)&=&\frac{1}{z}-1+\frac{z}{2}\, , \nn\\
{p}_{qg}^{0}(z)&=&\frac{1}{2}-z+z^{2}\, , \nn\\
{p}_{gg}^{0}(z)&=&2{\cal{D}}_{0}(z)+\frac{2}{z}-4+2z-2z^{2}+b_{0}\delta(1-z)\, ,\nn\\
{p}_{gg,F}^{0}(z)&=&b_{0,F}\delta(1-z)\, ,\label{eq:LOsplit}
\end{eqnarray}
with $b_0=11/6$ and $b_{0,F}=-1/3$.
Here we have labelled with $x_1$ and $x_2$ the momentum fractions carried by the initial-state partons and with $z$ the momentum fraction carried by the identified final-state parton.  

	\newpage

\end{appendix}

\bibliographystyle{JHEP}
\bibliography{IFfrag}

\providecommand{\href}[2]{#2}\begingroup\raggedright\begin{thebibliography}{10}

\bibitem{Field:1976ve}
R.D.~Field and R.P.~Feynman, \emph{{Quark Elastic Scattering as a Source of
  High Transverse Momentum Mesons}},
  \href{https://doi.org/10.1103/PhysRevD.15.2590}{\emph{Phys. Rev. D}
  {\bfseries 15} (1977) 2590}.

\bibitem{Field:1977fa}
R.D.~Field and R.P.~Feynman, \emph{{A Parametrization of the Properties of
  Quark Jets}}, \href{https://doi.org/10.1016/0550-3213(78)90015-9}{\emph{Nucl.
  Phys. B} {\bfseries 136} (1978) 1}.

\bibitem{Altarelli:1977zs}
G.~Altarelli and G.~Parisi, \emph{{Asymptotic Freedom in Parton Language}},
  \href{https://doi.org/10.1016/0550-3213(77)90384-4}{\emph{Nucl. Phys. B}
  {\bfseries 126} (1977) 298}.

\bibitem{Albino:2008gy}
S.~Albino, \emph{{The Hadronization of partons}},
  \href{https://doi.org/10.1103/RevModPhys.82.2489}{\emph{Rev. Mod. Phys.}
  {\bfseries 82} (2010) 2489}
  [\href{https://arxiv.org/abs/0810.4255}{{\ttfamily 0810.4255}}].

\bibitem{Metz:2016swz}
A.~Metz and A.~Vossen, \emph{{Parton Fragmentation Functions}},
  \href{https://doi.org/10.1016/j.ppnp.2016.08.003}{\emph{Prog. Part. Nucl.
  Phys.} {\bfseries 91} (2016) 136}
  [\href{https://arxiv.org/abs/1607.02521}{{\ttfamily 1607.02521}}].

\bibitem{Albino:2008fy}
S.~Albino, B.A.~Kniehl and G.~Kramer, \emph{{AKK Update: Improvements from New
  Theoretical Input and Experimental Data}},
  \href{https://doi.org/10.1016/j.nuclphysb.2008.05.017}{\emph{Nucl. Phys. B}
  {\bfseries 803} (2008) 42} [\href{https://arxiv.org/abs/0803.2768}{{\ttfamily
  0803.2768}}].

\bibitem{deFlorian:2014xna}
D.~de~Florian, R.~Sassot, M.~Epele, R.J.~Hern\'andez-Pinto and M.~Stratmann,
  \emph{{Parton-to-Pion Fragmentation Reloaded}},
  \href{https://doi.org/10.1103/PhysRevD.91.014035}{\emph{Phys. Rev. D}
  {\bfseries 91} (2015) 014035}
  [\href{https://arxiv.org/abs/1410.6027}{{\ttfamily 1410.6027}}].

\bibitem{Sato:2016wqj}
N.~Sato, J.J.~Ethier, W.~Melnitchouk, M.~Hirai, S.~Kumano and A.~Accardi,
  \emph{{First Monte Carlo analysis of fragmentation functions from
  single-inclusive $e^+ e^-$ annihilation}},
  \href{https://doi.org/10.1103/PhysRevD.94.114004}{\emph{Phys. Rev. D}
  {\bfseries 94} (2016) 114004}
  [\href{https://arxiv.org/abs/1609.00899}{{\ttfamily 1609.00899}}].

\bibitem{Anderle:2015lqa}
D.P.~Anderle, F.~Ringer and M.~Stratmann, \emph{{Fragmentation Functions at
  Next-to-Next-to-Leading Order Accuracy}},
  \href{https://doi.org/10.1103/PhysRevD.92.114017}{\emph{Phys. Rev. D}
  {\bfseries 92} (2015) 114017}
  [\href{https://arxiv.org/abs/1510.05845}{{\ttfamily 1510.05845}}].

\bibitem{Bertone:2017tyb}
{\scshape NNPDF} collaboration, \emph{{A determination of the fragmentation
  functions of pions, kaons, and protons with faithful uncertainties}},
  \href{https://doi.org/10.1140/epjc/s10052-017-5088-y}{\emph{Eur. Phys. J. C}
  {\bfseries 77} (2017) 516}
  [\href{https://arxiv.org/abs/1706.07049}{{\ttfamily 1706.07049}}].

\bibitem{Borsa:2022vvp}
I.~Borsa, R.~Sassot, D.~de~Florian, M.~Stratmann and W.~Vogelsang,
  \emph{{Towards a Global QCD Analysis of Fragmentation Functions at
  Next-to-Next-to-Leading Order Accuracy}},
  \href{https://doi.org/10.1103/PhysRevLett.129.012002}{\emph{Phys. Rev. Lett.}
  {\bfseries 129} (2022) 012002}
  [\href{https://arxiv.org/abs/2202.05060}{{\ttfamily 2202.05060}}].

\bibitem{AbdulKhalek:2022laj}
{\scshape MAP (Multi-dimensional Analyses of Partonic distributions)}
  collaboration, \emph{{Pion and kaon fragmentation functions at
  next-to-next-to-leading order}},
  \href{https://doi.org/10.1016/j.physletb.2022.137456}{\emph{Phys. Lett. B}
  {\bfseries 834} (2022) 137456}
  [\href{https://arxiv.org/abs/2204.10331}{{\ttfamily 2204.10331}}].

\bibitem{Furmanski:1980cm}
W.~Furmanski and R.~Petronzio, \emph{{Singlet Parton Densities Beyond Leading
  Order}}, \href{https://doi.org/10.1016/0370-2693(80)90636-X}{\emph{Phys.
  Lett. B} {\bfseries 97} (1980) 437}.

\bibitem{Almasy:2011eq}
A.A.~Almasy, S.~Moch and A.~Vogt, \emph{{On the Next-to-Next-to-Leading Order
  Evolution of Flavour-Singlet Fragmentation Functions}},
  \href{https://doi.org/10.1016/j.nuclphysb.2011.08.028}{\emph{Nucl. Phys. B}
  {\bfseries 854} (2012) 133}
  [\href{https://arxiv.org/abs/1107.2263}{{\ttfamily 1107.2263}}].

\bibitem{Altarelli:1979kv}
G.~Altarelli, R.K.~Ellis, G.~Martinelli and S.-Y.~Pi, \emph{{Processes
  Involving Fragmentation Functions Beyond the Leading Order in QCD}},
  \href{https://doi.org/10.1016/0550-3213(79)90062-2}{\emph{Nucl. Phys. B}
  {\bfseries 160} (1979) 301}.

\bibitem{Rijken:1996ns}
P.J.~Rijken and W.L.~van Neerven, \emph{{Higher order QCD corrections to the
  transverse and longitudinal fragmentation functions in electron - positron
  annihilation}},
  \href{https://doi.org/10.1016/S0550-3213(96)00669-4}{\emph{Nucl. Phys. B}
  {\bfseries 487} (1997) 233}
  [\href{https://arxiv.org/abs/hep-ph/9609377}{{\ttfamily hep-ph/9609377}}].

\bibitem{Mitov:2006ic}
A.~Mitov, S.~Moch and A.~Vogt, \emph{{Next-to-Next-to-Leading Order Evolution
  of Non-Singlet Fragmentation Functions}},
  \href{https://doi.org/10.1016/j.physletb.2006.05.005}{\emph{Phys. Lett. B}
  {\bfseries 638} (2006) 61}
  [\href{https://arxiv.org/abs/hep-ph/0604053}{{\ttfamily hep-ph/0604053}}].

\bibitem{deFlorian:1997zj}
D.~de~Florian, M.~Stratmann and W.~Vogelsang, \emph{{QCD analysis of
  unpolarized and polarized Lambda baryon production in leading and
  next-to-leading order}},
  \href{https://doi.org/10.1103/PhysRevD.57.5811}{\emph{Phys. Rev. D}
  {\bfseries 57} (1998) 5811}
  [\href{https://arxiv.org/abs/hep-ph/9711387}{{\ttfamily hep-ph/9711387}}].

\bibitem{Goyal:2023xfi}
S.~Goyal, S.-O.~Moch, V.~Pathak, N.~Rana and V.~Ravindran, \emph{{NNLO QCD
  corrections to semi-inclusive DIS}},
  \href{https://arxiv.org/abs/2312.17711}{{\ttfamily 2312.17711}}.

\bibitem{Bonino:2024qbh}
L.~Bonino, T.~Gehrmann and G.~Stagnitto, \emph{{Semi-inclusive deep-inelastic
  scattering at NNLO in QCD}},
  \href{https://arxiv.org/abs/2401.16281}{{\ttfamily 2401.16281}}.

\bibitem{Bonino:2024wgg}
L.~Bonino, T.~Gehrmann, M.~L\"ochner, K.~Sch\"onwald and G.~Stagnitto,
  \emph{{Polarized semi-inclusive deep-inelastic scattering at NNLO in QCD}},
  \href{https://arxiv.org/abs/2404.08597}{{\ttfamily 2404.08597}}.

\bibitem{Goyal:2024tmo}
S.~Goyal, R.N.~Lee, S.-O.~Moch, V.~Pathak, N.~Rana and V.~Ravindran,
  \emph{{NNLO QCD corrections to polarized semi-inclusive DIS}},
  \href{https://arxiv.org/abs/2404.09959}{{\ttfamily 2404.09959}}.

\bibitem{Aversa:1988vb}
F.~Aversa, P.~Chiappetta, M.~Greco and J.P.~Guillet, \emph{{QCD Corrections to
  Parton-Parton Scattering Processes}},
  \href{https://doi.org/10.1016/0550-3213(89)90288-5}{\emph{Nucl. Phys. B}
  {\bfseries 327} (1989) 105}.

\bibitem{Catani:1996vz}
S.~Catani and M.H.~Seymour, \emph{{A General algorithm for calculating jet
  cross-sections in NLO QCD}},
  \href{https://doi.org/10.1016/S0550-3213(96)00589-5}{\emph{Nucl. Phys. B}
  {\bfseries 485} (1997) 291}
  [\href{https://arxiv.org/abs/hep-ph/9605323}{{\ttfamily hep-ph/9605323}}].

\bibitem{Frixione:1995ms}
S.~Frixione, Z.~Kunszt and A.~Signer, \emph{{Three jet cross-sections to
  next-to-leading order}},
  \href{https://doi.org/10.1016/0550-3213(96)00110-1}{\emph{Nucl. Phys. B}
  {\bfseries 467} (1996) 399}
  [\href{https://arxiv.org/abs/hep-ph/9512328}{{\ttfamily hep-ph/9512328}}].

\bibitem{Gehrmann-DeRidder:2005btv}
A.~Gehrmann-De~Ridder, T.~Gehrmann and E.W.N.~Glover, \emph{{Antenna
  subtraction at NNLO}},
  \href{https://doi.org/10.1088/1126-6708/2005/09/056}{\emph{JHEP} {\bfseries
  09} (2005) 056} [\href{https://arxiv.org/abs/hep-ph/0505111}{{\ttfamily
  hep-ph/0505111}}].

\bibitem{Currie:2013vh}
J.~Currie, E.~Glover and S.~Wells, \emph{{Infrared Structure at NNLO Using
  Antenna Subtraction}},
  \href{https://doi.org/10.1007/JHEP04(2013)066}{\emph{JHEP} {\bfseries 04}
  (2013) 066} [\href{https://arxiv.org/abs/1301.4693}{{\ttfamily 1301.4693}}].

\bibitem{DelDuca:2016ily}
V.~Del~Duca, C.~Duhr, A.~Kardos, G.~Somogyi, Z.~Sz\H{o}r, Z.~Tr\'ocs\'anyi
  et~al., \emph{{Jet production in the CoLoRFulNNLO method: event shapes in
  electron-positron collisions}},
  \href{https://doi.org/10.1103/PhysRevD.94.074019}{\emph{Phys. Rev. D}
  {\bfseries 94} (2016) 074019}
  [\href{https://arxiv.org/abs/1606.03453}{{\ttfamily 1606.03453}}].

\bibitem{Catani:2007vq}
S.~Catani and M.~Grazzini, \emph{{An NNLO subtraction formalism in hadron
  collisions and its application to Higgs boson production at the LHC}},
  \href{https://doi.org/10.1103/PhysRevLett.98.222002}{\emph{Phys. Rev. Lett.}
  {\bfseries 98} (2007) 222002}
  [\href{https://arxiv.org/abs/hep-ph/0703012}{{\ttfamily hep-ph/0703012}}].

\bibitem{Czakon:2010td}
M.~Czakon, \emph{{A novel subtraction scheme for double-real radiation at
  NNLO}}, \href{https://doi.org/10.1016/j.physletb.2010.08.036}{\emph{Phys.
  Lett. B} {\bfseries 693} (2010) 259}
  [\href{https://arxiv.org/abs/1005.0274}{{\ttfamily 1005.0274}}].

\bibitem{Czakon:2014oma}
M.~Czakon and D.~Heymes, \emph{{Four-dimensional formulation of the
  sector-improved residue subtraction scheme}},
  \href{https://doi.org/10.1016/j.nuclphysb.2014.11.006}{\emph{Nucl. Phys. B}
  {\bfseries 890} (2014) 152}
  [\href{https://arxiv.org/abs/1408.2500}{{\ttfamily 1408.2500}}].

\bibitem{Gaunt:2015pea}
J.~Gaunt, M.~Stahlhofen, F.J.~Tackmann and J.R.~Walsh, \emph{{N-jettiness
  Subtractions for NNLO QCD Calculations}},
  \href{https://doi.org/10.1007/JHEP09(2015)058}{\emph{JHEP} {\bfseries 09}
  (2015) 058} [\href{https://arxiv.org/abs/1505.04794}{{\ttfamily
  1505.04794}}].

\bibitem{Cacciari:2015jma}
M.~Cacciari, F.A.~Dreyer, A.~Karlberg, G.P.~Salam and G.~Zanderighi,
  \emph{{Fully Differential Vector-Boson-Fusion Higgs Production at
  Next-to-Next-to-Leading Order}},
  \href{https://doi.org/10.1103/PhysRevLett.115.082002}{\emph{Phys. Rev. Lett.}
  {\bfseries 115} (2015) 082002}
  [\href{https://arxiv.org/abs/1506.02660}{{\ttfamily 1506.02660}}].

\bibitem{Caola:2017dug}
F.~Caola, K.~Melnikov and R.~R\"ontsch, \emph{{Nested soft-collinear
  subtractions in NNLO QCD computations}},
  \href{https://doi.org/10.1140/epjc/s10052-017-4774-0}{\emph{Eur. Phys. J. C}
  {\bfseries 77} (2017) 248}
  [\href{https://arxiv.org/abs/1702.01352}{{\ttfamily 1702.01352}}].

\bibitem{Magnea:2018hab}
L.~Magnea, E.~Maina, G.~Pelliccioli, C.~Signorile-Signorile, P.~Torrielli and
  S.~Uccirati, \emph{{Local analytic sector subtraction at NNLO}},
  \href{https://doi.org/10.1007/JHEP12(2018)107}{\emph{JHEP} {\bfseries 12}
  (2018) 107} [\href{https://arxiv.org/abs/1806.09570}{{\ttfamily
  1806.09570}}].

\bibitem{Bertolotti:2022aih}
G.~Bertolotti, L.~Magnea, G.~Pelliccioli, A.~Ratti, C.~Signorile-Signorile,
  P.~Torrielli et~al., \emph{{NNLO subtraction for any massless final state: a
  complete analytic expression}},
  \href{https://doi.org/10.1007/JHEP07(2023)140}{\emph{JHEP} {\bfseries 07}
  (2023) 140} [\href{https://arxiv.org/abs/2212.11190}{{\ttfamily
  2212.11190}}].

\bibitem{Devoto:2023rpv}
F.~Devoto, K.~Melnikov, R.~R\"ontsch, C.~Signorile-Signorile and
  D.M.~Tagliabue, \emph{{A fresh look at the nested soft-collinear subtraction
  scheme: NNLO QCD corrections to N-gluon final states in $ q\overline{q} $
  annihilation}}, \href{https://doi.org/10.1007/JHEP02(2024)016}{\emph{JHEP}
  {\bfseries 02} (2024) 016}
  [\href{https://arxiv.org/abs/2310.17598}{{\ttfamily 2310.17598}}].

\bibitem{Czakon:2021ohs}
M.L.~Czakon, T.~Generet, A.~Mitov and R.~Poncelet, \emph{{B-hadron production
  in NNLO QCD: application to LHC t$ \overline{t} $ events with leptonic
  decays}}, \href{https://doi.org/10.1007/JHEP10(2021)216}{\emph{JHEP}
  {\bfseries 10} (2021) 216}
  [\href{https://arxiv.org/abs/2102.08267}{{\ttfamily 2102.08267}}].

\bibitem{Gehrmann:2022cih}
T.~Gehrmann and R.~Sch\"urmann, \emph{{Photon fragmentation in the antenna
  subtraction formalism}},
  \href{https://doi.org/10.1007/JHEP04(2022)031}{\emph{JHEP} {\bfseries 04}
  (2022) 031} [\href{https://arxiv.org/abs/2201.06982}{{\ttfamily
  2201.06982}}].

\bibitem{Chen:2022gpk}
X.~Chen, T.~Gehrmann, E.W.N.~Glover, M.~H\"ofer, A.~Huss and R.~Sch\"urmann,
  \emph{{Single photon production at hadron colliders at NNLO QCD with
  realistic photon isolation}},
  \href{https://doi.org/10.1007/JHEP08(2022)094}{\emph{JHEP} {\bfseries 08}
  (2022) 094} [\href{https://arxiv.org/abs/2205.01516}{{\ttfamily
  2205.01516}}].

\bibitem{Daleo:2006xa}
A.~Daleo, T.~Gehrmann and D.~Maitre, \emph{{Antenna subtraction with hadronic
  initial states}},
  \href{https://doi.org/10.1088/1126-6708/2007/04/016}{\emph{JHEP} {\bfseries
  04} (2007) 016} [\href{https://arxiv.org/abs/hep-ph/0612257}{{\ttfamily
  hep-ph/0612257}}].

\bibitem{Gehrmann:2022pzd}
T.~Gehrmann and G.~Stagnitto, \emph{{Antenna subtraction at NNLO with
  identified hadrons}},
  \href{https://doi.org/10.1007/JHEP10(2022)136}{\emph{JHEP} {\bfseries 10}
  (2022) 136} [\href{https://arxiv.org/abs/2208.02650}{{\ttfamily
  2208.02650}}].

\bibitem{Gehrmann:2023dxm}
T.~Gehrmann, E.W.N.~Glover and M.~Marcoli, \emph{{The colourful antenna
  subtraction method}},
  \href{https://doi.org/10.1007/JHEP03(2024)114}{\emph{JHEP} {\bfseries 03}
  (2024) 114} [\href{https://arxiv.org/abs/2310.19757}{{\ttfamily
  2310.19757}}].

\bibitem{Gehrmann-DeRidder:2004ttg}
A.~Gehrmann-De~Ridder, T.~Gehrmann and E.W.N.~Glover, \emph{{Infrared structure
  of $e^+ e^-\to$ 2 jets at NNLO}},
  \href{https://doi.org/10.1016/j.nuclphysb.2004.05.017}{\emph{Nucl. Phys. B}
  {\bfseries 691} (2004) 195}
  [\href{https://arxiv.org/abs/hep-ph/0403057}{{\ttfamily hep-ph/0403057}}].

\bibitem{Gehrmann-DeRidder:2005svg}
A.~Gehrmann-De~Ridder, T.~Gehrmann and E.W.N.~Glover, \emph{{Quark-gluon
  antenna functions from neutralino decay}},
  \href{https://doi.org/10.1016/j.physletb.2005.02.039}{\emph{Phys. Lett. B}
  {\bfseries 612} (2005) 36}
  [\href{https://arxiv.org/abs/hep-ph/0501291}{{\ttfamily hep-ph/0501291}}].

\bibitem{Gehrmann-DeRidder:2005alt}
A.~Gehrmann-De~Ridder, T.~Gehrmann and E.W.N.~Glover, \emph{{Gluon-gluon
  antenna functions from Higgs boson decay}},
  \href{https://doi.org/10.1016/j.physletb.2005.03.003}{\emph{Phys. Lett. B}
  {\bfseries 612} (2005) 49}
  [\href{https://arxiv.org/abs/hep-ph/0502110}{{\ttfamily hep-ph/0502110}}].

\bibitem{Braun-White:2023sgd}
O.~Braun-White, N.~Glover and C.T.~Preuss, \emph{{A general algorithm to build
  real-radiation antenna functions for higher-order calculations}},
  \href{https://doi.org/10.1007/JHEP06(2023)065}{\emph{JHEP} {\bfseries 06}
  (2023) 065} [\href{https://arxiv.org/abs/2302.12787}{{\ttfamily
  2302.12787}}].

\bibitem{Braun-White:2023zwd}
O.~Braun-White, N.~Glover and C.T.~Preuss, \emph{{A general algorithm to build
  mixed real and virtual antenna functions for higher-order calculations}},
  \href{https://doi.org/10.1007/JHEP11(2023)179}{\emph{JHEP} {\bfseries 11}
  (2023) 179} [\href{https://arxiv.org/abs/2307.14999}{{\ttfamily
  2307.14999}}].

\bibitem{Fox:2023bma}
E.~Fox and N.~Glover, \emph{{Initial-final and initial-initial antenna
  functions for real radiation at next-to-leading order}},
  \href{https://doi.org/10.1007/JHEP12(2023)171}{\emph{JHEP} {\bfseries 12}
  (2023) 171} [\href{https://arxiv.org/abs/2308.10829}{{\ttfamily
  2308.10829}}].

\bibitem{Jakubcik:2022zdi}
P.~Jakub\v{c}\'\i{}k, M.~Marcoli and G.~Stagnitto, \emph{{The parton-level
  structure of $e^+e^-$ to 2 jets at N3LO}},
  \href{https://doi.org/10.1007/JHEP01(2023)168}{\emph{JHEP} {\bfseries 01}
  (2023) 168} [\href{https://arxiv.org/abs/2211.08446}{{\ttfamily
  2211.08446}}].

\bibitem{Chen:2023fba}
X.~Chen, P.~Jakub\v{c}\'\i{}k, M.~Marcoli and G.~Stagnitto, \emph{{The
  parton-level structure of Higgs decays to hadrons at N3LO}},
  \href{https://doi.org/10.1007/JHEP06(2023)185}{\emph{JHEP} {\bfseries 06}
  (2023) 185} [\href{https://arxiv.org/abs/2304.11180}{{\ttfamily
  2304.11180}}].

\bibitem{Chen:2023egx}
X.~Chen, P.~Jakub\v{c}\'\i{}k, M.~Marcoli and G.~Stagnitto, \emph{{Radiation
  from a gluon-gluino colour-singlet dipole at N3LO}},
  \href{https://doi.org/10.1007/JHEP12(2023)198}{\emph{JHEP} {\bfseries 12}
  (2023) 198} [\href{https://arxiv.org/abs/2310.13062}{{\ttfamily
  2310.13062}}].

\bibitem{Daleo:2009yj}
A.~Daleo, A.~Gehrmann-De~Ridder, T.~Gehrmann and G.~Luisoni, \emph{{Antenna
  subtraction at NNLO with hadronic initial states: initial-final
  configurations}}, \href{https://doi.org/10.1007/JHEP01(2010)118}{\emph{JHEP}
  {\bfseries 01} (2010) 118} [\href{https://arxiv.org/abs/0912.0374}{{\ttfamily
  0912.0374}}].

\bibitem{Gehrmann:2011wi}
T.~Gehrmann and P.F.~Monni, \emph{{Antenna subtraction at NNLO with hadronic
  initial states: real-virtual initial-initial configurations}},
  \href{https://doi.org/10.1007/JHEP12(2011)049}{\emph{JHEP} {\bfseries 12}
  (2011) 049} [\href{https://arxiv.org/abs/1107.4037}{{\ttfamily 1107.4037}}].

\bibitem{Gehrmann-DeRidder:2012too}
A.~Gehrmann-De~Ridder, T.~Gehrmann and M.~Ritzmann, \emph{{Antenna subtraction
  at NNLO with hadronic initial states: double real initial-initial
  configurations}}, \href{https://doi.org/10.1007/JHEP10(2012)047}{\emph{JHEP}
  {\bfseries 10} (2012) 047} [\href{https://arxiv.org/abs/1207.5779}{{\ttfamily
  1207.5779}}].

\bibitem{Chen:2022clm}
X.~Chen, T.~Gehrmann, E.W.N.~Glover and J.~Mo, \emph{{Antenna subtraction for
  jet production observables in full colour at NNLO}},
  \href{https://doi.org/10.1007/JHEP10(2022)040}{\emph{JHEP} {\bfseries 10}
  (2022) 040} [\href{https://arxiv.org/abs/2208.02115}{{\ttfamily
  2208.02115}}].

\bibitem{Chen:2022ktf}
X.~Chen, T.~Gehrmann, E.W.N.~Glover, A.~Huss and M.~Marcoli, \emph{{Automation
  of antenna subtraction in colour space: gluonic processes}},
  \href{https://doi.org/10.1007/JHEP10(2022)099}{\emph{JHEP} {\bfseries 10}
  (2022) 099} [\href{https://arxiv.org/abs/2203.13531}{{\ttfamily
  2203.13531}}].

\bibitem{Chetyrkin:1981qh}
K.G.~Chetyrkin and F.V.~Tkachov, \emph{{Integration by Parts: The Algorithm to
  Calculate beta Functions in 4 Loops}},
  \href{https://doi.org/10.1016/0550-3213(81)90199-1}{\emph{Nucl. Phys. B}
  {\bfseries 192} (1981) 159}.

\bibitem{Laporta:2000dsw}
S.~Laporta, \emph{{High precision calculation of multiloop Feynman integrals by
  difference equations}},
  \href{https://doi.org/10.1142/S0217751X00002159}{\emph{Int. J. Mod. Phys. A}
  {\bfseries 15} (2000) 5087}
  [\href{https://arxiv.org/abs/hep-ph/0102033}{{\ttfamily hep-ph/0102033}}].

\bibitem{vonManteuffel:2012np}
A.~von Manteuffel and C.~Studerus, \emph{{Reduze 2 - Distributed Feynman
  Integral Reduction}},  \href{https://arxiv.org/abs/1201.4330}{{\ttfamily
  1201.4330}}.

\bibitem{Gehrmann:1999as}
T.~Gehrmann and E.~Remiddi, \emph{{Differential equations for two loop four
  point functions}},
  \href{https://doi.org/10.1016/S0550-3213(00)00223-6}{\emph{Nucl. Phys. B}
  {\bfseries 580} (2000) 485}
  [\href{https://arxiv.org/abs/hep-ph/9912329}{{\ttfamily hep-ph/9912329}}].

\bibitem{Catani:1998bh}
S.~Catani, \emph{{The Singular behavior of QCD amplitudes at two loop order}},
  \href{https://doi.org/10.1016/S0370-2693(98)00332-3}{\emph{Phys. Lett. B}
  {\bfseries 427} (1998) 161}
  [\href{https://arxiv.org/abs/hep-ph/9802439}{{\ttfamily hep-ph/9802439}}].

\bibitem{Caletti:2024xaw}
S.~Caletti, A.~Gehrmann-De~Ridder, A.~Huss, A.R.~Garcia and G.~Stagnitto,
  \emph{{QCD predictions for vector boson plus hadron production at the LHC}},
  \href{https://arxiv.org/abs/2405.17540}{{\ttfamily 2405.17540}}.

\bibitem{Caletti:2022hnc}
S.~Caletti, A.J.~Larkoski, S.~Marzani and D.~Reichelt, \emph{{Practical jet
  flavour through NNLO}},
  \href{https://doi.org/10.1140/epjc/s10052-022-10568-7}{\emph{Eur. Phys. J. C}
  {\bfseries 82} (2022) 632}
  [\href{https://arxiv.org/abs/2205.01109}{{\ttfamily 2205.01109}}].

\bibitem{Czakon:2022wam}
M.~Czakon, A.~Mitov and R.~Poncelet, \emph{{Infrared-safe flavoured
  anti-k$_{T}$ jets}},
  \href{https://doi.org/10.1007/JHEP04(2023)138}{\emph{JHEP} {\bfseries 04}
  (2023) 138} [\href{https://arxiv.org/abs/2205.11879}{{\ttfamily
  2205.11879}}].

\bibitem{Gauld:2022lem}
R.~Gauld, A.~Huss and G.~Stagnitto, \emph{{Flavor Identification of
  Reconstructed Hadronic Jets}},
  \href{https://doi.org/10.1103/PhysRevLett.130.161901}{\emph{Phys. Rev. Lett.}
  {\bfseries 130} (2023) 161901}
  [\href{https://arxiv.org/abs/2208.11138}{{\ttfamily 2208.11138}}].

\bibitem{Caola:2023wpj}
F.~Caola, R.~Grabarczyk, M.L.~Hutt, G.P.~Salam, L.~Scyboz and J.~Thaler,
  \emph{{Flavored jets with exact anti-kt kinematics and tests of infrared and
  collinear safety}},
  \href{https://doi.org/10.1103/PhysRevD.108.094010}{\emph{Phys. Rev. D}
  {\bfseries 108} (2023) 094010}
  [\href{https://arxiv.org/abs/2306.07314}{{\ttfamily 2306.07314}}].

\bibitem{lewin:1981}
L.~Lewin, \emph{Polylogarithms and associated functions}, North Holland,
  Amsterdam (1981).

\bibitem{Curci:1980uw}
G.~Curci, W.~Furmanski and R.~Petronzio, \emph{{Evolution of Parton Densities
  Beyond Leading Order: The Nonsinglet Case}},
  \href{https://doi.org/10.1016/0550-3213(80)90003-6}{\emph{Nucl. Phys. B}
  {\bfseries 175} (1980) 27}.

\bibitem{Rijken:1996vr}
P.J.~Rijken and W.L.~van Neerven, \emph{{O$(\alpha_s^2)$ contributions to the
  longitudinal fragmentation function in $e^+ e^-$ annihilation}},
  \href{https://doi.org/10.1016/0370-2693(96)00898-2}{\emph{Phys. Lett. B}
  {\bfseries 386} (1996) 422}
  [\href{https://arxiv.org/abs/hep-ph/9604436}{{\ttfamily hep-ph/9604436}}].

\end{thebibliography}\endgroup

\end{document}